\documentclass[a4paper,10pt]{article}
\usepackage{a4wide,bm,epsfig,booktabs}
\usepackage{setspace} 
\usepackage{colortbl}
\usepackage{caption}
\usepackage{subcaption}
\usepackage{relsize} 
\usepackage{graphicx}
\usepackage{cancel}  
\usepackage{slashed}
\usepackage{verbatim}   
\usepackage{mathtools}
\usepackage[flushmargin,hang]{footmisc} 
\usepackage{amsmath,enumerate}
\usepackage[ntheorem]{defoqua} 
\usepackage{stmaryrd}  
\usepackage{cite}
\usepackage{fancyheadings} 
\usepackage[arrow, matrix,curve]{xy}
\usepackage{array} 
\usepackage{color}
\usepackage[usenames,dvipsnames]{xcolor}
\usepackage{hyperref}

\let\origenumerate\enumerate
\def\enumerate{\origenumerate\itemsep0pt}
\let\origitemize\itemize
\def\itemize{\origitemize\itemsep0pt} 

\input xypic

\newcommand{\itspace} {\vspace{0pt}}
\newcommand{\itspacec} {\vspace{0pt}}

\newcommand{\murs} {\mathfrak{z}}

\newcommand{\Homm} {\mathrm{Hom}}
\newcommand{\zd} {k}  

\newcommand{\Gel} {\mathcal{G}}  

\newcommand{\dd} {\mathrm{d}}
\newcommand{\RB} {\mathbb{R}_{\mathrm{Bohr}}}
\newcommand{\RR} {\mathbb{R}}
\newcommand{\CCC} {\mathbb{C}}
\newcommand{\NN} {\mathbb{N}}

\newcommand{\SU} {\mathrm{SU}(2)}
\newcommand{\SO} {\mathrm{SO}(3)}

\newcommand{\csim} {\sim_\Con}
\newcommand{\isim} {\sim_\im}
\newcommand{\psim} {\sim_{\mathrm{par}}}

\newcommand{\w} {\omega}
\newcommand{\Con} {\mathcal{A}}
\newcommand{\AR} {\Con_{\mathrm{red}}}
\newcommand{\A} {\ovl{\Con}}
\newcommand{\mg} {\mathfrak{g}}

\newcommand{\vv} {\vec{v}}

\newcommand{\IndA} {\vartheta}
\newcommand{\Paths} {\mathcal{P}}
\newcommand{\mm} {\tau}
\newcommand{\parall}[2] {\mathcal{P}_{#1}^{#2}}
\newcommand{\Cyls}[1] {\gG_{\mathcal{P}_{#1}}}
\newcommand{\Cylsk} {\gG_{\mathcal{P}}}
\newcommand{\Cylk} {\cC_{\mathcal{P}}}
\newcommand{\Cyl}[1] {\cC_{#1}}

\newcommand{\AL} {\mathrm{AL}}
\newcommand{\uberl} {\varrho}
\newcommand{\uberll}[2] {#1(#2)}
\newcommand{\Co}[1] {\alpha_{#1}}
\newcommand{\ic}[1] {\omega^c}
\newcommand{\gc}[4] {\gamma_{\vec{#1},\vec{#2}}^{#3,#4}}
\newcommand{\gcc}[4] {\gamma_{#1,#2}^{#3,#4}}
\newcommand{\ishomc} {\mathfrak{v}}
\newcommand{\homm} {\varepsilon}
\newcommand{\hommm} {\epsilon}
\newcommand{\Autt} {\mathrm{Aut}}
\newcommand{\Iso} {\mathrm{Mor}}
\newcommand{\AF} {\Iso_{\mathrm{F}}}

\newcommand{\Ge}{E}
\newcommand{\Gee} {\RR^3 \rtimes_\uberl \SU} 
\newcommand{\pc}[1] {\beta_{#1}}
\newcommand{\GB} {G_{\mathrm{Bohr}}}
\newcommand{\CAP} {C_{\mathrm{AP}}} 

\newcommand{\iB} {i_{\mathrm{B}}}
\newcommand{\Homeo} {\mathrm{Homeo}}

\newcommand{\CC}[1] {C^{#1}}
\newcommand{\X} {\ovl{X}}
\newcommand{\x} {\ovl{x}}
\newcommand{\oRR} {Y}

\newcommand{\qR} {\ovl{\RR}}
\newcommand{\lin} {\mathrm{l}}
\newcommand{\mc} {\mathrm{c}}
\newcommand{\dt}[2] {\frac{\dd}{\dd#1}\Big|_{#1=#2}}
\newcommand{\Cstar} {C^*}

\newcommand{\vg} {\widetilde{\gamma}}
\newcommand{\vw} {\widetilde{\w}}
\newcommand{\vp} {\widetilde{p}}

\newcommand{\vpdotw} {\dot{\wt{\gamma}}\hspace{-5.5pt}\phantom{\wt{\gamma}}_p^\w(t)}

\newcommand{\aA} {\mathfrak{A}}
\newcommand{\bB} {\mathfrak{B}}
\newcommand{\cC} {\mathfrak{A}}
\newcommand{\gG} {\mathfrak{G}}
\newcommand{\dD} {\mathfrak{D}}

\newcommand{\wt}[1] {\widetilde{#1}}

\newcommand{\ovl}[1]{\overline{#1}} 
\newcommand{\dom}{\mathrm{dom}} 
\newcommand{\im}{\mathrm{im}} 
\newcommand{\cp} {\circ}
\newcommand{\Spec}{\mathrm{Spec}}

\newcommand{\Borel}{\mathfrak{B}}

\newcommand{\Lzw}[2] {L^{2}\left(#1,#2\right)}

\newcommand{\BRq}{\Borel\raisebox{0.2ex}{$($}\raisebox{-0.1ex}{$\qR$}\raisebox{0.2ex}{$)$}}

\begin{document} 

\title{Invariant Connections in Loop Quantum Gravity}
\author{Maximilian Hanusch\thanks{e-mail:
    {\tt hanuschm@fau.edu}}\\   
  \\
  {\normalsize\em Institut f\"ur Mathematik}\\[-0.15ex]
  {\normalsize\em Universit\"at Paderborn}\\[-0.15ex]
  {\normalsize\em Warburger Stra\ss e 100}\\[-0.15ex]
  {\normalsize\em 33098 Paderborn}\\[-0.15ex]
  {\normalsize\em Germany}} 
\date{March 9, 2016}
\maketitle

\frenchspacing

\begin{abstract}
Given a group $G$ and an abelian $\Cstar$-algebra $\aA$, the antihomomorphisms $\Theta\colon G\rightarrow \Autt(\aA)$ are in one-to-one with those left actions $\Phi\colon G\times \Spec(\aA)\rightarrow \Spec(\aA)$ whose translation maps $\Phi_g$ are continuous; whereby  continuities of $\Theta$ and $\Phi$ turn out to be equivalent if $\aA$ is unital.  
In particular, a left action $\phi\colon G \times X\rightarrow X$ can be uniquely extended to the spectrum of a $\Cstar$-subalgebra $\aA$ of the bounded functions on $X$ if $\phi_g^*(\aA)\subseteq \aA$ holds for each $g\in G$. 
In the present paper, we apply this to the framework of loop quantum gravity. We show that, on the level of the configuration spaces, quantization and reduction in general do not commute, i.e., that the symmetry-reduced quantum configuration space is (strictly) larger than the quantized configuration space of the reduced classical theory. Here, the quantum-reduced space has the advantage to be completely characterized by a simple algebraic relation, whereby the quantized reduced classical space is usually hard to compute.
\end{abstract}

\tableofcontents 
\section{Introduction}
\label{Intro}
Invariant connections on principal fibre bundles play a key role in 
symmetry reduction in the framework of loop quantum gravity (LQG). There, 
symmetries usually are represented by Lie groups of automorphisms of the $\SU$-bundle of interest; and traditionally the set of invariant connections is quantized in order to obtain the quantum configuration space of the reduced theory. Here, as for full LQG, quantization of the configuration space just means to consider the spectrum of a separating $\Cstar$-subalgebra of the bounded functions thereon.\footnote{For works addressing the interaction of symmmetry reduction and quantization in the broader sense involving also state spaces, operators thereon, and constraints, see for example \cite{JonEng0,JonEng1,JonEng2,TKOS}, as well as forthcoming works by the author.}
These algebras are usually generated by matrix entries of parallel transports along certain curves in the base manifold, so that one has to take the following issues into account:
\begingroup
\setlength{\leftmargini}{12pt}
\begin{itemize}
\item
 \itspacec
  Let $\aA$ and $\aA_{\mathrm{red}}$ denote the $\Cstar$-algebras used to define the quantum configuration spaces of the full and that of the reduced theory, respectively. Then, if one wants the reduced quantum configuration space 
  to be canonically embedded into that of full LQG, one has to ensure that the restriction of the $\Cstar$-algebra $\aA$ to the set of invariant connections $\Con_{\mathrm{red}}$, is contained in $\aA_{\mathrm{red}}$. \cite{ChrisSymmLQG} 
  Thus, here the most natural thing to do is to define $\aA_{\mathrm{red}}:=\ovl{\aA|_{\Con_{\mathrm{red}}}}$; the closure of  $\aA|_{\AR}$ in the bounded functions on $\AR$ w.r.t.\ the uniform norm.
\item 
  \itspace
  The quantized spaces depend very sensitively on the set of curves used to define the $\Cstar$-algebras $\aA$ and $\aA_{\mathrm{red}}$. For instance, in  homogeneous isotropic loop quantum cosmology (LQC), traditionally the set of linear curves is used to define the reduced quantum configuration space. \cite{Ashtekar2008} This leads to the Bohr compactification $\RB$ of $\RR$; and, adjoining one single circular curve,  
  already inflates this space to $\RR\sqcup \RB$ (cf.\ Lemma \ref{lemma:Circular}) on which  a Haar measure no longer exists, cf.\ Theorem \ref{theorem:noGroupStruc}.

{\it Comment: 
  During the evaluation of this article, the techniques developed here allowed to single out the Bohr measure on both $\RB$ and $\RR\sqcup\RB$ -- hence, the standard kinematical Hilbert space of homogeneous isotropic loop quantum cosmology (LQC) -- by means of the same invariance condition. \cite{MEAS}}
\item    
\itspace   
There are two non-trivial steps to execute. First of all, one has to calculate the set of invariant connections that correspond to the Lie group of automorphisms representing the symmetry of interest.  
  Here, Wang's theorem \cite{Wang} can be applied if the action induced on the base manifold is transitive; and in the general case, \cite{InvConn} can be used. Although, especially in the general case, this can be very difficult,  it is usually even much more complicated to calculate the spectrum of $\aA_{\mathrm{red}}$, which is the second step. For homogeneous isotropic LQC, this is easy if one only takes linear curves into account. But, then there exists no canonical embedding of $\RB$ into the quantum configuration space of the full theory. This arises from the fact that the latter space is defined by using all embedded analytic curves instead of only just the linear ones. \cite{Brunnhack} Now, if one uses embedded analytic curves also for the reduced space, the calculation is narrowly performable, because $\Con_{\mathrm{red}}\cong \RR$ holds. \cite{ChrisSymmLQG} Also for the homogeneous case, this seems not to be completely hopeless, because there $\Con_{\mathrm{red}}$ is in bijection with $\mathrm{End}(\RR^3)$, which is still a finite dimensional vector space. 
  However, already in the spherically symmetric (invariance under rotations) or in the semi-homogeneous case (invariance under translations w.r.t.\ a fixed subspace of $\RR^3$), things seem to be impossible, as $\Con_{\mathrm{red}}$ is parametrized by functions in this case. \cite{InvConn}
\end{itemize}
\endgroup
\noindent
Moreover, since the above procedure means to quantize a reduced classical space, the question arises, whether there might be some loss of information in comparison with reducing the quantum space directly.
Thus, also in view of the last two points, one might look for an alternative, more handy approach, allowing for a symmetry reduction directly on the quantum level. 

In this paper, we provide such a reduction concept by investigating under which assumptions a left action of a group on a set can be extended to an action on the spectrum of a $\Cstar$-subalgebra of the bounded functions on this set. 
The crucial observation then is that a Lie group of automorphisms $G$ of a principal fibre bundle, defines a left action $\phi$ on the corresponding set of smooth connections $\Con$ in such a way that 
\begin{align*}
	\Con_{\mathrm{red}}=\{\w\in \Con\: |\: \phi(g,\w)=\w\:\:\:\forall\:g\in G\}
\end{align*}
holds, i.e., that $\AR$ is the fix point set of $\phi$. 
Then, if $\Phi$ is an extension of this action to $\Spec(\aA)$, we can define the reduced quantum space $\ovl{\Con}_{\mathrm{red}}$ to consist of all invariant generalized connections, i.e., to be the fix point set of
\begin{align*}
\hspace{25pt}\ovl{\Con}_{\mathrm{red}}:=\{\ovl{\w}\in \Spec(\aA)\:|\: \Phi(g,\ovl{\w})=\ovl{\w}\:\:\:\forall\:g\in G\}
\end{align*}
of $\Phi$.
We show that this is possible in a unique way, if the set of curves that is used to define $\aA$, is invariant in a certain sense. 
We will see that this procedure always leads to a larger reduced quantum configuration space, which is even strictly larger in the analytic  cosmological case.\footnote{More precisely, this will be proven for homogeneous isotropic LQC. For homogeneous LQC, non-commutativity already holds on the level of linear curves, cf.\ Example 5.17 in \cite{MAXTH}.} Then, $\ovl{\Con}_{\mathrm{red}}\subseteq \A$ is  characterized by a simple invariance condition, leading to the notion of an invariant homomorphism when using the standard identification of elements in $\A$ with homomorphisms of paths. 
Along the way, we also show that the quantized classical configuration space $\ovl{\AR}$ occurs to be the closure of $\AR$ in $\A$, providing an alternative description of this space.
\vspace{6pt}

\noindent
This paper is organized as follows:
\begingroup
\setlength{\leftmargini}{12pt}
\begin{itemize}
\item
  Section \ref{sec:prelimin} contains the notations, basic definitions, and some of their elementary implications. 
\item
  In Section \ref{sec:SPecExtGr}, we discuss, under which conditions a left action $\phi\colon G\times X\rightarrow X$ of a group $G$ on a set $X$ extends to an action on the spectrum of a $\Cstar$-subalgebra of the bounded functions on $X$. For this, we first investigate the relations between (continuous) group actions $\Phi\colon G\times \Spec(\aA)\rightarrow \Spec(\aA)$ and
  $\Cstar$-dynamical systems $(\aA,G,\Theta)$, for $\aA$ an abelian $\Cstar$-algebra, and an (continuous) antihomomorphism $\Theta\colon G\rightarrow \Autt(\aA)$. 
  Then, we adapt this to the framework of loop quantum gravity, where $X$ is the set of smooth connections on a principal fibre bundle $P$ with compact structure group, $\aA$ is some $\Cstar$-algebra of cylindrical functions defined by some invariant set of curves in the base manifold, and where $\phi$ arises from some Lie group of automorphisms of $P$.
\item
  In Section 4, we show that quantization and reduction of the LQG configuration space in general do not commute, and    
prove that there cannot exist any Haar measure on the LQC configuration space $\RR\sqcup \RB$ introduced in \cite{ChrisSymmLQG}. In addition to that, we discuss some further measure theoretical aspects of the reduced spaces. 
\end{itemize}
\endgroup
\section{Preliminaries}
\label{sec:prelimin}
Let us start with fixing the notations.
\subsection{Notations}
Manifolds are always assumed to be smooth or analytic. 
If $M$,$N$ are manifolds, and $f\colon M\rightarrow N$ a smooth map, then $\dd f\colon TM\rightarrow TN$ denotes the differential map between their tangent manifolds.
The map $f$ is said to be an immersion iff for each $x\in M$, the restriction $\dd_xf:=\dd f|_{T_xM}\colon T_xM\rightarrow T_{f(x)}N$ is injective.
If $V$ is a finite dimensional vector space, 
a $V$-valued 1-form on a manifold $N$, is a smooth map $\w\colon TN\rightarrow V$ whose restriction $\w_y:=\w|_{T_yN}$ is linear for all $y\in N$.
The pullback of $\w$ by $f$ is the map
\begin{align*}
f^*\w\colon TM\rightarrow V,\quad \vv_x\rightarrow \w_{f(x)}(\dd_x f(\vv_x)).
\end{align*}
A curve is a continuous map $\gamma\colon D\rightarrow M$, for $D\subseteq \RR$ an interval, i.e., of the form $(a,b)$, $(a,b]$, $[a,b)$ or $[a,b]$ for some $a<b$. Then, $\gamma$ is said to be embedded analytic if $\gamma=\gamma'|_D$ holds for an analytic immersive embedding (in the sense of maps between manifolds) $\gamma'\colon I\rightarrow M$.\footnote{In the following, $D$ will always denote an arbitrary, $I,J$ an open, and $K$ some compact interval.} Similarly, $\gamma$ is said to be of class 
$\CC{k}$ iff $\gamma'|_{D}=\gamma$ holds for a $\CC{k}$-map $\gamma'\colon I\rightarrow M$ defined on an open interval $I$. Here, we will allow $k\in \{\mathbb{N}_{\geq 1},\infty,\w\}$, whereby $\w$ means analytic. 
For the case that $k\neq \infty$, or $t$ is not contained in the interior of $D$, we define the tangent vector $\dot\gamma(t) \in T_{\gamma(t)}M$ in the canonical way. Finally, a curve $\gamma$ is called {\bf piecewise} $\CC{k}$ or {\bf $\CC{k}$-path} iff there exist real numbers $a=t_0 <{\dots} <t_k=b$, such that $\gamma_i:=\gamma|_{[t_i,t_{i+1}]}$ is of class $\CC{k}$ for each $0\leq i\leq k-1$. The family $\{\gamma_i\}_{0\leq i\leq k-1}$ will be called {\bf decomposition} of $\gamma$.

If $\Xi\colon G\times X\rightarrow X$ is a left action of a group $G$ on a set $X$, we define the maps 
\begin{align*}
&\Xi_g\colon X\rightarrow X,\quad x\mapsto \Xi(g,x)\qquad\text{for each}\qquad g\in G,\\[2pt]
&\Xi_x\colon G\rightarrow X,\quad \hspace{1pt}g\mapsto\Xi(g,x)\qquad\hspace{1pt}\text{for each}\qquad x\in X,
\end{align*}
and, if it is clear which action is meant, we will usually write $g\cdot x$ instead of $\Xi(g,x)$. 

If nothing different is said, a tuple $(G,\Xi)$ will always denote a left action of a Lie group $G$ on a manifold that will be specified in the particular cases. 
We will denote the Lie algebra of $G$ by $\mg$, and define $\Co{g}\colon G\rightarrow G$,\: $h\mapsto g \cdot h\cdot g^{-1}$. The differential $\dd_e\Co{g}\colon \mathfrak{g}\rightarrow \mathfrak{g}$ of $\Co{g}$ at the identity $e\in G$, will be denoted by $\Ad(g)$. 

\subsection{Principal fibre bundles and invariant connections}
Let $\pi\colon P\rightarrow M$ be a smooth map between the manifolds $P$ and $M$, and denote by $F_x:=\pi^{-1}(x)\subseteq P$ the fibre over $x\in M$ in $P$. Moreover, let $S$ be a Lie group that acts via $R\colon P\times S\rightarrow P$ from the right on $P$. 
If there exists an open covering $\{U_\alpha\}_{\alpha\in I}$ of $M$, and a family $\{\phi_\alpha\}_{\alpha\in I}$ of diffeomorphisms
$\phi_\alpha\colon \pi^{-1}(U_\alpha)\rightarrow U_\alpha\times S$ with
\begin{align}
  \label{eq:bundlemaps}
  \phi_\alpha(p\cdot s)=\big(\pi(p),[\pr_2\cp\phi_\alpha](p)\cdot s\big) \qquad \forall\:p\in \pi^{-1}(U_\alpha),\quad \forall\: s\in S,
\end{align}
then $(P,\pi,M,S)$ is called {\bf principal fibre bundle} with total space $P$, projection map $\pi$, base manifold $M$, and structure group $S$. Here, $\pr_2$ denotes the projection onto the second factor $S$; and it follows from \eqref{eq:bundlemaps} that $\pi$ is surjective, and that
\begingroup
\setlength{\leftmargini}{12pt}
\begin{itemize}
\item
  $R_s(F_x)\subseteq F_x$ holds for each $x\in M$ and all $s\in S$.
\item
  If $x\in M$ and $p,p'\in F_x$, then $p'=p\cdot s$ holds for a unique element $s\in S$.
\end{itemize} 
\endgroup
\noindent 
The subspace $Tv_pP:=\ker[\dd_p\pi]\subseteq T_pP$ is called {\bf vertical tangent space} at $p\in P$, and the 
{\bf fundamental vector field} w.r.t.\ $\vec{s}\in\mathfrak{s}$ is defined by 
\begin{align*}
  \wt{s}(p):=\frac{\dd}{\dd t}\Big|_{t=0}\: p\cdot \exp(t\cdot \vec{s})\in Tv_pP\qquad\forall\: p\in P.
\end{align*}
The map $\mathfrak{s}\ni\vec{s}\rightarrow \wt{s}(p)\in Tv_pP$ is a vector space isomorphism for each $p\in P$. 
Then, complementary to that, a {\bf connection} $\w$ is an $\mathfrak{s}$-valued 1-form on $P$ with 
\begingroup
\setlength{\leftmargini}{12pt}   
\begin{itemize}
\item
  \hspace{17.2pt}$R_s^*\w= \Ad(s^{-1})\cp \w$\hspace{5pt} for each $s\in S$,
\item
  $\w_p(\wt{s}(p))=\vec{s}$ \hspace{52.5pt}for each\hspace{1pt} $\vec{s}\in \mathfrak{s}$. 
\end{itemize}
\endgroup
\noindent
The subspace $Th_pP:=\ker[\w_p]\subseteq T_pP$ is called the {\bf horizontal vector space} at $p$ (w.r.t.\ $\w$), and we have $\dd R_s(Th_p P)=Th_{p\cdot s}P$ for each $s\in S$. In addition to that, $T_pP= Th_pP\oplus Tv_pP$ holds for each $p\in P$. We denote the set of smooth connections on $P$ by $\Con$.

A diffeomorphism $\varphi\colon P\rightarrow P$ is said to be an  {\bf automorphism} iff we have
\begin{align*}
\varphi(p\cdot s)=\varphi(p)\cdot s\qquad\forall\: p\in P,\quad\forall\:s\in S,
\end{align*} 
whereby $\varphi$ is called a {\bf gauge transformation} iff additionally $\pi\cp \varphi =  \pi$ holds.

It is easy to see that an $\mathfrak{s}$-valued 1-form $\w$ on $P$ is a connection iff this is the case for $\varphi^*\w$, for each automorphism $\varphi$ of $P$.
A Lie group $(G,\theta)$ that acts on $P$ is said to be a {\bf Lie group of automorphisms} of $\boldsymbol{P}$ iff $\theta_g$ is an automorphism for each $g\in G$. This is equivalent to say that $\theta(g,p\cdot s)=\theta(g,p)\cdot s$ holds for each $p\in P$, each $g\in G$, and each $s\in S$. In this situation, we oftenly write $g\cdot p\cdot s$, instead of $(g\cdot p)\cdot s=g\cdot(p\cdot s)$. 
\begin{definition}[Induced actions -- Invariant connections]
\label{def:Connections}
If $(G,\theta)$ is a Lie group of automorphisms of $P$, we define the left actions
\begingroup
\setlength{\leftmargini}{12pt}
\begin{itemize}
\item 
$\IndA\colon G\times M \rightarrow M$,\:\:\:$(g,x)\mapsto (\pi\cp\theta)(g, p_x)$\quad for\quad $p_x\in \pi^{-1}(x)$\quad arbitrary,
\item 
   $\phi\colon G\times \hspace{2pt}\Con \rightarrow \hspace{1pt}\Con$,\hspace{9.5pt}$(g,\w)\mapsto \theta_{g^{-1}}^*\w$.
\end{itemize}
\endgroup
\noindent 
Then, $\w\in \Con$ is called $G$-invariant iff $\phi(g,\w)=\w$ holds for all $g\in G$, and the set of all such connections is denoted by $\Con_G$ in the following.
\end{definition}
The most relevant case for our later considerations is discussed in
\begin{example}[Homogeneous isotropic connections]
  \label{ex:LQC} 
  Let $P$ denote the trivial principal fibre bundle $\RR^3\times \SU$, and consider the representation
  \begin{align*}
    \uberl \colon \SU&\rightarrow \Autt\left(\RR^3\right)\\
    \sigma &\mapsto \murs^{-1} \cp \Ad(\sigma)\cp \murs,
  \end{align*}
  for $\murs\colon \RR^3\rightarrow \mathfrak{su}(2)$ defined by 
\begin{align*}  
  \textstyle\murs(\vv):= \sum_{i=1}^3 v^i \tau_i\qquad\quad\text{for}\qquad\quad \vv=\sum_{i=1}^3 v^i \vec{e}_i,
  \end{align*}
  with $\{\vec{e}_1,\vec{e}_2,\vec{e}_3\}$ the standard basis in $\RR^3$,  
  and matrices
  \begin{align*}
    \tau_1:=\begin{pmatrix} 0 & -\I  \\ -\I & 0  \end{pmatrix}\qquad\qquad \tau_2:=\begin{pmatrix} 0 & -1  \\ 1 & 0  \end{pmatrix}\qquad\qquad \tau_3:=\begin{pmatrix} -\I & 0  \\ 0 & \I  \end{pmatrix}.
  \end{align*}
  Recall that each $\sigma\in \SU$ can be written as\footnote{Here, $\mathbb{1}$ denotes the identity in $\SU$, which we will usually denote by $e$ in the following.} 
  \begin{equation} 
    \label{eq:expSU2} 
    \sigma= \cos(\alpha/2)\cdot \mathbb{1} + \sin(\alpha/2)\cdot \murs(\vec{n})=\exp\big(\alpha/2\cdot\murs(\vec{n})\big)
  \end{equation}
  for some $\|\vec{n}\|=1$ and $\alpha\in [0,2\pi)$, whereby  
  $\uberl(\sigma)$ rotates $x\in \RR^3$ by the angle $\alpha$ w.r.t.\ the axis $\vec{n}$; i.e., $\uberl\colon \SU\rightarrow \SO$ is the universal covering map. Then, for simplicity, we will write $\sigma(x)$ instead of $\uberl(\sigma)(x)$ in the following.
  
Now, consider the semi direct product $\Ge:=\Gee$, whose multiplication is given by 
\begin{align*}
	(v,\sigma)\cdot_\uberl (v',\sigma'):=(v+\sigma(v'),\sigma\cdot \sigma')\qquad\forall\: (v,\sigma),(v',\sigma')\in \Ge.
\end{align*}  
We let $\theta\colon E\times P\rightarrow P$,\:\: $(q,p)\mapsto q\cdot_\uberl p$, which is well defined, because $\Ge$ and $P$ do equal as a set. Then, $\Ge$ is a Lie group that resembles the euclidean one, and using Wang's theorem\cite{Wang}, one can deduce that the connections of the form
  \begin{align}
    \label{eq:euklconns}
    \ic{c}_{(x,s)}(\vv_x,\vec{\sigma}_s)= c\hspace{1pt} \Ad(s^+)(\hspace{1pt}\murs(\vv_x))+s^+\vec{\sigma}_s \qquad\forall\: (\vv_x,\vec{\sigma}_s)\in T_{(x,s)}P 
  \end{align}  
  are exactly the $\Ge$-invariant ones, cf.\ Appendix \ref{subsec:InvIsoHomWang}. Here, $c$ runs over $\RR$, and $s^+\in \SU$ denotes the adjoint of $s\in \SU$. The
  pullbacks of these connections by the global section $s\colon x\rightarrow (x,e)$, are given by $\wt{\w}^c=c\cdot \sum_{i=1}^3\tau_i\: \dd x^i$. These are the connections used in loop quantum cosmology to describe homogeneous  isotropic universe. \cite{MathStrucLQG, ChrisSymmLQG} 
  It is straightforward to see that $\varphi^*\w^{c} = \w^{d}$ for a gauge transformation $\varphi\colon P\rightarrow P$, implies that $c=d$ and $\varphi=\id_P$ holds. For this reason, it is not necessary to effort the concept of gauge transformations in this context. \hspace*{\fill}$\ddagger$
\end{example} 
\noindent
Finally, let $\gamma\colon [a,b]\rightarrow M$ be a $\CC{1}$-curve in $M$, and $\w$ a connection on $P$. Then, for each $p\in \pi^{-1}(\gamma(a))$, there exists a unique $\CC{1}$-curve $\gamma_p^\w\colon [a,b]\rightarrow P$ with
\begin{align*}
\pi\cp\gamma_p^\w=\gamma\qquad\qquad \gamma_p^\w(a)=p\qquad\qquad\text{as well as}\qquad\qquad \dot\gamma_p^\w(t)\in Th_{\gamma_p^\w(t)}P\quad\forall\: t\in [a,b],
\end{align*}
hence $\w_{\gamma_p^\w(t)}(\dot\gamma_p^\w(t))=0$ for each $t\in [a,b]$. \cite{KobNomiz}
This curve is called {\bf horizontal lift} of $\gamma$ w.r.t.\ $\w$ in the point $p$, and
the morphism\footnote{This means that $\parall{\gamma}{\w}(p\cdot s)=\parall{\gamma}{\w}(p)\cdot s$ holds for all $p\in F_{\pi(p)}$, and each $s\in S$.}
\begin{align*}
\parall{\gamma}{\w}\colon \pi^{-1}(\gamma(a))\rightarrow \pi^{-1}(\gamma(b)),\quad p\mapsto \gamma_p^\w(b) 
\end{align*}
is called {\bf parallel transport} along $\gamma$ w.r.t.\ $\w$. Then, for a $\CC{1}$-path $\gamma$, one defines the parallel transport by $\parall{\gamma}{\w}:= \parall{\gamma_0}{\w}\cp{\dots}\cp\parall{\gamma_{k-1}}{\w}$, whereby $\gamma_i$ are $\CC{1}$-curves with $\gamma_i=\gamma|_{[\tau_i,\tau_{i+1}]}$ for $0\leq i\leq k-1$ and $a=t_0<{\dots}< t_k=b$. It is straightforward to see that this definition is independent of the explicit choice of the decomposition of $\gamma$.
 
\begin{definition}
\label{def:Connectionss}
A family $\mathbf{\nu}=\{\nu_x\}_{x\in M}\subseteq P$ with $\nu_x \in F_x$ for each $x\in M$ is called a {\bf section} in the following. In this case, we denote by $\psi_x\colon \pi^{-1}(x)\rightarrow S$ the morphism, uniquely determined by $p=\nu_x\cdot \psi_x(p)$, and for a $\CC{k}$-path $\gamma$ in $M$, we define the map 
\begin{align}
\label{holos}
  h_\gamma^\nu\colon \Con \rightarrow S,\quad \w \mapsto \psi_{\gamma(b)} \cp \parall{\gamma}{\w}\big(\nu_{\gamma(a)}\big).
\end{align}
\end{definition}

\subsection{Maps on spectra}
\label{subsec:GrAcOnSp}
For an abelian Banach algebra $\aA$, we denote the corresponding norm by $\|\cdot\|_\aA$, and define $\|f\|_a:=|\chi(a)|$ for each $a\in \aA$, and a function $f\colon \aA\rightarrow \CCC$. Moreover, we denote by
\begingroup
\setlength{\leftmargini}{12pt}
\begin{itemize}
\item
$\Spec(\aA)$ the set of all non-zero multiplicative $\CCC$-valued functionals on $\aA$.
 \item
 $C_0(\Spec(\aA))$ the set of all complex-valued continuous functions on $\Spec(\aA)$ that {\bf vanish at infinity}.\footnote{This means that for each $f\in C_0(\Spec(\aA))$, and each $\epsilon >0$, there exists a compact $K_\epsilon\subseteq \Spec(\aA)$, such that $|f(\chi)|\leq \epsilon$ holds for each $\chi\in \Spec(\aA)\backslash K_\epsilon$.}
\item
 	$\Gel\colon \aA \rightarrow C_0(\Spec(\aA))$,\:\: $a\mapsto [\hspace{1pt}\hat{a}\colon f\mapsto f(a)]$ the Gelfand transformation.	
\end{itemize}
\endgroup
\noindent
The Gelfand-Naimark theorem then states that $\Gel$ is an isometric $^*$-isomorphism if $\aA$ is a $\Cstar$-algebra, and we obtain
\begin{lemma} 
\label{lemma:homzuspec}
\begingroup
\setlength{\leftmargini}{16pt}
\begin{enumerate}
\item
\label{lemma:homzuspec1}
    If $\lambda\colon \aA\rightarrow \bB$ is a homomorphism of abelian Banach algebras $\aA$ and $\mathfrak{B}$, then 
\begin{align*}    
    \ovl{\lambda}\colon \Spec(\bB)\rightarrow \Spec(\aA),\quad \chi \mapsto \chi \cp \lambda
\end{align*}
    is continuous if it is well-defined. This is the case, e.g., if $\lambda$ is surjective or unital.
\item
\label{lemma:homzuspec2}
    If $\aA$ is a $\Cstar$-algebra, then $\eta\colon \Autt(\aA)\rightarrow \Homeo(\Spec(\aA))$, $\lambda\mapsto\ovl{\lambda}$  is a group antiisomorphism.
\end{enumerate}
\endgroup
\end{lemma}
\begin{proof}
\begingroup
\setlength{\leftmargini}{16pt}
\begin{enumerate}
\item
      Since the image of $\ovl{\lambda}$ consists of homomorphisms, the only case in which well-definedness fails, is when $\ovl{\lambda}(\chi)=0$ holds for some $\chi\in \Spec(\bB)$. 
      
      If $\lambda$ is unital, we have $\ovl{\lambda}(\chi)(1_{\aA})=(\chi\cp\lambda)(1_{\aA})=\chi(1_\bB)=1\neq 0$, so that $\ovl{\lambda}(\chi)\neq 0$ holds for each $\chi\in \Spec(\bB)$. 
      Now, for each $\chi\in \Spec(\bB)$, we have $\chi(b)\neq 0$  for some $b\in \bB$. Thus, if $\lambda$ is surjective, we find $a\in\aA$ with $\lambda(a)=b$, with what $\ovl{\lambda}(\chi)(a)=\chi(b)\neq 0$, hence $\ovl{\lambda}(\chi)\neq 0$ holds. 
      
      For continuity, let $\Spec(\bB)\supseteq\{\chi_\alpha\}_{\alpha\in I}\rightarrow \chi$ be a converging net. Then, for $a\in \aA$ and $\epsilon> 0$, we find $\alpha_\epsilon\in I$, such that $\|\chi_\alpha-\chi\|_{\lambda(a)}\leq \epsilon$ holds for all $\alpha\geq \alpha_\epsilon$. Thus, for $\alpha\geq \alpha_\epsilon$, we have
      \begin{align*}
       \|\ovl{\lambda}(\chi_\alpha)-\ovl{\lambda}(\chi)\|_a=|\chi_\alpha(\lambda(a))-\chi(\lambda(a))|=\|\chi_\alpha-\chi\|_{\lambda(a)}\leq \epsilon,
      \end{align*} 
      hence $\Spec(\aA)\supseteq\{\ovl{\lambda}(\chi_\alpha)\}_{\alpha\in I}\rightarrow \ovl{\lambda}(\chi)$, showing continuity of $\ovl{\lambda}$.
    \item
      Since each $\lambda\in \Autt(\aA)$ is a surjective homomorphism, the image of $\eta$ consists of well defined and continuous maps by Part \ref{lemma:homzuspec1}). Moreover, $\eta$ is an antihomomorphism by
	\begin{align*}
		\eta(\lambda\cp \lambda')(\chi)=(\chi \cp \lambda)\cp \lambda'=\eta(\lambda')(\chi\cp \lambda)=\eta(\lambda')(\eta(\lambda)(\chi)),
	\end{align*}      
     and since $\lambda^{-1}\in \Autt(\aA)$ exists, $\eta(\lambda)^{-1}=\eta(\lambda^{-1})$ is continuous as well. This means that the image of $\eta$ consists of homeomorphisms, and shows well-definedness of this map. 
     
     For injectivity of $\eta$, assume that $\eta(\lambda)=\eta(\lambda')$ holds  for $\lambda,\lambda'\in \Autt(\aA)$. Then, for  
      $a\in\aA$, and each $\chi\in \Spec(\aA)$, we have
      \begin{align*}
        \Gel(\lambda(a))(\chi)=(\chi\cp\lambda)(a)=\eta(\lambda)(\chi)(a)
        =\eta(\lambda')(\chi)(a)=(\chi\cp\lambda')(a)
        =\Gel(\lambda'(a))(\chi),
      \end{align*}
     	hence $\Gel(\lambda(a))=\Gel(\lambda'(a))$. Since $\Gel$ is injective, this implies that $\lambda(a)=\lambda'(a)$ holds for all $a\in \aA$, hence $\lambda=\lambda'$. For surjectivity of $\eta$, we let 
      \begin{align}
        \label{eq:tautau}  
        \begin{split} 
          \tau\colon\Homeo(\Spec(\aA))&\rightarrow\Autt(\aA)\\
          h&\mapsto \big[a\mapsto \Gel^{-1}[\hspace{1pt}\Gel(a)\cp h]\big],
        \end{split}
      \end{align}
      which is well defined, because $\Gel(a)\cp h\in C_0(\Spec(\aA))$ holds for each $a\in \aA$. In fact, 
      $\tau(h)$ is a homomorphism because $\Gel$ and $\Gel^{-1}$ are homomorphism; and then  
      \begin{align*}
        \left[\tau(h^{-1})\cp \tau(h)\right](a)&=\Gel^{-1}\left[\Gel(\tau(h)(a))\cp h^{-1}\right]
        =\Gel^{-1}\left[\Gel(a)\cp h \cp h^{-1}\right]=a\qquad\forall\: a\in \aA
      \end{align*}
      shows that $\tau(h)\in \Autt(\aA)$ holds.
      Then, for $\chi\in \Spec(\aA)$, and each $a\in \aA$, we have
      \begin{align*}
        \eta(\tau(h))(\chi)(a)&=\ovl{\tau(h)}(\chi)(a)=(\chi\cp \tau(h))(a)
        =\chi(\Gel^{-1}[\Gel(a)\cp h])\\
        &=[\Gel(a)\cp h](\chi)=\Gel(a)(h(\chi))=h(\chi)(a),
      \end{align*}
      hence $\eta\cp \tau = \id_{\Homeo(\Spec(\aA))}$, which shows surjectivity of $\eta$.
\end{enumerate}
\endgroup
\end{proof}

\subsection{Bounded functions}
\label{subsec:boundedfun}
For a set $X$, we denote by $B(X):=\{f\colon X\rightarrow \CCC\:|\:\|f\|_{\infty}< \infty\}$ the algebra of bounded complex-valued functions on $X$, which is an abelian $\Cstar$-algebra w.r.t.\ the uniform norm $\|\cdot\|_\infty$.

Now, in general, if $U$ is a subset of a topological space, then $\ovl{U}$ will denote its closure therein. But, in order to the keep notations as simple as possible, we will use the following
\begin{convention}
\label{conv:Boundedfunc}
Let $\aA\subseteq B(X)$ denote some fixed $\Cstar$-subalgebra, and let $\upsilon\colon Y\rightarrow X$ be some map.
\begingroup
\setlength{\leftmargini}{12pt}
\begin{itemize} 
  \item
    The spectrum of $\aA$ is denoted by $\ovl{X}$ in the following. This is motivated by the first part of the next lemma; and since $X$ is not assumed to carry any topology in the following, this is not in conflict the our notation concerning closures of subsets of topological spaces. 
  \item
    We define $\aA_\upsilon:=\ovl{\upsilon^*(\aA)}$ for the $^*$-algebra $\upsilon^*(\aA):=\{f\cp \upsilon\:|\: f\in \aA\}\subseteq B(Y)$, as well as $\ovl{Y_\upsilon}:= \Spec(\aA_\upsilon)$ in accordance with the previous point.
  \item  
    	We let $X_\aA$ denote the set of all $x\in X$, whose image under 
	\begin{align*} 
	   	\iota_X\colon X &\rightarrow \mathrm{Hom}(\aA,\CCC)\\
	   		x&\mapsto [f\mapsto f(x)]
    \end{align*}
    	is non-zero. Thus, we have $X_\aA=\{x\in X\: |\:\exists\: f\in \aA : f(x)\neq 0\}$, hence $x\in X_\aA$ iff $\iota_X(x)\in \Spec(\aA)$ holds. 
  \item
    The set $\wt{Y_\upsilon}\subseteq \ovl{X}$ is defined to be the closure $\wt{Y_\upsilon}:=\ovl{\iota_X\big(X_\aA\cap\upsilon(Y)\big)}\subseteq \Spec(\aA)$. This is motivated by the third part of the next lemma, which shows that $\wt{Y_\upsilon}\cong \ovl{Y_\upsilon}$ holds if $\aA$ is unital.
\end{itemize}
\endgroup
\end{convention}
The first part of the next lemma is a slight variation of Proposition 2.1 in \cite{ChrisSymmLQG}, and originates from \cite{Rendall}. The second part can also be derived from Corollary 2.19 in \cite{ChrisSymmLQG}.
\begin{lemma}
  \label{lemma:dicht}
  \begingroup
\setlength{\leftmargini}{16pt}
  \begin{enumerate}
  \item
    \label{lemma:dicht1}
    If $X$ is a set and $\aA\subseteq B(X)$ a $\Cstar$-subalgebra, then $\iota_X(X_\aA)\subseteq \Spec(\aA)$ is dense, i.e., $\ovl{X}=\ovl{\iota_X(X_\aA)}$ holds. Moreover, $\iota_X$ is injective iff $\aA$ separates the points in $X$. 
  \item
    \label{lemma:dicht2}
    Let $\aA$ be unital, $Y$ a set, and $\upsilon\colon Y \rightarrow X$ a map.
    Then, $\ovl{\upsilon^*}\colon \ovl{Y_\upsilon}\rightarrow \ovl{X}$ is
    the unique continuous map which extends $\upsilon$ in the sense that the following diagram is commutative:
    \begin{center}
      \makebox[0pt]{
        \begin{xy}
          \xymatrix{
            \ovl{Y_\upsilon} \ar@{->}[r]^-{\ovl{\upsilon^*}}   &  \ovl{X}  \\
            Y\ar@{.>}[u]^{\iota_{Y}}\ar@{->}[r]^-{\upsilon} & X  \ar@{.>}[u]^{\iota_X}
          }
        \end{xy}
      }
    \end{center}    
    The map $\ovl{\upsilon^*}$ is an embedding, i.e., a homeomorphism onto its image equipped with the relative topology.
  \item
    \label{lemma:dicht3}
    In the situation of Part \ref{lemma:dicht2}), we have $\ovl{\upsilon^*}(\ovl{Y_\upsilon})=\ovl{\iota_X(\upsilon(Y))}=\wt{Y_\upsilon}\subseteq \ovl{X}$.
  \item
    \label{lemma:dicht4}
    If $\rho \colon X\rightarrow X$ is a map, then
    $\aA\subseteq \rho^*(\aA)$ implies $\rho(X_\aA)\subseteq X_\aA$. 
  \end{enumerate}
 \endgroup
 \end{lemma}
  \begin{proof}
    \begingroup
\setlength{\leftmargini}{16pt}
    \begin{enumerate}
    \item 
    The second statement is clear from the definitions.
    
	Moreover, if $U:=\Spec(\aA)\backslash \ovl{\iota(X_\aA)}$ in non-empty, it is an open neighbourhood of some $\chi\in \Spec(\aA)$. Since $\Spec(\aA)$ is locally compact Hausdorff, by Urysohn's lemma, we find a continuous function $\hat{f}\colon \Spec(\aA)\rightarrow [0,1]$ having compact support contained in $U$, such that $\hat{f}(\chi)=1$ holds. 
      Since $\hat{f}\in C_0(\Spec(\aA))$ holds, we have $\hat{f}=\Gel(f)$ for some $f\in \aA$, hence
\begin{align*}
	f(x)=\iota(x)(f)=\hat{f}(\iota(x))=0\qquad \forall\: x\in X_\aA
\end{align*}      
 by construction, as well as $f(x)=0$ for all $x\in X\backslash X_\aA$ by definition of $X_\aA$. Thus, $f=0$, hence $\hat{f}=\Gel(f)=0$ holds, which contradicts that $\hat{f}(\chi)=1$.
    \item
      Since $\aA$ and $\aA_\upsilon$ are unital, we have $X_\aA=X$ and $Y_{\aA_\upsilon}=Y$. 
      Moreover, $\upsilon^*\colon \aA \rightarrow \aA_\upsilon$ is a unital algebra homomorphism, so that $\ovl{\upsilon^*}\colon \Spec(\aA_\upsilon)\rightarrow \Spec(\aA)$ is well-defined and continuous by Lemma \ref{lemma:homzuspec}.\ref{lemma:homzuspec1}.\footnote{Observe that there is no ad hoc reason for the $^*$-algebra $\upsilon^*(\aA)$ to be closed in $B(Y)$, i.e., $\upsilon^*$ is not necessarily surjective. For this reason, we are forced to assume unitality, in order to guarantee well-definedness of $\ovl{\upsilon^*}$.} Then, $\ovl{\upsilon^*}\cp \iota_Y = \iota_X\cp \upsilon$ is obvious, and the uniqueness statement is clear from denseness of $\im[\iota_Y]$ in $\Spec(\aA_\upsilon)$, as well as continuity of $\ovl{\upsilon^*}$. 
      
     For the last statement, it suffices to show that $\ovl{\upsilon^*}$ is injective, because $\Spec(\aA_\upsilon)$ is compact, and $\im[\ovl{\upsilon^*}]$ is Hausdorff w.r.t.\ the relative topology inherited from $\Spec(\aA)$. Thus, let us assume that $\ovl{\upsilon^*}(\chi)=\ovl{\upsilon^*}(\chi')$ holds for $\chi,\chi'\in \Spec(\aA_\upsilon)$. Then, we have 
     \begin{align*}
      \chi(\upsilon^*(f))=\ovl{\upsilon^*}(\chi)(f)=\ovl{\upsilon^*}(\chi')(f)=\chi'(\upsilon^*(f))\qquad \forall\: f\in \aA,
     \end{align*} 
      hence $\chi|_{\upsilon^*(\aA)}=\chi'|_{\upsilon^*(\aA)}$, so that $\chi=\chi'$ follows from continuity of $\chi,\chi'$ as $\upsilon^*(\aA)$ is dense in $\aA_\upsilon$.
    \item
      We have $\ovl{\upsilon^*}(\iota_Y(Y))=\iota_X(\upsilon(Y))$ by Part \ref{lemma:dicht2}),  hence
      \begin{align*}
        \ovl{\upsilon^*}(\ovl{Y_\upsilon})=\ovl{\upsilon^*}\big(\:\ovl{\iota_Y(Y)}\:\big)\subseteq \ovl{\ovl{\upsilon^*}(\iota_Y(Y))}=\ovl{\iota_X(\upsilon(Y))}.
      \end{align*}
      Here, the first step is clear from Part \ref{lemma:dicht1}) applied to $Y$ and $\aA_\upsilon$, and the second one is due to continuity of $\ovl{\upsilon^*}$. Now, $\iota_X(\upsilon(Y))=\ovl{\upsilon^*}(\iota_Y(Y))$ also implies 
      \begin{align*}
        \iota_X(\upsilon(Y))\subseteq\ovl{\upsilon^*}\big(\:\ovl{\iota_Y(Y)}\:\big)\stackrel{\text{Part} \ref{lemma:dicht1})}{=}\ovl{\upsilon^*}(\hspace{1pt}\ovl{Y_\upsilon}\hspace{1pt})\qquad \Longrightarrow\qquad \ovl{\iota_X(\upsilon(Y))}\subseteq \ovl{\upsilon^*}(\ovl{Y_\upsilon})
      \end{align*}
      as $\ovl{\upsilon^*}(\ovl{Y_\upsilon})$ is compact.
    \item
      For $x\in X_\aA$, we have $f(x)\neq 0$ for some $f\in \aA$, whereby $f=g\cp \rho$ holds for some $g\in \aA$ by assumption. Thus, we have $g(\rho(x))=f(x)\neq 0$, hence $\rho(x)\in X_\aA$.
    \end{enumerate}
    \endgroup
\end{proof}

\section{Spectral Extension of Group Actions}
\label{sec:SPecExtGr}
In this section, we are going to lift Lie groups of automorphisms of principal fibre bundles with compact structure groups to spectra of $\Cstar$-algebras of cylindrical functions. In a first step, we will use the concept of a $\Cstar$-dynamical system, in order to extend a left action $\phi\colon G\times X\rightarrow X$ of a group $G$ on a set $X$ to the spectrum of a $\Cstar$-subalgebra of the bounded functions on $X$.
Then, we adapt this to the case where $X$ equals the set of smooth connections on a principal fibre bundle with compact structure group. For this, it will be crucial that 
the set of paths used for the definition of the cylindrical functions fulfills a certain invariance property.
 
Let us start with some general statements concerning group actions on spectra of abelian $\Cstar$-algebras.
\subsection{Group actions on spectra}
\label{subsec:graonsp} 
First recall that a $\Cstar$-dynamical system is a triple $(\aA,G,\Theta)$  consisting of a $\Cstar$-algebra $\aA$, a group $G$, and an antihomomorphism $\Theta\colon G\rightarrow \Aut(\aA)$. If $G$ is a topological group, then $\Theta$ is said to be continuous iff for each $a\in \aA$, the map $g\mapsto \Theta(g)(a)$ is continuous. Due to  \cite{0922.46050}, each $\Cstar$-dynamical system with $G$ locally compact and $\Theta$ continuous, gives rise to a continuous left action of $G$ on $\Spec(\aA)$.
In the next lemma, we will discuss this assignment for the abelian case; whereby, in the first part, we will drop any continuity assumptions. Here, commutativity of $\aA$ ensures injectivity of this assignment, which will provide us with a necessary condition for continuity of the corresponding action $\Phi$.  
\begin{lemma} 
  \label{lemma:leftactionsvsautos}
  Let $\aA$ be an abelian $\Cstar$-algebra. 
    \begingroup
\setlength{\leftmargini}{16pt}
  \begin{enumerate}
  \item
    \label{lemma:leftactionsvsautos1}
    The $\Cstar$-dynamical systems $(\aA,G,\Theta)$ are in one-to-one with such left actions $\Phi\colon G\times \Spec(\aA)\rightarrow \Spec(\aA)$, for which $\Phi_g$ is continuous for all $g\in G$.
  \item
    \label{lemma:leftactionsvsautos2}
    If $G$ is a topological group, then continuity of $\Theta$ implies continuity of $\Phi$. The converse implication holds if $\aA$ is unital.
  \end{enumerate}
  \endgroup
  \end{lemma}
  \begin{proof}
      \begingroup
\setlength{\leftmargini}{16pt}
    \begin{enumerate}
    \item
      Let $\Theta\colon G\rightarrow \Autt(\aA)$ be given, and define the corresponding left action by $\Phi(g,\chi):=\eta(\Theta(g))(\chi)$. Then, $\Phi_g$ is well-defined and continuous by Lemma \ref{lemma:homzuspec}.\ref{lemma:homzuspec2}; and since $\eta$ and $\Theta$ are antihomomorphisms, the left action property follows by
      \begin{align*}
        \Phi(g\cdot h,\chi)&=\eta(\Theta(g\cdot h))(\chi)=\eta(\Theta(h)\cp\Theta(g))(\chi)\\ 
        &=\eta(\Theta(g))\left(\eta(\Theta(h))(\chi)\right)
        =\Phi(g,\eta(\Theta(h))(\chi))=\Phi(g,\Phi(h,\chi)).
      \end{align*}
      Conversely, if $\Phi\colon G\times \Spec(\aA)\rightarrow \Spec(\aA)$ is given, then $\Phi_g\in \mathrm{Homeo}(\Spec(\aA))$ holds for each $g\in G$ by assumption, so that $\Theta(g):=\tau(\Phi_g)$ is an element in $\Autt(\aA)$, for $\tau\colon \mathrm{Homeo}(\Spec(\aA))\rightarrow \Autt(\aA)$ defined by  \eqref{eq:tautau}. Now, since $\eta$ is an antiisomorphism, the same is true for $\tau=\eta^{-1}$, and we obtain
      \begin{align*}
        \Theta(g\cdot h)&=\tau(\Phi_{g\cdot h})=\tau(\Phi_g\cp \Phi_h)
        =\tau(\Phi_h)\cp \tau(\Phi_g)=\Theta(h)\cp \Theta(g).
      \end{align*}
    \item
      Let $\Theta$ be continuous, and $G\times \Spec(\aA)\supseteq \{(g_\alpha,\chi_\alpha)\}_{\alpha\in I}\rightarrow (g,\chi)\in G\times \Spec(\aA)$ a converging net. Then, $\{g_\alpha\}_{\alpha\in I}\rightarrow g$ and $\{\chi_\alpha\}_{\alpha\in I}\rightarrow \chi$ are converging nets as well. 
      By continuity of $\Theta$, for $a\in \aA$ and $\epsilon>0$, we find $\alpha_\epsilon\in I$, such that $\|\Theta(g)(a)-\Theta(g_\alpha)(a)\|_\aA\leq \frac{\epsilon}{2}$ holds for all $\alpha \geq \alpha_\epsilon$. Thus, since $\|\chi_\alpha\|_{\mathrm{op}}\leq 1$ holds for each $\alpha\in I$, we have
      \begin{align*}
       \textstyle |\chi_\alpha(\Theta(g)(a)-\Theta(g_\alpha)(a))|\leq \|\Theta(g)(a)-\Theta(g_\alpha)(a)\|_\aA\leq\frac{\epsilon}{2}\qquad \forall\:\alpha \geq \alpha_\epsilon.
      \end{align*}
      Moreover, since $\{\chi_\alpha\}_{\alpha\in I}\rightarrow \chi$ holds, we find $\alpha'_{\epsilon}\in I$ with  
      $\|\chi-\chi_\alpha\|_{\Theta(g)(a)}\leq \frac{\epsilon}{2}$
      for each $\alpha\geq \alpha'_\epsilon$, so that for 
       $\alpha\in I$ with $\alpha\geq \alpha_\epsilon,\alpha'_\epsilon$, we have
      \begin{align*}
        \|\Phi(g,\chi)-\Phi(g_\alpha,\chi_\alpha)\|_a
        &=\|\eta(\Theta(g))(\chi)-\eta(\Theta(g_\alpha))(\chi_\alpha)\|_{a}
        =|\chi(\Theta(g)(a))-\chi_\alpha(\Theta(g_\alpha)(a))|\\
        &\leq \|\chi-\chi_\alpha\|_{\Theta(g)(a)} + |\chi_\alpha(\Theta(g)(a)-\Theta(g_\alpha)(a))|\\
        &\textstyle\leq \frac{\epsilon}{2}+\frac{\epsilon}{2}=\epsilon.
      \end{align*}
      Now, if $\aA$ is unital and $\Phi$ continuous, we fix $a\in \aA$, and consider the continuous function
      \begin{align*}
        \alpha((g,\chi),(g',\chi')):= |(\Gel(a)\cp \Phi)(g,\chi)-(\Gel(a)\cp \Phi)(g',\chi')|. 
      \end{align*}
       Then, $\alpha^{-1}(B_\epsilon(0))$ is open and contains $((g,\chi),(g,\chi))$ for each $(g,\chi)\in G\times \Spec(\aA)$. Now, if $g\in G$ is fixed, for each $\chi\in \Spec(\aA)$, we find open neighbourhoods $B_\chi\subseteq G$ and $U_\chi \subseteq \Spec(\aA)$ of $g$ and $\chi$, respectively, such that $B_\chi\times U_\chi \times B_\chi\times U_\chi\subseteq \alpha^{-1}(B_\epsilon(0))$ holds.  
      Then, by compactness of $\Spec(\aA)$, we find $\chi_1,\dots, \chi_n\in \Spec(\aA)$, such that the corresponding sets $U_{\chi_j}$ cover $\Spec(\aA)$. Now, $B_g:= B_{\chi_1} \cap {\dots} \cap B_{\chi_n}$ is an open neighbourhood of $g$, and we have
      \begin{align*}
        \alpha((g,\chi),(g',\chi)) < \epsilon \qquad \forall\: \chi \in \Spec(\aA),\quad\forall\:  g'\in B_g.
      \end{align*}
      Consequently, $\|\Gel(a)\cp\Phi_g -\Gel(a)\cp\Phi_{g'}\|_{\infty}\leq \epsilon$ holds for each $g'\in B_g$, hence
      \begin{align*}
        \|\Theta(g)(a)-\Theta(g')(a)\|_{\aA}&= \|\tau(\Phi_g)(a)-\tau(\Phi_{g'})(a)\|_{\aA}\\
        &=\|\Gel^{-1}[\Gel(a)\cp\Phi_g]-\Gel^{-1}[\Gel(a)\cp\Phi_{g'}]\|_{\aA}\\
        &=\|\Gel^{-1}[\Gel(a)\cp\Phi_g-\Gel(a)\cp\Phi_{g'}]\|_{\aA}\\
        &=\left\|\Gel(a)\cp\Phi_g-\Gel(a)\cp\Phi_{g'}\right\|_{\infty}\\
        &\leq \epsilon
      \end{align*}
      for each $g'\in B_g$, which shows continuity of $\Theta(\cdot)(a)$ at $g\in G$. 
    \end{enumerate}
    \endgroup
  \end{proof}

\begin{definition}
  \label{def:ixschlange}
  Let $\phi$ be a left action of a group $G$ on the set $X$, and let $\aA\subseteq B(X)$ be a $\Cstar$-subalgebra. 
      \begingroup
\setlength{\leftmargini}{16pt}
  \begin{enumerate}
  \item
    \label{def:ixschlange1}
    Then, $\aA$ is called $\phi$-invariant iff $\phi_g^*(\aA)\subseteq \aA$ holds for all $g\in G$, with what $(\aA,G,\Theta)$ is a $\Cstar$-dynamical system for $\Theta(g)(f):=\phi_g^*(f)$.
  \item
    \label{def:ixschlange2}
    We define the set of $G$-invariant elements in $X$ by
	\begin{align*}    
    X_G:=\left\{x\in X\:|\:\phi(g,x)=x\:\:\: \forall\: g\in G \right\}, 
	\end{align*}    
    and denote by $\ovl{X_G}$ the spectrum of the $\Cstar$-algebra $\aA_G:=\ovl{\aA|_{X_G}}$. 
    Then, $\aA_G$ is the restriction $\Cstar$-algebra w.r.t.\ the inclusion $i_{X_G}\colon X_G\rightarrow X$ introduced in Convention \ref{conv:Boundedfunc}. 
  \item
    \label{def:ixschlange3}
    According to the last point in Convention \ref{conv:Boundedfunc}, we define  $\wt{X_G}:=\ovl{\iota_X(X_\aA\cap X_G)}$. 
  \end{enumerate}
  \endgroup
\end{definition}
Then, we have
\begin{proposition}
  \label{prop:autspec} 
  Let $\phi$ be a left action of a group $G$ on a set $X$, and $\aA\subseteq B(X)$ a $\phi$-invariant 
  $\Cstar$-subalgebra. Then,
   \begingroup
\setlength{\leftmargini}{16pt} 
  \begin{enumerate}
  \item
    \label{prop:autspec1} 
    There is a unique left action $\Phi\colon G\times \X \rightarrow \X$,
    such that
    \begin{enumerate}
    \item[$\mathrm{(a)}$]
      $\Phi_g$ is continuous for each $g\in G$,
    \item[$\mathrm{(b)}$]
      $\Phi$ extends $\phi$ in the sense that on $X_\aA$, we have
      \begin{align}
        \label{eq:extens}
        \Phi_g\cp\iota_X=\iota_X\cp\phi_g\qquad\forall\: g\in G.
      \end{align}
    \end{enumerate}
    This action is explicitly given by $\Phi(g,\ovl{x})= \ovl{x}\cp \phi_g^*$. 
  \item
    \label{prop:autspec2} 
    If $G$ is a topological group, then $\Phi$ is continuous if $\phi^*_\bullet f\colon G\rightarrow \aA$,\: $g \mapsto f\cp \phi_g$ is continuous for each $f\in \aA$. The converse implication holds if $\aA$ is unital.
  \item
    \label{prop:autspec3} 
    The set of invariant elements  
	\begin{align*}
		\ovl{X}_G=\left\{\x \in \X\:\big|\: \Phi(g,\x)=\x \:\:\: \forall\:g\in G\right\}
	\end{align*}    
     is closed in $\ovl{X}$, and we have $\wt{X_G}\subseteq \ovl{X}_G$.
  \item
    \label{prop:autspec4} 
    If $\aA$ is unital, then $\ovl{i_{X_G}^*}\colon \ovl{X_G}\rightarrow \wt{X_G}$ is a homeomorphism, and
    the following diagram is commutative:
    \begin{center}
      \makebox[0pt]{
        \begin{xy}
          \xymatrix{
            \ovl{X_G}\: \ar@{->}[r]^-{\ovl{i_{X_G}^*}}_-{\cong}   & \: \wt{X_G}\:  \ar@{->}[r]^-{\subseteq}   & \:\ovl{X}_G\: \ar@{->}[r]^-{\subseteq} & \:\ovl{X} \\
            & \:X_G\: \ar@{.>}[ul]^{\iota_{X_{G}}} \ar@{.>}[u]^{\iota_X} \ar@{^{(}->}[r]^{i_{X_G}}  &  X \ar@{.>}[ur]^{\iota_{X}}.    
          }
        \end{xy}
      }
    \end{center}
  \end{enumerate}
  \endgroup
  \end{proposition}
  \begin{proof}
          \begingroup
\setlength{\leftmargini}{16pt} 
    \begin{enumerate}
    \item
      First observe that \eqref{eq:extens} makes sense, because $\phi_{g^{-1}}^*(\aA)\subseteq \aA$ implies $\aA = \phi_{g}^*(\phi_{g^{-1}}^*(\aA))\subseteq \phi_{g}^*(\aA)$, hence $\phi_g\colon X_\aA\rightarrow X_\aA$ by Lemma \ref{lemma:dicht}.\ref{lemma:dicht4}.
      Now, for uniqueness, let $\Phi'$ be another action fulfilling (a) and (b).  
      Then, \eqref{eq:extens} shows
      \begin{align*}
      	\Phi'_g|_{\iota_X(X_\aA)}=\Phi_g|_{\iota_X(X_\aA)}\qquad\Longrightarrow\qquad \Phi'_g=\Phi_g\qquad\quad\forall\:g\in G
      \end{align*}
       by (a) and denseness of $\iota_X(X_\aA)$ in $\Spec(\aA)$. For existence, consider the $\Cstar$-algebraical system $(\aA,G,\Theta)$ for $\Theta(g):= \phi_g^*$.
      Then, Lemma \ref{lemma:leftactionsvsautos}.\ref{lemma:leftactionsvsautos1} provides us with a left action $\Phi\colon G\times \Spec(\aA)\rightarrow \Spec(\aA)$, such that $\Phi_g$ is continuous for each $g\in G$, and which is defined by
      \begin{align*}
        \Phi(g,\ovl{x})=\eta(\Theta(g))(\ovl{x})=\ovl{x}\cp \Theta(g)=\ovl{x}\cp \phi^*_g.
      \end{align*}
     	Then, for $x\in X_\aA$ and each $f\in \aA$, we have 
      \begin{align*}
        \Phi_g(\iota_X(x))(f)
        =\iota_X(x)(\phi_g^*f)
        =(f\cp \phi_g)(x)=f(\phi_g(x))=(\iota_X\cp\phi_g)(x)(f),
      \end{align*}
      hence \eqref{eq:extens}.
    \item
      We have $\Theta(\cdot)(f)=\phi_\bullet^*f$, so that the continuity statement is clear from Lemma \ref{lemma:leftactionsvsautos}.\ref{lemma:leftactionsvsautos2}.
    \item
      Let $\ovl{X}_G\supseteq \{\x_\alpha\}_{\alpha\in I}\rightarrow \x\in \X$ be a converging net. Then, for each $g\in G$, continuity of $\Phi_g$ shows 
      \begin{align*}
        \textstyle\Phi(g,\x)=\Phi(g,\lim_\alpha\x_\alpha)=\lim_\alpha\Phi(g,\x_\alpha)=\lim_\alpha \x_\alpha=\x,
      \end{align*}
      so that $\ovl{X}_G$ is closed. Thus, $\wt{X_G}\subseteq \ovl{X}_G$ is clear, because $\iota_X(X_\aA\cap X_G)\subseteq \ovl{X}_G$ holds by \eqref{eq:extens}.
    \item
      This follows from the Parts \ref{lemma:dicht2}) and \ref{lemma:dicht3}) of Lemma \ref{lemma:dicht}, if we define $\upsilon:=i_{X_G}$.
    \end{enumerate}
    \endgroup
  \end{proof}
For an application of this proposition to the Bohr compactification of $\RR$, cf.\  Corollary \ref{cor:bohrmultimitlambda} and \cite{{MEAS}}.
\begin{remark}[Unitality]
  If $\aA$ is unital, Proposition \ref{prop:autspec}.\ref{prop:autspec1} can also be derived from Corollary 2.19 in \cite{ChrisSymmLQG} by extending each $\phi_g^*\colon \aA\rightarrow \aA$ uniquely to a continuous map $\Phi_g\colon \Spec(\aA)\rightarrow \Spec(\aA)$. In fact, then it follows from the uniqueness property of these maps that $\Phi_{g\cdot h}=\Phi_g\cp \Phi_h$ holds for all $g,h\in G$, with what $\Phi\colon (g,\ovl{x})\rightarrow \Phi_g(\ovl{x})$ is a well defined group action with the properties from Proposition \ref{prop:autspec}.\ref{prop:autspec1}.   
\end{remark} 

\subsection{Invariant generalized connections}
\label{sec:InvGenConnes}
We now are going to apply the previous subsection to $\Cstar$-algebras of cylindrical functions.  
Thus, in the following, let $(P,\pi,M,S)$ denote some principal fibre bundle with compact structure group, $(G,\theta)$ be some Lie group of automorphisms of $P$, and $\Paths$ some set of $\CC{k}$-paths in $M$. 

We choose a section $\nu=\{\nu_x\}_{x\in M}\subseteq P$ 
with maps $\psi_x$ as in Definition \ref{def:Connectionss}, and consider the functions of the form $\rho_{ij} \cp h^\nu_\gamma\colon \Con\rightarrow \CCC$, for $h_\gamma^\nu$ defined by \eqref{holos}. Here, $\gamma$ runs over all elements in $\Paths$, and $\rho_{ij}$ are entries of any irreducible matrix representation $\rho$ of $S$. 
Let $\Cylsk$ denote the $^*$-algebra generated by these functions, and observe that then $\Cylsk\subseteq B(\Con)$ holds by compactness of $S$. Then, the closure $\Cylk$ of $\Cylsk$ in $B(\Con)$ is well defined, and always unital because $\rho\cp h^\nu_\gamma=1$ holds for the trivial representation $\rho$ of $S$.

Moreover, from the morphism properties of parallel transports and the maps $\psi_x$, it follows that this definition does not depend on the explicit choice of the section $\nu$. In fact, if $\nu'=\{\nu'_x\}_{x\in M}\subseteq P$ is another section, then for $\Paths\ni \gamma\colon [a,b]\rightarrow M$, we have
\begin{align}
  \label{eq:indep}
  h^{\nu'}_\gamma(\w)=\psi'_{\gamma(b)}\big(\nu_{\gamma(b)}\big)\cdot h^{\nu}_\gamma(\w)\cdot\psi_{\gamma(a)}\big(\nu'_{\gamma(a)}\big).
\end{align}
We will call $\Cylk$ the $\Cstar$-algebra of cylindrical functions w.r.t.\ $\Paths$, and denote its spectrum by $\ovl{\Con}$. The elements $\ovl{\w}\in \ovl{\Con}$ will be called {\bf generalized connections} in the following. 
 Now, according to Definition \ref{def:ixschlange}.\ref{def:ixschlange2},  we have 
\begin{align*}
\Con_G=\{\w\in\Con\:|\: \phi(g,\w)=\w\:\:\:\forall\:g\in G\}\qquad\quad\text{as well as}\qquad\quad \ovl{\Con_G}=\Spec\big(\hspace{1pt}\ovl{\Cylk|_{\Con_G}}\hspace{1pt}\big)
\end{align*}
for the action $\phi$ from Definition \ref{def:Connections}. Then, if    
the set of curves $\Paths_\alpha$ under consideration carries an index $\alpha$, we will write $\ovl{\Con}_\alpha$ instead of $\A$, as well as $\ovl{\Con_{G,\alpha}}$ instead of $\ovl{\Con_G}$. The same conventions will be used for the space $\A_G$ introduced in Corollary \ref{cor:CylSpecAction}.
\begin{definition}
  \label{def:InvPfade}
  A set $\Paths$ of $\CC{k}$-paths in $M$ is said to be $G$-invariant iff for each $g\in G$, and each $\gamma \in \Paths$, the curve 
\begin{align*}
 g \cdot\gamma\colon  \dom[\gamma]\rightarrow M,\quad t\mapsto \IndA(g,\gamma(t))
\end{align*}  
  is an element in $\Paths$ again.  
\end{definition}
\begin{remark}
  Let $\Paths$ be a collection of $\CC{k}$-paths in $M$, and assume that $\IndA$ is of class $\CC{k}$. We consider the set $\langle \Paths\rangle$ of all translates $g\cdot \gamma$ with $g\in G$ and $\gamma \in \Paths$, and observe that $\langle\Paths\rangle$ is $G$-invariant and consists of $\CC{k}$-paths.
\end{remark}
\begin{lemma} 
  \label{lemma:CylSpecActionvorb}
  If $\Paths$ is $G$-invariant, then $\phi^*_g(\Cylk)\subseteq \Cylk$ holds for each $g\in G$, for the left action $\phi\colon G\times \Con \rightarrow \Con$ from Definition \ref{def:Connections}. 
  \end{lemma}
  \begin{proof}
    The elementary but technical proof is given in Appendix \ref{subsec:InvCyl}.
  \end{proof}
In the next section, we will show that the inclusion $\wt{X_G}\subseteq \ovl{X}_G$ from Proposition \ref{prop:autspec} 
even can be proper. For this, we first reformulate this proposition for cylindrical functions, which is the content of the next corollary. 
Then, we will use the isomorphism from Lemma \ref{lemma:SpeczuHomm}, in order to obtain a more practicable description of the space $\ovl{\Con}_G$ by means of invariant homomorphisms of paths, which will be done in Subsection \ref{subsec:InvHoms}.
\begin{corollary}
  \label{cor:CylSpecAction}
  If $\Paths$ is $G$-invariant, then 
  $\Phi\colon G\times \ovl{\Con}\rightarrow \ovl{\Con}$, $(g,\ovl{\w})\mapsto \ovl{\w}\cp \phi_g^*$ 
  is the unique left action,  
  for which $\Phi_g$ is continuous for each $g\in G$, and which
  extends $\phi$ in the sense that $\Phi_g\cp \iota_\Con=\iota_\Con\cp \phi_g$ holds for all $g\in G$.
  The subset $\ovl{\Con}_G\subseteq \A$ is compact, and the following diagram is commutative:
  \begin{align}
    \label{eq:inclusionsdiag}
    \begin{split}
      \makebox[0pt]{
        \begin{xy}
          \xymatrix{
            \ovl{\Con_G}\: \ar@{->}[r]^-{\ovl{i_{\Con_G}^*}}_-{\cong}   &  \:\wt{\Con_G}\: \ar@{->}[r]^-{\subseteq}   & \:\ovl{\Con}_G\:\ar@{->}[r]^-{\subseteq} &\:\ovl{\Con} \\
            & \:\Con_G\: \ar@{.>}[ul]^{\iota_{\Con_{G}}} \ar@{.>}[u]^{\iota_\Con} \ar@{^{(}->}[r]^{i_{\Con_G}}  &  \Con \ar@{.>}[ur]^{\iota_{\Con}}.    
          }
        \end{xy}
      }
    \end{split}
  \end{align}
  Here, $\ovl{i_{\Con_G}^*}\colon \ovl{\Con_G}\rightarrow \wt{\Con_G}=\ovl{\iota_\Con(\Con_G)}$ is a homeomorphism, and $\Phi$ is continuous iff for all generators $f=\rho_{ij}\cp h_\gamma^\nu$ of $\Cylk$, the map $\theta_f\colon G\rightarrow \Cylk,\:\:g \mapsto f\cp \theta_{g^{-1}}^*$ is continuous.
  \end{corollary}
  \begin{proof}
    $\Cylk$ is unital, and $\phi_g^*\big(\Cylk\big)\subseteq \Cylk$ holds for each $g\in G$ by Lemma \ref{lemma:CylSpecActionvorb}. Thus, the claim is clear from Proposition \ref{prop:autspec}, because  
     $\phi_\bullet^*\colon g\mapsto \phi_g^*f=f\cp \phi_g=f\cp \theta_{g^{-1}}^*$ is continuous for each $f\in \Cylk$ iff this is the case for all generators $f=\rho_{ij}\cp h^\nu_\gamma$ of $\Cylk$. 
  \end{proof}
The elements of $\ovl{\Con}_G$ are called $G$-invariant generalized connections in the following, and the next example shows that the action $\Phi$ is not continuous for the case of homogeneous isotropic LQC.
\begin{Example}
  \label{ex:Eukl}
  Let $P=\RR^3 \times \SU$, $\Ge=\Gee$, and $\theta$ be as in Example \ref{ex:LQC}. Define $\nu=\{\nu_x\}_{x\in M}$ by $\nu_x:=(x,e)$ for each $x\in M=\RR^3$, and let $\Paths$ contain all the linear curves $[0,1]\ni t\mapsto x+ t\cdot \vv\in \RR^3$ for some $x,\vv\in \RR^3$. Consider the connections $\w^r$ defined by the right invariant geometric distribution, specified by the following smooth sections $\mathcal{E}^r_i\colon P \rightarrow TP$ for $1\leq i\leq 3$:
  \begin{align*}
  	\mathcal{E}^r_1(x,s)&:=(\vec{e}_1,r x_2\cdot\tau_2\cdot s)\in T_{(x,s)}P \qquad\text{ for }\qquad x=(x_1,x_2,x_3)\in \RR^3
  	\\
  	\mathcal{E}^r_{i}(x,s)&:=(\vec{e}_{i},\vec{0})\hspace{106pt}\text{for }\qquad i=2,3
  \end{align*}
  with $\tau_2\cdot s:=\dd_{e} R_s(\tau_2)\in T_s\SU$.
\vspace{6pt}

\noindent
Now, let us consider the linear curve $\gamma_y\colon[0,1]\rightarrow \RR^3,\:\: \gamma_y(t)=y\cdot\vec{e}_2 + t\cdot\vec{e}_1$. 
  Then, its horizontal lift $\gamma_y^{r}\colon [0,1]\rightarrow \RR^3\times \SU$ w.r.t.\ $\w^r$ in $(y\cdot \vec{e}_2,e)$, is given by 
  $\gamma_y^{r}(t)=\left(\gamma_y(t),\exp(try\cdot \tau_2)\right)$, because  
 $\pi \cp \gamma_y^{r}=\gamma_y$, as well as
   \begin{align*}
    \dot\gamma_y^{r}(t)=(\vec{e}_1,ry\cdot \tau_2\cdot \exp(try\cdot \tau_2))=\mathcal{E}^r_1(\gamma_y(t),\exp(try\cdot\tau_2))=\mathcal{E}^r_1(\gamma_y^{r}(t))\qquad\forall\: t\in [0,1]
  \end{align*}
  holds. Then, by the choice of $\nu$, for each $y\in \RR$, we have
  \begin{align*}
    h_{\gamma_y}^\nu(\w^r)=\pr_2\cp \gamma_y^{r}(1)=\exp(ry\cdot \tau_2)\stackrel{\eqref{eq:expSU2}}{=}\begin{pmatrix} \cos(ry) & -\sin(ry)  \\ \sin(ry) & \cos(ry)\end{pmatrix}.
  \end{align*}
  Moreover, 
  $\phi_g^*\hspace{1pt}h^\nu_{\gamma_y}=h^\nu_{\gamma_{\lambda +y}}$ holds for $g:=(-\lambda \cdot \vec{e}_2,e)$,\footnote{This follows, e.g., from \eqref{eq:trafogenerators} for $\gamma'=g^{-1}\cdot \gamma_y=\gamma_{\lambda+y}$, because in our situation  the factors $\delta_1,\delta_2$ just equal $e\in \SU$.} so that we have
  \begin{align*}
    \label{eq:Absch}
      \|\phi_e^*(\rho_{11}\cp h^\nu_{\gamma_0})-\phi_{g}^*(\rho_{11}\cp h^\nu_{\gamma_0})\|_{\infty}
      &\geq |(\rho_{11}\cp h^\nu_{\gamma_0})(\w^r)-\phi_g^*(\rho_{11}\cp h^\nu_{\gamma_0})(\w^r)|\\
      &=|(\rho_{11}\cp h^\nu_{\gamma_0})(\w^r)-(\rho_{11}\cp h^\nu_{\gamma_\lambda})(\w^r)|\\
      &=|1-\cos(r\lambda)|.
  \end{align*} 
  Since this equals $1$ for $r=\pi/(2\lambda)$ if $\lambda>0$, the map
	\begin{align*}
	\theta_f\colon g\mapsto f\cp \theta_{g^{-1}}^*\qquad\text{for}\qquad f:=\rho_{11}\cp h^\nu_{\gamma_0}
	\end{align*}
	is not continuous   
  at $(0,e)\in E$, which, in turn, implies discontinuity of the extended action $\Phi$ on $\Spec(\Cylk)$ by Corollary \ref{cor:CylSpecAction}. 
  
  In fact, each neighbourhood of $(0,e)$ contains $h_\lambda=(-\lambda\cdot \vec{e}_2,e)$ for some $\lambda> 0$, so that $g\rightarrow 0$ does not imply $\theta_f(g)\rightarrow \theta_f(e)$, because we have shown that 
\begin{align*}
\|\theta_f(e)-\theta_f(h_\lambda)\|_{\infty}= \|\phi_e^*f-\phi^*_{h_\lambda}f\|_{\infty}  \geq |1-\cos(\pi/2)|=1
\end{align*}  
  holds. \hspace*{\fill} {$\ddagger$}  
\end{Example}

\subsection{Invariant homomorphisms}
\label{subsec:InvHoms}
So far, we have shown that a Lie group of automorphisms $(G,\theta)$ on $P$,  and a $G$-invariant set of paths $\Paths$, give rise to the reduced space $\ovl{\Con}_G\subseteq \A= \Spec\left(\Cylk\right)$. 
In this subsection, we will present a more down-to-earth characterization of $\ovl{\Con}_G$ that, in particular, will be useful 
for our investigations concerning the inclusion relations between the spaces  $\ovl{\Con_G}$ and $\ovl{\Con}_G$. 
For this, we first give a short introduction into homomorphisms of paths and highlight their relation to the space $\ovl{\Con}$. We will only state the important facts at this point; the details can be found in Appendix \ref{sec:HomPaths}.
 
Thus, for the rest of this section, let $\Paths$ be always stable under inversions and decompositions of its elements, and recall that two paths $\gamma,\gamma'\in \Paths$ are said to be equivalent (write $\gamma\csim \gamma'$) iff $\Paths_{\gamma}^\w=\Paths_{\gamma'}^\w$ holds for all $\w\in \Con$.
We define $\AF:=\bigsqcup_{x,x'\in M}\Iso(x,x')$, for $\Iso(x,x')$ the set of maps $\varphi\colon F_x\rightarrow F_{x'}$ with $\varphi(p\cdot s)=\varphi(p)\cdot s$  for each $p\in F_x$, and each $s\in S$.
 
Then, a homomorphism of paths is a map $\homm \colon \Paths\rightarrow \Iso_F$, such that (cf. Definition \ref{def:hompaths})
\begingroup
\setlength{\leftmargini}{12pt}
\begin{itemize}
\item
  $\homm(\gamma)\in \Iso(\gamma(a),\gamma(b))$ holds for $\gamma\in \Paths$ with $\dom[\gamma]=[a,b]$,
\item
  $\homm$ is compatible w.r.t.\ inversion and decomposition of the elements in $\Paths$,
\item
  $\homm(\gamma)=\homm(\gamma')$ holds for all $\gamma,\gamma'\in \Paths$ with $\gamma\csim \gamma'$.
\end{itemize}
\endgroup
\noindent
For the second point, observe that each $\varphi\in \AF$ is invertible, and that we have $\varphi_2\cp\varphi_1 \in \Iso(x,z)$ if $\varphi_1 \in \Iso(x,y)$ and $\varphi_2 \in \Iso(y,z)$ holds.
In particular, for $\w\in\Con$, the map $\gamma \mapsto \Paths_\gamma^\w\in \Iso(\gamma(a),\gamma(b))$ is an element in $\AF$.

The above definition differs from the traditional one \cite{Ashtekar2008} in the point that we require $\homm$ to be compatible w.r.t.\ decompositions of paths, and not w.r.t.\ their concatenations. This is to avoid technicalities, as it allows to restrict to embedded analytic curves, instead of considering all the piecewise ones, cf. Section \ref{sec:symmred}. In fact, since both sets give rise to the same $\Cstar$-algebra of cylindrical functions, they define the same set $\A$ of generalized connections.

Now, due to denseness of $\iota_\Con(\Con)$ in $\ovl{\Con}$, for each $\ovl{\w}\in \ovl{\Con}$, there exists a net $\{\w_\alpha\}_{\alpha\in I}\subseteq \Con$ with $\{\iota_\Con(\w_\alpha)\}_{\alpha\in I}\rightarrow \ovl{\w}$. Then, 
we obtain a well-defined injective map $\kappa\colon \ovl{\Con}\rightarrow \Hom(\Paths,\Iso_F)$ if we define, cf.\ Lemma \ref{lemma:SpeczuHomm}
\begin{align*}
	\kappa(\ovl{\w})(\gamma)(p):=\textstyle\lim_\alpha \Paths_\gamma^{\w_\alpha}(p)\qquad \text{for}\qquad \dom[\gamma]=[a,b]\quad\text{and}\quad p\in F_{\gamma(a)}.
\end{align*}
This map turns out to be surjective if the set $\Paths$ is in addition independent. This means that for each finite collection $\{\gamma_1,\dots,\gamma_l\}\subseteq \Paths$, there exists a finite collection $\{\delta_1,\dots,\delta_n\}\subseteq \Paths$, such that:
\begingroup
\setlength{\leftmargini}{18pt}
\begin{itemize}
\item[1.)]
  For each $1\leq j\leq l$, there exists a decomposition $\{(\gamma_j)_i\}_{1\leq i\leq k_j}$ of $\gamma_j$, such that each $(\gamma_j)_i$ is equivalent to one of the paths $\delta_r,\delta_r^{-1}$ for some $1\leq r\leq n$.
\item[2.)]
  For each choice $s_1,\dots,s_n\in S$, there exists some $\w\in \Con$, such that $h^\nu_{\w}(\delta_i)= s_i$ holds for all $1\leq i\leq n$.
\end{itemize}
\endgroup
\noindent
Thus, in in the following, let $\Paths$ be additionally invariant and independent. 
We now will use the bijection $\kappa$, to obtain a more concrete description of the space $\ovl{\Con}_G$.
For this, we define the set of $G$-invariant homomorphisms by  
$\Homm_{G}(\Paths,\AF):=\kappa\big(\ovl{\Con}_G\big)$, and conclude
\begin{lemma}
  \label{lemma:ShapeInvHomms}
  If $\Paths$ is $G$-invariant and independent, and $\homm\in \Homm(\Paths,\AF)$, then 
  $\homm\in \Homm_G(\Paths,\AF)$ holds iff we have
  \begin{align}
  \label{osdpsdfpsd}
    \homm(g \cdot \gamma)=\theta_g\cp \homm(\gamma)\cp \theta_{g^{-1}}\qquad \forall\: g\in G,\quad\forall \:\gamma\in \Paths.
  \end{align}
 \end{lemma}
  \begin{proof}
    Let $\{\w_\alpha\}_{\alpha\in I}\subseteq \Con$ be a net with $\{\iota_\Con(\w_\alpha)\}_{\alpha\in I}\rightarrow \ovl{\w}:=\kappa^{-1}(\homm)$.
    Then, from
    \begingroup
\setlength{\leftmargini}{17pt}
    \begin{itemize}
    \item[a)]
      continuity of $\Phi_g$ by Corollary \ref{cor:CylSpecAction},
    \item[b)]
     $\Phi_g\cp \iota_\Con = \iota_\Con \cp \phi_g$ by Corollary \ref{cor:CylSpecAction},
    \item[c)]
     \eqref{gdfgfgdgf} in the proof of Lemma \ref{lemma:CylSpecActionvorb},
    \item[d)]
      continuity of $\theta_g$ for each $g\in G$,
    \item[e)]
      the definition of $\kappa$ (as well as $\{\iota_\Con(\phi(g,\w_\alpha))\}_{\alpha\in I}\rightarrow \Phi_g(\ovl{\w})$ by b), for the second line below),
    \end{itemize}
    \endgroup
    \noindent
    for each $g\in G$, we obtain
    \begin{align}
      \label{eq:InvHomTau}
      \begin{split}
    \textstyle  \kappa( \Phi(g,\ovl{\w}))(g\cdot \gamma)(p) 
      &\stackrel{\mathrm{a)}}{=}\textstyle\kappa(\lim_\alpha (\Phi_g\cp \iota_\Con)(\w_\alpha))(g\cdot \gamma)(p)\\
      &\stackrel{\mathrm{b)}}{=}\textstyle\kappa(\lim_\alpha (\iota_\Con\cp\phi_g)(\w_\alpha))(g\cdot \gamma)(p)\stackrel{\mathrm{e)}}{=}\lim_\alpha \parall{g\cdot \gamma}{\phi(g,\w_\alpha)}(p)\\
      &\stackrel{\mathrm{c)}}{=}\textstyle\lim_\alpha \theta\big(g,\parall{ \gamma}{\w_\alpha}(g^{-1}\cdot p)\big)\stackrel{\mathrm{d)}}{=}\theta\big(g,\lim_\alpha\parall{\gamma}{\w_\alpha}(g^{-1}\cdot p)\big)\\
      &\stackrel{\mathrm{e)}}{=}\textstyle\theta\big(g,\kappa(\ovl{\w})(\gamma)(g^{-1}\cdot p)\big)=\theta(g,\homm(\gamma)(g^{-1}\cdot p)),
      \end{split}
    \end{align}
    from which the claim is clear.
  \end{proof}
Now,
\begin{remdef}
  \label{rem:euklrem}
   \begingroup
\setlength{\leftmargini}{16pt}
  \begin{enumerate}
  \item
    \label{rem:euklrem1}
    If $P=M\times S$ is trivial, then $\Homm(\Paths,\AF)$ can be canonically  identified with the set $\Homm(\Paths, S)$ of all maps $\hommm\colon \Paths \rightarrow S$ that fulfil 
        the algebraic properties \textrm{(b)}-\textrm{(d)} from Definition \ref{def:hompaths}, whereby the composition in b) has to be replaced by the structure group multiplication. The corresponding bijection 
    $\Omega\colon \Homm(\Paths, \AF)\rightarrow \Homm(\Paths, S)$ then is explicitly given by
    \begin{align*}
      \Omega(\homm)(\gamma)=\pr_2\left( \homm(\gamma)(\gamma(a),e)\right)\qquad \text{for}\qquad \dom[\gamma]=[a,b]
    \end{align*}
    with inverse map
    $\Omega^{-1}(\hommm)(\gamma)(\gamma(a),s):=(\gamma(b),\hommm\big(\gamma)\cdot s\big)$. For such a trivial bundle, we define $\Homm_G(\Paths,S):=\Omega(\Homm_G(\Paths,\AF))$. 
  \item
    \label{rem:euklrem2}
    Assume that we are in the situation of Example \ref{ex:LQC}. Then, for $g=(v,\sigma)$, we have $g^{-1}=(-\uberll{\sigma^+}{v},\sigma^+)$, so that for $\gamma\in \Paths$ with $\dom[\gamma]=[a,b]$ as well as $\homm\in \Homm_E(\Paths,\AF)$, we obtain
    \begin{align*}
      \Omega(\homm)(\gamma)& =\hspace{3.8pt}(\Omega\cp \kappa)\left(\Phi(g^{-1},\kappa^{-1}(\homm))\right)(\gamma)\\
      &=\hspace{4pt}  \pr_2\cp \left(\kappa\left(\Phi(g^{-1},\kappa^{-1}(\homm))\right)(\gamma)(\gamma(a),e)\right)\\[-2pt]
      &\hspace{-3pt} \stackrel{\eqref{eq:InvHomTau}}{=}\hspace{4.2pt}[\pr_2\cp\theta]\left((-\uberll{\sigma^+}{v},\sigma^+),\homm\big(v+\uberll{\sigma}{\gamma}\big)(v+\uberll{\sigma}{\gamma(a)},\sigma)\right)\\[2pt]
      &=\hspace{3.8pt}[\Co{\sigma^+}\cp \pr_2]\big(\homm(v+\uberll{\sigma}{\gamma})(v+\uberll{\sigma}{\gamma(a)},e)\big)\\[2pt]
      &=\hspace{3.8pt}\left[\Co{\sigma^+}\cp \Omega(\homm)\right](v+\uberll{\sigma}{\gamma}).
    \end{align*}
    Consequently, $\hommm \in \Homm_E(\Paths,S)$ holds iff we have 
    \begin{align}
      \label{eq:InvGenConnRel}
      \hommm(v+\uberll{\sigma}{\gamma)}=(\Co{\sigma}\cp \hommm)(\gamma)\qquad \forall\:(v,\sigma)\in \Ge,\quad \forall\:\gamma \in \Paths.
    \end{align}
    In particular, this means that the value of $\hommm$ is independent on the starting point of the path $\gamma$. Then, it is straightforward to see that for the spherically symmetric and then (semi-)homogeneous case, we have
    \begin{align}
      \begin{array}{lrclcl}
        \label{eq:algrels}
        R \colon  & \hommm(\uberll{\sigma}{\gamma})\!\!\!&=&\!\!\!\Co{\sigma}\cp \hommm(\gamma) && \forall\:\sigma\in \SU,\quad\forall\:\gamma \in \Paths, \\
        V\colon&  \hommm(v+\gamma)\!\!\! & =&\!\!\! \hommm(\gamma)&& \forall\:\vv\in V,\hspace{31pt}\forall\:\gamma \in \Paths. 
      \end{array}
    \end{align}
    Here, $R:=\SU$ acts via $\theta_R(\sigma,(x,s)):=(\uberll{\sigma}{x},\sigma\cdot s)$, and for the linear subspace $V\subseteq \RR^3$ we define the action $\theta_V\colon V\times P\rightarrow P$, $(v,(x,s))\mapsto (v+x,s)$.
  \item
    \label{rem:euklrem3}
    Let $l\in \RR_{>0}$ and $x,\vv$ in $\RR^3$ with $\|\vv\|=1$,
    and consider the curve $\gamma\colon [0,l]\rightarrow \RR^3$, $t\mapsto  t\cdot \vv$ which we assume to be contained in $\Paths$. Moreover, assume that $\hommm \in \Hom(\Paths,\SU)$ is an element, for which the first relation in \eqref{eq:algrels} holds for all $\sigma\in \SU$. Then, for $t\in \RR$, we let  
    $\sigma_{t}:=\exp(t\cdot\murs(\vv))$ for $\murs\colon \RR^3\rightarrow \mathfrak{s}$ defined as in Example \ref{ex:LQC}. Then, $\sigma_{t}(\gamma)=\gamma$ holds for all $t\in \RR$, because we have $\gamma(0)=0$, and since $\sigma_t$ corresponds to a rotation in $\RR^3$ around the axis determined by $\vv=\dot\gamma(0)$. It follows that
    \begin{align*}
      \hommm(\gamma)=\hommm(\sigma_{t}(\gamma))\stackrel{\eqref{eq:algrels}}{=}\alpha_{\sigma_t} (\hommm(\gamma))=\sigma_t\cdot \hommm(\gamma)\cdot\sigma_t^+\qquad\quad \forall\: t\in \RR,
    \end{align*}
    and differentiation at zero gives
    \begin{align}
      \label{eq:kommu}
      0&=\frac{\dd}{\dd t}\Big|_{t=0}\sigma_t\cdot \hommm(\gamma)\cdot\sigma_t^+
      =\murs(\vv)\cdot\hommm(\gamma)-\hommm(\gamma)\cdot\murs(\vv).
    \end{align}
    Since $\hommm(\gamma) \in \SU$ holds, we have
    \begin{align*}
      \hommm(\gamma)=\begin{pmatrix} a & b  \\ -\ovl{b} & \ovl{a}  \end{pmatrix}\quad\text{ for some }\quad a,b\in\CCC\text{ with }|a|^2+|b|^2=1,
    \end{align*}
    and in combination this with \eqref{eq:kommu}, for $\vv=\vec{e}_1$, we obtain $\hommm(\gamma)\in H_{\vec{e}_1}$. Here,
    for $\vv\in \RR^3\backslash\{0\}$, the Torus $H_{\vv}\subseteq \SU$ is defined by
    \begin{align}
      \label{eq:Torus}
      H_{\vv}:=\{\exp(t\cdot\murs(\vv))\:|\: t\in [0,2\pi\backslash \|\vv\|)\}.
    \end{align}
    Now, if $\vv\in \RR^3$ with $\|\vv\|=1$ is arbitrary, we find some $\sigma\in \SU$, such that $\vv=\uberll{\sigma}{\vec{e}_1}$ holds. Then, for $\gamma_0\colon [0,l]\ni t\mapsto t\cdot \vec{e}_1$, we have $\hommm(\gamma_0)=\exp(t_l\cdot \murs(\vec{e}_1))$ for some $t\in [0,2\pi)$,
    and the first relation in \eqref{eq:algrels} gives
    \begin{align}  
      \label{eq:linearinToruss}
      \begin{split}
        \hommm(\gamma)&=\hommm(\uberll{\sigma}{\gamma_0})
        =\Co{\sigma}(\hommm(\gamma_0))
        =\Co{\sigma}(\exp(t_l\cdot\murs(\vec{e}_1)))
        =\exp(t_l\cdot \murs(\uberll{\sigma}{\vec{e}_1}))=\exp(t_l\cdot \vv)\in H_{\vv}.
      \end{split}
    \end{align}
    This will become important for the proof of the second part of Theorem \ref{th:propersubset} in Subsection \ref{subsec:QuantLevel}.\hspace*{\fill}{ $\ddagger$} 
  \end{enumerate}  
  \endgroup
\end{remdef}
Finally, a closer look at \eqref{eq:linearinToruss}, leads to the following statement (Corollary \ref{cor:AshtLew}) concerning the Ashtekar-Lewandowski measures $\mu_{\AL}$  
on $\ovl{\Con}_\w$; the quantum configurations space that corresponds to the set $\Paths_\w$ of embedded analytic curves in $\RR^3$ with compact domain, for $P=\RR^3\times \SU$. \cite{ProjTechAL} (see also Appendix B in \cite{MeasuresOnRquer}) Recall that $\mu_{\AL}$ can be characterized by the following property:\newline
\vspace{-2ex}
\newline
\noindent
Let $\alpha=(\gamma_1,\dots,\gamma_k)\subseteq \Paths_\w$ be a finite subset, such that 
\begin{align*}
  \gamma_i\cap \gamma_j\subseteq \{\gamma_i(a_i),\gamma_i(b_i),\gamma_j(a_j),\gamma_j(b_j)\}\qquad \text{holds for}\qquad 1\leq i\neq j\leq k
\end{align*} 
with $\dom[\gamma_i]=[a_i,b_i]$ for $i=1,\dots,k$.
Then, the push forward of $\mu_{\AL}$ by
\begin{align*}	
  \pi_\alpha\colon \mu_{\AL}&\rightarrow \SU^k\\
  \ovl{\w}&\mapsto \left((\Omega\cp \kappa) (\ovl{\w})(\gamma_1),{\dots},(\Omega\cp \kappa) (\ovl{\w})(\gamma_k)\right)
\end{align*}
equals the Haar measure $\mu_k$ on $\SU^{|\alpha|}$.
\begin{corollary}
  \label{cor:AshtLew}
  Let $E=\Gee$ and $R=\SU$ be as in Remark and Definition \ref{rem:euklrem}.\ref{rem:euklrem2}. Then, the Borel sets $\ovl{\Con}_{\Ge,\w}, \ovl{\Con}_{R,\w}\subseteq \ovl{\Con}_\w$ 
  are of measure zero w.r.t. $\mu_{\AL}$. 
  \end{corollary}
  \begin{proof}
    $\ovl{\Con}_{\Ge,\w}, \ovl{\Con}_{R,\w}$ are Borel sets, because they are compact by Corollary \ref{cor:CylSpecAction}. Moreover, it follows from the algebraic relations in Remark and Definition \ref{rem:euklrem}.\ref{rem:euklrem2} that $\ovl{\Con}_{\Ge,\w}\subseteq \ovl{\Con}_{R,\w}$ holds.
    Then, for $\gamma\colon [0,l]\rightarrow \RR^3$, $t\mapsto  t\cdot \vv$  and $\pi_\gamma\colon \ovl{\Con}_\w\rightarrow \SU,\:\:\ovl{\w} \mapsto (\Omega\cp \kappa)(\ovl{\w})(\gamma)$, we have
    \begin{align*}
      \mu_{\AL}\big(\ovl{\Con}_{R,\w}\big)\stackrel{\eqref{eq:linearinToruss}}{\leq} \mu_{\AL}\big(\pi_\gamma^{-1}(H_{\vec{n}})\big)=\mu_{0}(H_{\vec{n}})=0
    \end{align*}
    for the Haar measure $\mu_{0}$ on $\SU$. 
  \end{proof}

\section{Symmetry Reduction in LQG}
\label{sec:symmred}
Reduction of the LQG-configuration space, traditionally means to consider the spectrum of a $\Cstar$-algebra of the form $\aA_G=\ovl{\Cylk|_{\Con_G}}\subseteq B(\Con_G)$. \cite{MathStrucLQG,ChrisSymmLQG} Here, $G$ is a Lie group that acts via automorphisms on a $\SU$-bundle, and $\Paths$ is some set of $\CC{k}$-paths in the corresponding base manifold, such that $\aA_G$ separates the points in $\Con_G$. Conceptually, this means first to reduce the classical configuration space $\Con$, and then to quantize by considering $\ovl{\Con_G}=\Spec(\aA_G)$. 

Anyhow, the results from Subsection \ref{sec:InvGenConnes} now allow to reduce the quantum configuration space directly, and the fact that then $\ovl{\Con_G}\cong \wt{\Con_G} \subset \ovl{\Con}_G$ can hold, just says that quantization and reduction in general do not commute; i.e., that there can be some loss of information when first reducing and then quantizing. 

In the present section, this will be worked out for homogeneous isotropic LQC. This means that $P=\RR^3\times \SU$ and $\Ge=\Gee$ are as in Example \ref{ex:LQC}, and that $\Paths$ essentially denotes the set of all linear, or all embedded analytic curves in $\RR^3$. 
For this, we will need the concept of the Bohr compactification of a locally compact abelian group (LCA group), being reviewed in the next short subsection, cf. Chapter $1$ in \cite{RudinFourier}.

\subsection{Bohr compactification}
If $G$ is an LCA group, the dual group $\Gamma$ of $G$ is the set of continuous homomorphisms $\chi\colon G\rightarrow U(1)\subseteq \CCC$, endowed with the group structure
\begin{align*}
  (\chi*\chi')(g):=\chi(g)\cdot\chi'(g),\qquad\quad \chi^{-1}(g):=\ovl{\chi(g)},\qquad \quad 1_\Gamma\colon g\mapsto 1 \qquad\quad \forall\:g\in G.
\end{align*}
Then, $\Gamma$ becomes an LCA group, when equipped with the topology generated by the sets 
\begin{align*}
B_{K,\epsilon}(\chi):=\{\chi'\in \Gamma\:|\: |\chi(g)-\chi'(g)|< \epsilon\quad\forall\:g\in K\}.
\end{align*}
Here, $K\subseteq G$ is compact, and we have $\chi\in \Gamma$, and $\epsilon>0$.
 The Pontryagin duality then states that for $\widehat{\Gamma}$ the dual of $\Gamma$,
the map $j\colon G\rightarrow \widehat{\Gamma}$,\: $j(g)\colon \chi\mapsto\chi(g)$ is an isomorphism and a homeomorphism.

Now, if we equip $\Gamma$ with the 
discrete topology, we obtain a further LCA group $\Gamma_d$. The Bohr compactification $\GB$ of $G$ is defined to be the dual of $\Gamma_d$. Then, $\GB$ is compact, because this is always the case for duals of discrete LCA groups. Moreover, we have $\widehat{\Gamma}\subseteq \GB$, because $\GB$ equals the set of all homomorphisms $\psi\colon \Gamma\rightarrow U(1)$, whereby $\widehat{\Gamma}$ consists of the continuous (w.r.t.\ the  topology on $\Gamma$) ones. 
One then can show that the map $i_{\mathrm{B}}\colon G\rightarrow \GB$ defined as $j$ above, is a continuous isomorphism to the dense subgroup $\iB(G)\subseteq \GB$. Now, since $\Gamma_d$ is discrete, each compact set is finite, so that  
the topology on $\GB$ is generated by the sets
\begin{align*}
B_{\chi,\epsilon}(\psi):=\{\psi'\in \GB\:|\: |\psi(\chi)-\psi'(\chi)|< \epsilon\}\qquad\text{for}\quad\epsilon>0\quad\text{and}\quad \chi\in \Gamma_d.
\end{align*}
Moreover, it is not hard to see that $\GB$ is isomorphic and homeomorphic to the spectrum of the $\Cstar$-algebra $\CAP(G)$ of almost periodic functions\footnote{This is the $\Cstar$-subalgebra of $B(G)$ that is generated by the elements of $\Gamma$.} on $G$ via, cf., e.g., Lemma 3.8 in \cite{MAXTH}
\begin{align}
\label{osdapopsao}
	\mathrm{r}\colon \Spec(\CAP(G))\rightarrow \GB,\quad \psi\mapsto\psi|_{\Gamma_d},
\end{align}
with what the group structure on $\GB$ carries over to $\Spec(\CAP(G))$ in a suitable way. Then, under the above identification, the map $i_B$ corresponds to the map $\iota_G$ from Convention \ref{conv:Boundedfunc} with $G=G_\aA$ for $\aA=\CAP(G)$, i.e., is a group homomorphism. 

If $G$ is the additive group of real numbers, then $\Gamma$ consists of the homomorphism
$\chi_\tau\colon x\mapsto e^{\I\tau x}$ for $\tau\in \RR$.
\subsection{Group structures and actions}
\label{subsec:NoGoes}
Let $X$ be a locally compact Hausdorff space, and $C_b(X)\subseteq B(X)$ the set of continuous bounded  functions on $X$. For an open subset $Y\subseteq X$, let $\aA_0:=C_{0,Y}\subseteq C_b(X)$ denote the set of the continuous functions on $X$ that vanish at infinity and outside $Y$. Finally, assume that $\aA_1\subseteq C_b(X)$ is some unital $\Cstar$-subalgebra, such that $\aA_0\cap \aA_1=\emptyset$ holds, and $\aA:=\aA_0 \oplus \aA_1\subseteq C_b(X)$  is closed. Equip the space $Y\sqcup \Spec(\aA_1)$ with the topology generated by the sets of the following types, cf. \cite{ChrisSymmLQG}:
\begin{align*}
  \begin{array}{lcrclcl}
    \text{Type 1:} && V & \!\!\!\sqcup\!\!\! & \emptyset 
    && \text{with open $V \subseteq Y$,} \\
    \text{Type 2:} && K^c & \!\!\!\sqcup\!\!\! & \Spec(\aA_1)
    && \text{with compact $K \subseteq Y$,} \\
    \text{Type 3:} && f^{-1}(U) & \!\!\!\sqcup\!\!\! & \Gel(f)^{-1}(U) 
    && \text{for open $U \subseteq \CCC$ and $f \in \aA_1$.}
  \end{array}
\end{align*}
Then, Proposition 3.4 in \cite{ChrisSymmLQG} states that $\Spec(\aA)\cong Y\sqcup \Spec(\aA_1)$ holds via the homeomorphism $\xi\colon Y\sqcup \Spec(\aA_1)\rightarrow \Spec(\aA)$, defined by
\begin{equation}
  \label{eq:Ksiii}
  \xi(\psi) := 
  \begin{cases} 
    f\mapsto f(\psi) &\mbox{if } \psi\in Y\\
    f_0\oplus f_1\mapsto \psi(f_1)  & \mbox{if } \psi\in \Spec(\aA_1)
  \end{cases}
\end{equation}
for $f_0\in \aA_0$ and $f_1\in \aA_1$. Moreover, observe that the corresponding relative topologies of $Y$ and $\Spec(\aA_1)$  coincide with their usual ones.

We now will consider the case where $X=\RR$ and $\aA_1=\CAP(\RR)$ holds, and 
 where $Y\subseteq \RR$ is a non-empty open subset. In this case, Corollary B.2 in \cite{ChrisSymmLQG} shows that $\aA_0\cap \aA_1=\emptyset$ holds, and that $\aA=\aA_0 \oplus \aA_1\subseteq C_b(X)$ is closed. Then, we have
\begin{theorem}
  \label{theorem:noGroupStruc}
  There is no continuous group structure on $\qR=Y\sqcup \RB$. 
\end{theorem}
\begin{proof}
    Assume there is such a group structure with multiplication $\star$ and identity $e$.
    In a first step, we show that there is some $\psi\in \qR$, for which the continuous\footnote{Recall that the relative topology of $\RB$ w.r.t.\ that on $\qR$, equals its usual one.} 
    restriction
    $\star[\psi,\cdot\:]|_{\RB}$ 
    takes at least one value in $\oRR$. 
    
    In fact, elsewise, $\star[\psi,\RB]\subseteq \RB$ holds for each $\psi \in \qR$, so that for $\psi\in \RB\neq \emptyset$, we obtain $e=\star\big[\psi^{-1},\psi\big]\in\RB$, hence $\qR=\star\big[\qR,e\big]\subseteq \RB$, which is impossible as $Y\neq \emptyset$ holds. 
    
    Thus, we find $\psi \in \qR$ and $\psi\in \RB$, such that $\psi\star \psi\in \oRR$ holds.
    Then, the preimage $U$ of $Y$ under the continuous map $\star[\psi,\cdot\:]|_{\RB}$  is a non-empty open subset of $\RB$, and since $\RB$ is compact, finitely many translates of the form $\psi+ U$ cover $\RB$. Thus, for the cardinality of $\RB$, we have
    \begin{align*}
      | \RB |&=\left|\bigcup_{i=0}^n\psi_i+U\right|\leq \left|\bigsqcup_{i=0}^n U\right| 
      =\left|\bigsqcup_{i=0}^n \star[\psi,U]\right|\leq n|\RR|=|\RR|.
    \end{align*}
    But, this is impossible, because $|\RB|> |\RR|$ holds. 
    In fact, if $\{\tau_\alpha\}_{\alpha\in I}\subseteq \RR$ is a $\mathbb{Q}$-base of $\RR$, then $|I|=|\RR|$ holds,\footnote{$\RR$ equals the set $F$ of all finite subsets of $\mathbb{Q}\times I$. Then, $|F|=|\mathbb{Q}\times I|$ holds since $\mathbb{Q}\times I$ is infinite. Similarly, we have $|\mathbb{Q}\times I|=|I|$, because $I$ is infinite.} and we obtain an injective map $\tau\colon 2^I\rightarrow \RB$ in the following way.
    For $J\subseteq I$, we let 
    \begin{align*}
      \delta_J(\alpha):=\left\{
	\begin{array}{ll}
          0  & \mbox{if } \alpha\in J  \\
          \frac{2\pi}{\tau_\alpha} & \mbox{if }  \alpha\notin J,
	\end{array}\right. 
    \end{align*}
    and define\footnote{Due to \eqref{osdapopsao}, it suffices to specify the values of some element of $\RB$ on the subset $\Gamma\subseteq \CAP(\RR)$.} $\tau(J)\colon \Gamma\rightarrow U(1)$ by $\tau(J)(\chi_0):=1$, as well as
    \begin{align*}    	
      \textstyle\tau(J)\left(\chi_{\tau}\right):=\prod_{i=1}^n \chi_{q_i\cdot \tau_i}(\delta_J(\alpha_i))\qquad \text{for} \qquad\tau=\sum_{i=1}^l q_i\cdot \tau_{\alpha_i}\quad \text{with}\quad q_1,\dots,q_l \in \mathbb{Q}.
    \end{align*}
    Then, $\tau$ is injective, because
    $\tau(J)(\chi_{q\cdot \tau_\alpha})=1$ holds for each $q\in \mathbb{Q}$ iff we have $\alpha\in J$. This shows 
    $|\RB|\geq |\mathcal{P}(I)|=|\mathcal{P}(\RR)|>|\RR|$.
  \end{proof}
In the sequel, we will only be concerned with the case where $Y=\RR$ holds, so that we let $\qR:=\RR\sqcup \RB$ in the following. Observe that then $\qR$ is homeomorphic via \eqref{eq:Ksiii} to the spectrum of $C_0(\RR)\oplus \CAP(\RR)$.

We close this subsection
with an application of
Proposition \ref{prop:autspec} to $\RB$; a further application can be found in  \cite{{MEAS}}. For this, let $\phi\colon \RR_{\neq 0}\times \RR\rightarrow \RR$ denote the multiplicative action, and conclude that
\begin{corollary}
  \label{cor:bohrmultimitlambda}
  There exists a unique left action $\Phi\colon \RR_{\neq 0}\times \RB \rightarrow \RB$, such that $\Phi_\lambda\in \Autt(\RB)$ is continuous for each $\lambda\in\RR_{\neq 0}$, and for which we have
  \begin{equation}
    \label{eq:Bohrmult}
    \Phi(\lambda,\iota_{\RR}(x))=\iota_\RR(\lambda\cdot x) \qquad \forall\: \lambda\in\RR_{\neq 0},\quad \forall\: x\in \RR.
  \end{equation}
  This action is not continuous.
\end{corollary}
  \begin{proof}
  $\aA=\CAP(\RR)$ is $\phi$-invariant because $\phi_\lambda^*\chi_\tau= \chi_{\lambda \tau}$ holds, so that $\Phi$ is the unique left action from Proposition \ref{prop:autspec}.
    This action is not continuous, because $\phi_\bullet^*\chi_\tau\colon \lambda\mapsto \chi_{\lambda \tau}$ is not continuous for $\tau\neq 0$. Then, $\Phi_\lambda\in \Autt(\RB)$ is clear from \eqref{eq:Bohrmult} and continuity of $\Phi_\lambda$, because $\phi_\lambda\in \Autt(\RR)$ holds, and since $\iota_\RR$ is a homomorphism. 
  \end{proof}

\subsection{Loop quantum cosmology}
\label{subsec:LQC}
In this subsection, we will discuss the configuration spaces that have been used for homogeneous isotropic LQC so far. At the same time, we will provide the facts and definitions that we will need to prove the inclusion result announced in the beginning of this section.

Thus, for what follows, let $P=\RR^3\times \SU$, $(\Ge,\theta)$, and $\Con_{\Ge}$ be as in Example \ref{ex:LQC}.
Moreover, let 
$\Paths_\w$ denote the set of {\bf embedded analytic  
curves} in $\RR^3$, i.e., the set of all curves with compact domain, which are the restriction of an analytic immersive embedded curve, defined on an open interval.
 Let us first emphasize the most relevant types.
\begin{definition}[Linear and circular curves]
  \label{def:lincirccurves}
     \begingroup
\setlength{\leftmargini}{16pt}
  \begin{enumerate}
  \item
    \label{def:lincirccurves1}
    Let $\Paths_\lin$ consist of all linear curves of the form  $\gamma \colon [a,b]\rightarrow \RR^3$, $t\mapsto x+ (t-a)\cdot \vv$, for some $x,\vv\in \RR^3$ with $\|\vv\|\neq 0$, i.e.,   
    $\gamma$ starts at $x$ and ends at $x+(b-a)\cdot\vv$. 
    Of course, $\Paths_\lin$ is closed under inversion and decomposition of its elements, and each $\gamma\in \Paths_\lin$ is equivalent to one curve of the form 
\begin{align*}
	x+\gamma_{\vv,l}\qquad\text{for}\qquad x, \vv\in \RR^3 \qquad\text{with}\qquad \|\vv\|=1 \qquad\text{and}\qquad  \gamma_{\vv,l}\colon [0,l]\rightarrow \RR,\:\: t\mapsto t \cdot \vv.
\end{align*}    
  \item
    \label{def:lincirccurves2}
    For $\vec{n},\vec{r},x\in \RR^3$ with $\|\vec{n}\|=1$ and $0< \tau< 1$, define
    \begin{align}
    \label{fggffdg}
    \begin{split}
      \gc{n}{r}{x}{\tau} \colon [0,2\pi \tau]&\rightarrow \RR^3\\
      t&\mapsto x + \cos(t)\cdot\vec{r} + \sin(t) \cdot\vec{n}\times \vec{r}.  
      \end{split}
    \end{align}
    This is the curve which starts at $x+ \vec{r}$, has absolute velocity $\|\dot\gamma_{\vec{n},\vec{r}}^{x,\tau}(t)\|=\|\vec{r}\|$ for all $0\leq t\leq 2\pi \tau$, and  traverses in the plane orthogonal to $\vec{n}$ on a circular orbit with center $x$ and winding number $\tau$ in counterclockwise rotation w.r.t.\ $\vec{n}$. The set of all such curves will be denoted by $\Paths_{\mc}$ in the following. Obviously, $\Paths_{\mc}$ is closed under decomposition and inversion of its elements.
  \item	
    \label{def:lincirccurves3}
    We define $\Paths_{\lin\mc}:=\Paths_\lin\sqcup\Paths_\mc$.
  \end{enumerate}
  \endgroup
\end{definition}
Then,
\begin{Remark}[Special holonomies]
  \label{rem:paralltransp}
   \begingroup
\setlength{\leftmargini}{16pt}
  \begin{enumerate}
  \item
    \label{rem:paralltransp1}
    If $\gamma\colon [a,b]\rightarrow \RR^3$ is a curve with $\dot\gamma(t)=\vv$ for each $t\in [a,b]$, its horizontal lift in $(\gamma(a),e)$ w.r.t.\ $\w^c\in \Con_E$ is given by $\wt{\gamma}^c_e(t)=\left(\gamma(t),\exp\left(-c\cdot [t-a] \cdot\murs(\vv)\right)\right)$, so that we have
    \begin{align}
      \label{eq:paralleukl}
      \pr_2\cp \parall{\gamma}{\w^c}(\gamma(a),e)= \cos(c\cdot [b-a]\cdot \|\vv\|)\cdot \mathbb{1} - \sin(c\cdot [b-a]\cdot\|\vv\|)\cdot \murs\left(\vv\slash\|\vv\|\right).
    \end{align}
  \item
    \label{rem:paralltransp2}  
    Let $\gamma_\tau:=\gamma_{\vec{e}_3,r\cdot \vec{e}_1}^{0,\tau}\in \Paths_\mc$, hence 
 	\begin{align}
 	\label{oppopasdsa}
 		 \gamma_\mm\colon [0,\mm]\rightarrow \RR,\quad t\mapsto r\cdot (\cos(t)\cdot\vec{e}_1+\sin(t)\cdot \vec{e}_2)
 	\end{align}    
    for $0<\mm<2\pi$. 
    Then, it can be deduced from\footnote{Alternatively, confer Lemma 5.6.1 in \cite{MAXTH}.} Subsection 5.1 in \cite{ChrisSymmLQG} that for $\pc{c}:= \sqrt{c^2r^2+\frac{1}{4}}$, we have
    \begin{align}
      \label{eq:paralllc}
      \begin{split}
        \pr_2\cp \parall{\gamma_\mm}{\w^c}(\gamma_\mm(0),e)
        &=\begin{pmatrix} \e^{-\frac{\I}{2}\mm}\left[\cos(\pc{c} \mm)+\frac{i}{2\pc{c}}\sin(\pc{c} \mm)\right] &\frac{cr}{\pc{c}} \e^{-\frac{\I}{2}\mm}\sin(\pc{c} \mm)  \\ -\frac{cr}{\pc{c}}\e^{\frac{\I}{2}\mm}\sin(\pc{c} \mm) & \e^{\frac{\I}{2}\mm}\left[\cos(\pc{c} \mm)-\frac{i}{2\pc{c}}\sin(\pc{c} \mm)\right]  \end{pmatrix}\\[5pt]
        &=\exp\left(\textstyle\frac{\tau}{2}\cdot  \tau_3\right)
        \cdot
        \underbrace{\begin{pmatrix} \cos(\pc{c} \mm)+\frac{i}{2\pc{c}}\sin(\pc{c} \mm) &\frac{cr}{\pc{c}} \sin(\pc{c} \mm)  \\ -\frac{cr}{\pc{c}}\sin(\pc{c} \mm) & \cos(\pc{c} \mm)-\frac{i}{2\pc{c}}\sin(\pc{c} \mm)  \end{pmatrix}}_{A(\tau,c)}.
      \end{split} 
    \end{align} 
    Then, if $\gamma
    \in \Paths_{\mc}$ is arbitrary, we find $x\in \RR^3$ and $\sigma\in \SU$, such that $\gamma=x+\uberll{\sigma}{\gamma_\mm}$ holds for some $0<\mm<1$. Then, property \eqref{eq:InvGenConnRel} of $(\Omega\cp \kappa)(\iota_\Con(\w^c)) \colon\gamma\mapsto \left(\pr_2\cp \parall{\gamma}{\w^c}\right)(\gamma(0),e)$ shows that  
    \begin{align*}
      \pr_2\cp \parall{\gamma}{\w^c}(x,e)&\stackrel{\eqref{eq:InvGenConnRel}}{=}\Co{\sigma}\big(\pr_2\cp \parall{\gamma_\mm}{\w^c}(\gamma_\mm(0),e)\big)
      \stackrel{\eqref{eq:paralllc}}{=}\Co{\sigma}\big( \exp\left(\textstyle\frac{\tau}{2}\cdot\tau_3\right)
        \cdot A(\tau,c) \big)\\
      &\hspace{3.3pt}=\hspace{3.3pt}\Co{\sigma}\left(\exp(\textstyle\frac{\tau}{2}\cdot\tau_3)\right)\cdot \Co{\sigma}(A(\tau,c)) 
      =\exp(\textstyle\frac{\tau}{2}\cdot \murs(\vec{n}))\cdot \Co{\sigma}(A(\tau,c)).
    \end{align*} 
    For this, recall that $(\Omega\cp \kappa)(\iota_\Con(\w^c))\in \Hom_E(\Paths_\w,\SU)$ holds, because we have $\ovl{\iota_\Con(\Con_E)}\subseteq \ovl{\Con}_E$ by \eqref{eq:inclusionsdiag}.
     \hspace*{\fill}{ $\ddagger$} 
  \end{enumerate}
  \endgroup
\end{Remark} 
Now, for a set of paths $\Paths_\alpha$, let $\iota_\Con$, $\ovl{i^*_{\Con_\Ge}}$ denote the maps from \eqref{eq:inclusionsdiag}, and define
\begin{center} 
  \makebox[0pt]{
    \extrarowheight5pt
    \begin{tabular}{|c|c|c|c|c|c|c|} \hline
      $\Cyl{\Paths_\alpha}$ &  $\ovl{\cC_\alpha|_{\Con_\Ge}}$ &  $\Spec(\cC_\alpha)$& $\Spec(\aA_{E,\alpha})$ &$\ovl{\iota_\Con(\Con_\Ge)}$& $\ovl{\Con}_\Ge$& $\ovl{i^*_{\Con_\Ge}}$
      \\[2pt] \hline
      $\cC_\alpha$ &  $\aA_{E,\alpha}$ &  $\ovl{\Con}_\alpha$&  $\ovl{\Con_{\Ge,\alpha}}$ & $\wt{\Con_{\Ge,\alpha}}$&$\ovl{\Con}_{\Ge,\alpha}$ & $\ovl{i_\alpha^*}$\\[2pt]\hline
    \end{tabular}
  }
\end{center}
Then, $\ovl{\Con_{\Ge,\alpha}}\cong \wt{\Con_{\Ge,\alpha}}\subseteq \ovl{\Con}_{\Ge,\alpha}\subseteq \ovl{\Con}_\alpha$ holds by \eqref{eq:inclusionsdiag}, whereby the isomorphism is the map $\ovl{i_\alpha^*}\colon \ovl{\Con_{\Ge,\alpha}}\rightarrow \wt{\Con_{\Ge,\alpha}}$. In the following, we will restrict to the cases where $\alpha\in\{\w,\lin\mc,\lin\}$ holds, whereby the relations of the corresponding spaces are sketched in the next diagram.	
\newpage
\begin{figure}[h]
  \begin{center}
    \makebox[0pt]{
      \begin{xy}
        \xymatrix{
          \qR\ar@{->}[r]^-{\cong} &\wt{\Con_{\Ge,\w}}\ar@{->}[r]^-{\subset}\ar@{|->}[d]_-{\ovl{j_1^*}}^-{\cong}  &**[r] \ovl{\Con}_{\Ge,\w} \ar@{|->}[d]^-{\text{restr.}} \ar@{->}[r]^-{\subset} &  \ovl{\Con}_\w\ar@{|->}[d]^-{\text{restr.}}\\
          \qR\ar@{->}[r]^-{\cong} &\wt{\Con_{\Ge,{\lin\mc}}}\ar@{|->}[d]_-{\ovl{j_2^*}}\ar@{->}[r]^-{\subset} & **[r]\ovl{\Con}_{\Ge,{\lin\mc}} \ar@{|->}[d]^-{\text{restr.}}\ar@{->}[r]^-{\subset}&\ovl{\Con}_{\lin\mc} \ar@{|->}[d]^-{\text{restr.}}\\
          \RB\ar@{->}[r]^{\cong}& \wt{\Con_{\Ge,{\lin}}}\ar@{->}[r]^-{=}    & **[r]\ovl{\Con}_{\Ge,{\lin}}\ar@{->}[r]^-{\subset}&\ovl{\Con}_{\lin}. 
        }
      \end{xy}
    }
  \end{center}
  \vspace{-4ex}
\end{figure}

\noindent
Now, to explain this diagram, let us identify $\Con_{\Ge}\subseteq \Con$ with $\RR$ via the map $\ishomc\colon \RR\rightarrow \Con_E$,\:\:$c \mapsto \w^c$. This will allow us to consider the spaces on the left hand side of the diagram as spectra of certain $\Cstar$-subalgebras of the bounded functions on $\RR$.
\begingroup
\setlength{\leftmargini}{12pt}
\begin{itemize}
\item	
  The restrictions on the right hand side can easily be understood in terms of homomorphisms of paths. For this observe that the sets $\Paths_\w$, $\Paths_{\lin\mc}$, $\Paths_\lin$ are independent by Lemma \ref{lemma:analytCurvesIndepetc}.\ref{lemma:analytCurvesIndepetc5}, so that $\ovl{\Con}_\alpha  \cong \Homm(\Paths_{\alpha},\Iso_F)$ holds for each $\alpha \in\{ \w,\lin\mc,\lin\}$. 
  Thus, for $\hommm\in \Hom(\Paths_\w,\Iso_F)$, the restriction of $\hommm$ to $\Paths_{\lin\mc}$ is an element in $\Hom(\Paths_{\lin\mc},\Iso_F)$, and the analogous statement holds for $\lin\mc$ and $\lin$ instead of $\w$ and $\lin\mc$, respectively. Moreover, if $\hommm$ is $\Ge$-invariant, the same is obviously true for the respective restriction.
\item
The equality $\wt{\Con_{E,\lin}}=\ovl{\Con_{E,\lin}}$ and the proper inclusion $\wt{\Con_{E,\w}}\subset\ovl{\Con_{E,\w}}$ are shown in Theorem \ref{th:propersubset}. The proper inclusion for $\alpha=\lin\mc$, then follows by the same argument as in the embedded analytic case, but we will not prove this in the following.
\item
To see that the inclusions $\ovl{\Con}_{\Ge,\alpha}\subseteq \ovl{\Con}_\alpha$ are indeed proper, it suffices to consider $\hommm\in \Homm(\Paths_\w,\SU)$, defined by\footnote{Here, well-definedness follows by the same arguments as in Theorem \ref{th:propersubset}, so that we will omit the proof at this point. Basically, one has to show that an embedded analytic curve $\gamma$ is already equivalent to some linear curve if this is the case for a subcurve of $\gamma$.}
  \[ 
  \hommm(\gamma):= 
  \begin{cases} 
    \exp\big(l \cdot \langle\vv,\vec{e}_1\rangle \cdot \tau_1\big) &\quad\mbox{if } \gamma \csim [\hspace{0.5pt}x+ \gamma_{\vv,l}] \in \Paths_{\lin}\\
    e & \quad\mbox{else}. 
  \end{cases}
  \] 
 In fact, applying a rotation by $\pi/2$ around $\vec{e}_2$, we see that already the restriction of $\epsilon$ to $\Paths_\lin$ cannot be $E$-invariant.
\item
  \itspace
  The maps $\ovl{j_1^*}$ and $\ovl{j_2^*}$
  on the vertical arrows on the left hand side arise from Lemma \ref{lemma:homzuspec}.\ref{lemma:homzuspec1} applied to the canonical injections $j_1\colon \aA_{E,\lin\mc}\rightarrow \aA_{E,\w}$ and $j_2\colon \aA_{E,\lin}\rightarrow \aA_{E,\lin\mc}$, so that 
  \begin{align*}
  	\ovl{j_1^*}\colon \chi\mapsto \chi|_{\aA_{E,\lin\mc}}\qquad\qquad\text{and}\qquad\qquad 	\ovl{j_2^*}\colon \chi\mapsto \chi|_{\aA_{E,\lin}}
  \end{align*}
 are just restriction maps.
  The isomorphism property of $\ovl{j_1^*}$ then corresponds to the fact 
  that linear and circular curves are sufficient to generate $\aA_{E,\w}$, i.e., that we have $\aA_{E,\w}=\aA_{E,\lin\mc}$. This was proven in \cite{ChrisSymmLQG}, 
  by computation of $\ishomc^*(\aA_{E,\w})$ and $\ishomc^*(\aA_{E,\lin\mathrm{c}})$, both turning out to equal $C_{0}(\RR) \oplus \CAP(\RR)$ whose spectrum is homeomorphic to $\qR=\RR\sqcup \RB$ equipped with the topology defined in Subsection \ref{subsec:NoGoes}. \cite{ChrisSymmLQG} 
\item
  \itspace
  We have $\wt{\Con_{E,\lin}}\cong \ovl{\Con_{E,\lin}}=\Spec(\aA_{E,\lin})\cong\RB$ as, by 
  Remark \ref{rem:paralltransp}.\ref{rem:paralltransp1}, the $^*$-algebra $\ishomc^*(\Cyls{\lin})\subseteq B(\RR)$ is generated by the homomorphisms $\chi_\tau$. In fact, then we have
  \begin{align*}  
    \ishomc^*(\aA_{E,\lin})&= \ishomc^*\big(\hspace{1pt}\ovl{\cC_\lin|_{\Con_{\Ge}}}\hspace{1pt}\big)=\ishomc^*\big(\hspace{1pt}\ovl{\Cyl{\Paths_\lin}|_{\Con_{\Ge}}}\hspace{1pt}\big)= \ishomc^*\big(\hspace{1pt}\ovl{\Cyls{\lin}|_{\Con_{\Ge}}}\hspace{1pt}\big) =\ovl{\ishomc^*\left(\Cyls{\lin}\right)}=\CAP(\RR),
  \end{align*}
  hence $\ovl{\Con_{E,\lin}}=\Spec(\aA_{E,\lin})\cong \Spec(\ishomc^*(\aA_{E,\lin}))=\Spec(\CAP(\RR))= \RB$.\hspace*{\fill} {$\ddagger$}
\end{itemize}
\endgroup
\noindent
Now, in the framework of LQG, the configuration space of the full gravitational theory is given by $\ovl{\Con}_\w=\Spec(\cC_\w)$, whereby that of homogeneous isotropic LQC is traditionally chosen to be $\ovl{\Con_{\Ge,\lin}}\cong \RB$. \cite{MathStrucLQG} 
\vspace{6pt}

\noindent
Then, in \cite{Brunnhack}, it was shown that there is no possibility to extend the injection $\ishomc$ to an embedding of $\RB$ into $\ovl{\Con}_\w$, i.e., that there is no embedding $\varsigma_0\colon \RB\rightarrow \ovl{\Con}_\w$, such that {\bf Diagram I)} commutes.
This arises from the fact that, for the definition of $\RB\cong \Spec(\aA_{E,\lin})$ and $\ovl{\Con}_\w =\Spec(\cC_\w)$ different sets of curves are used. In fact, due to \cite{ChrisSymmLQG}, it is a necessary condition for existence of $\varsigma_0$ that $\ishomc^*(\cC_\w)\subseteq \CAP(\RR)$ holds. But, this is not the case as there exist functions in $\cC_\w$, whose pullbacks by $\ishomc$ are elements in $C_0(\RR)$.
In contrast to that, there exists such an embedding $\varsigma_\lin$ for $\ovl{\Con}_{\Ge,\lin}=\Spec(\aA_{E,\lin})$, just by Lemma \ref{lemma:dicht}.\ref{lemma:dicht3}, cf.\ {\bf Diagram II)}.\footnote{In this diagram, $\ovl{i_\lin^*}\colon \Spec(\aA_{E,\lin})\rightarrow \Spec(\cC_\lin)$ denotes the embedding that corresponds to the injection $i\colon \Con_E \rightarrow \Con$, and analogously for the isomorphism $\ovl{\ishomc^*}$.}
\begin{figure}[h]
  \begin{minipage}[h]{0,42\textwidth}
    \begin{center}
      \makebox[0pt]{
        \begin{xy}
          \xymatrix{
            \RB  \ar@{->}[r]^-{\varsigma_0}   &  \ovl{\Con}_\w  \\
            \RR \ar@{->}[r]^-{\ishomc}_-{\phantom{\cong}}\ar@{->}[u]^{\iota_{\RR}}
            &  \Con  \ar@{->}[u]^{\iota_\Con}
          }
        \end{xy}
      }
    \end{center}	
    \vspace{-4ex}
    \caption*{{\bf I)}\: $\ovl{\Con}_\w= \Spec(\cC_\w)$}
    \label{lincurves}
    \subcaption*{\textbf{no such embedding} $\varsigma_0$}
  \end{minipage}
  \begin{minipage}[h]{0,56\textwidth}
    \begin{center}
      \makebox[0pt]{
        \begin{xy}
          \xymatrix{
            \RB  \ar@{->}[r]^-{\ovl{\ishomc^*}}_-{\cong}   &  \ovl{\Con_{E,\lin}}\ar@{->}[r]^-{\ovl{i_\lin^*}} & \ovl{\Con}_\lin  \\
            \RR \ar@{->}[r]^-{\ishomc}_-{\cong}\ar@{->}[u]^{\iota_{\RR}} 
            & \: \Con_E\:  \ar@{->}[u]^{\iota_{\Con_E}}\ar@{^{(}->}[r]^{i}   &\:\Con\: \ar@{->}[u]^-{\iota_\Con} 
          }
        \end{xy}
      }
    \end{center}	
    \vspace{-4ex}
    \caption*{{\bf II)}\: $\ovl{\Con}_\lin= \Spec(\cC_\lin)$}
    \label{lincurves}
    \subcaption*{\textbf{embedding}  $\varsigma_\lin:= \ovl{i_\lin^*}\cp \ovl{\ishomc^*}\colon \RB\rightarrow \ovl{\Con}_\lin$}
  \end{minipage}
\end{figure}

\noindent
Then, in order to obtain a reduced configuration space that is canonically embedded into $\ovl{\Con}_\w$, in \cite{ChrisSymmLQG}, the restriction $\Cstar$-algebra $\ishomc^*(\aA_{E,\w})=\ovl{(i_\w\cp\ishomc)^*(\cC_\w)}=C_0(\RR) \oplus \CAP(\RR)$ has been  introduced, and the embedding property of the map
\begin{align*}
\ovl{i_\w^*}\cp \ovl{\ishomc^*}\colon \Spec(\ishomc^*(\aA_{E,\w})) \rightarrow \ovl{\Con}_\w,
\end{align*}
hence, of the map $\varsigma=\ovl{i^*_\w}\cp \ovl{\ishomc^*}\cp \xi\colon \qR\rightarrow \A_\w$ 
has been verified, cf.\ {\bf Diagram III)}.
\begin{figure}[h]
  \begin{minipage}[h]{\textwidth}
    \begin{center}
      \makebox[0pt]{
        \begin{xy}
          \xymatrix{
            \ovl{\RR} \ar@{->}[r]^-{\xi}_-{\cong} &\Spec(\ishomc^*(\aA_{E,\w})) \ar@{->}[r]^-{\ovl{\ishomc^*}}_-{\cong}   &   \ovl{\Con_{E,\w}}\ar@{->}[r]^-{\ovl{i_\w^*}} & \ovl{\Con}_\w\\ 
            & \RR \ar@{->}[u]^-{\iota_\RR} \ar@{->}[r]^-{\ishomc}_-{\cong}    &\: \Con_E\: \ar@{->}[u]^-{\iota_{\Con_E}} \ar@{^{(}->}[r]^{i}   &\Con \ar@{->}[u]^-{\iota_\Con} 
          }
        \end{xy}
      }
    \end{center}
    \vspace{-3.5ex}
    \caption*{{\bf III)}\: \textbf{embedding} $\varsigma:=\ovl{i_\w^*}\cp\ovl{\ishomc^*}\cp \xi\colon \qR\rightarrow \ovl{\Con}_\w$}
    \label{anaccurves}
  \end{minipage}
\end{figure}

\noindent
 In the next subsection, we will use this embedding, in order to show that $\wt{\Con_{E,\w}}\subset \A_\w$ holds; and for this, the following lemma will be crucial. 
\begin{lemma}
  \label{lemma:nullBohr}
  Let $\xi\colon \qR\rightarrow \Spec(C_{0}(\RR) \oplus \CAP(\RR))$ be defined by \eqref{eq:Ksiii}. Then,  
  $\xi(\psi)(\chi_\tau)=1$ holds for each $\tau\in \RR$ iff we have $\psi\in \{0_\RR,0_{\mathrm{Bohr}}\}$.   
  \end{lemma}
  \begin{proof}
    Obviously, $\xi(\psi)(\chi_\tau)=1$ holds for all $\tau\in \RR$ if we have $\psi\in \{0_\RR,0_{\mathrm{Bohr}}\}$. Moreover, if $\xi(\psi)(\chi_\tau)=1$ holds for each $\tau\in \RR$ for $\psi=y\in \RR$, we must have $y= 0$, because $\xi(y)(\chi_{\pi/2y})=\I$ holds for $y\neq 0$. Finally, if we have $\psi\in \RB$, then $\psi(\chi_\tau)=1=0_{\mathrm{Bohr}}(\chi_\tau)$ for each $\tau \in \RR$ already implies $\psi=0_{\mathrm{Bohr}}$, because  the functions $\chi_\tau$ generate $\CAP(\RR)$.
  \end{proof}
So far, we have illustrated that, in contrast to $\RB$,
the space $\qR$ has the advantage to be canonically embedded (via $\varsigma$) into the quantum configuration space $\ovl{\Con}_\w$ of full LQG. 
Then, the next step would be to construct a reasonable measure on $\qR$, defining a suitable kinematical $L^2$-Hilbert space. 
Motivated by $\RB$, one might first look for Haar measures on $\qR$; but, 
 due to Theorem \ref{theorem:noGroupStruc}, no such measure can exist. Thus, one should lower the ambitions, and investigate all finite Radon\footnote{This means an inner regular and locally finite measure, defined on the Borel $\sigma$-algebra of $\qR$.} measures on $\qR$. The next lemma is supposed to serve as a starting point for this, cf.\ also Conclusions in \cite{ChrisSymmLQG}
\begin{lemma}
  \label{lemma:Radon}
  Let $\BRq$, $\Borel(\RR)$, and $\Borel(\RB)$ denote the Borel $\sigma$-algebras of the topological spaces $\qR$, $\RR$, and $\RB$, respectively. Then,
  \begingroup
\setlength{\leftmargini}{16pt}
  \begin{enumerate}
  \item	
    \label{lemma:Radon1}
    We have $\BRq=\Borel(\RR)\sqcup\Borel(\RB)$.
  \item
    \label{lemma:Radon2}
    If $\mu$ is a finite Radon measure on $\BRq$, then $\mu|_{\Borel(\RR)}$ and $\mu|_{\Borel(\RB)}$ are finite Radon measures as well. Conversely, if $\mu_\RR$ and $\mu_B$ are finite Radon measures on $\Borel(\RR)$ and $\Borel(\RB)$, respectively, then 
    \begin{align}
      \label{eq:RadonMeasures}	
      \mu(A):=\mu_\RR(A\cap \RR)+ \mu_B(A\cap \RB)\qquad \forall\: A\in \BRq	
    \end{align}
    is a finite Radon measure on $\BRq$. 
  \end{enumerate}
  \endgroup
\end{lemma}  
  \begin{proof}
    \begingroup
\setlength{\leftmargini}{16pt}
    \begin{enumerate}
    \item
     The right hand side is a $\sigma$-algebra, and contains the $\sigma$-algebra on left hand side because $U\cap \RR\in \Borel(\RR)$ as well as $U\cap \RB\in \Borel(\RB)$ holds for each open subset $U\subseteq \qR$. Now, $\Borel(\RR)\subseteq \Borel(\qR)$ holds, because if $U\subseteq \RR$ is open, then it is open in $\qR$ as well. Finally, $\Borel(\RB)\subseteq \qR$ holds  because if $A\subseteq \RB$ is closed, it is compact, hence compact (closed) in $\qR$. 
    \item
      The restrictions are well defined by the first part, and obviously finite. Their inner regularity is clear, because a subset of $\RR$ or $\RB$ is compact iff it is compact w.r.t.\ the topology on $\qR$.      
      Now, for the second statement, let $\mu$ be defined by \eqref{eq:RadonMeasures}. Then, $\mu$ is a finite Borel measure by the first part, and its inner regularity follows by a simple $\epsilon\slash 2$ argument from the inner regularities of $\mu_\RR$ and $\mu_B$.
    \end{enumerate}
    \endgroup		
  \end{proof}
Lemma \ref{lemma:Radon} shows that each normalized Radon measure on $\qR$ can be written as
\begin{align*}
  \mu_t(A):=t\:\mu_\RR(A\cap \RR)+ (1-t)\:\mu_B(A\cap \RB) \qquad \forall\: A\in \BRq
\end{align*}
for $t\in [0,1]$, and normalized Radon measures $\mu_\RR$ and $\mu_B$ on $\Borel(\RR)$ and $\Borel(\RB)$, respectively. 
Thus, it remains to fix the measures $\mu_\RR$ and $\mu_B$, as well as the parameter $t$. For $\mu_\RR$ and $\mu_B$, this might be done by writing $\qR$ as a projective limit in a suitable way;\footnote{During the evaluation of the present article, this has been done in \cite{MeasuresOnRquer}. Moreover, in \cite{{MEAS}}, Proposition \ref{prop:autspec} has been used, in order to single out the Bohr measure by means of the same invariance property on both the standard cosmological configuration space $\RB$ as well as on $\qR$.}
and for the dependence of the induced Hilbert space structure on the parameter $t$, observe that for fixed $\mu_\RR$, $\mu_B$, and $t_1,t_2\in (0,1)$, the spaces $\Lzw{\qR}{\mu_{t_1}}$ and $\Lzw{\qR}{\mu_{t_2}}$ are isometrically isomorphic. 
In fact, for $A\in \BRq$, let $\chi_A$ denote the respective characteristic function, and define the map
\begin{align*}
  \varphi \colon \Lzw{\qR}{\mu_{t_1}}&\rightarrow \Lzw{\qR}{\mu_{t_2}}\\
  \psi & \mapsto  \sqrt{\frac{t_1}{t_2}} \:\chi_{\RR}\cdot \psi + \sqrt{\frac{(1-t_1)}{(1-t_2)}}\:\chi_{\RB}\cdot \psi.
\end{align*}
Then, it is immediate that $\varphi$ is an isometric isomorphism, so that the parameter $t$ gives rise to at most $3$ different Hilbert space structures in this case.

\subsection{Symmetry reduction on quantum level}
\label{subsec:QuantLevel}
In this final subsection, we want to clarify the inclusion relations between $\RB$ and $\ovl{\Con}_{E,\lin}$, as well as that between $\qR$ and $\ovl{\Con}_{E,\w}$, cf.\ the {\bf Diagrams II)} and {\bf III)}, respectively. 

As already mentioned in the previous subsection, the sets $\Paths_\alpha$ are independent for $\alpha=\{\w,\lin\mc,\lin\}$ by Lemma \ref{lemma:analytCurvesIndepetc}.\ref{lemma:analytCurvesIndepetc5}, so that the corresponding map $\kappa\colon\ovl{\Con}_\alpha  \rightarrow \Homm(\Paths_\alpha,\AF)$ is bijective. 
Now, define $\nu:=\{(x,e)\}_{x\in \RR^3}$, and let $\Omega$ be as in Remark and Definition  \ref{rem:euklrem}.\ref{rem:euklrem1}. Then, for $\gamma\in \Paths_\alpha$ with $\dom[\gamma]=[a,b]$ and $\ovl{\w}\in \ovl{\Con}_\alpha$, we have
\begin{align}
  \label{eq:hilfseq}
  \begin{split}
    \textstyle(\Omega\cp \kappa)(\ovl{\w})(\gamma)&=\textstyle\pr_2(\kappa(\ovl{\w})(\gamma)(\gamma(a),e))=\pr_2\big(\lim_\beta\parall{\gamma}{\w_\beta}(\gamma(a),e)\big)\\
    &= \textstyle\lim_\beta h_\gamma^\nu(\w_\beta)=\lim_\alpha \left(\iota_\Con(\w_\beta) \left([h_\gamma^\nu]_{ij}\right)\right)_{ij}\\
    &= \left(\ovl{\w}([h_\gamma^\nu]_{ij})\right)_{ij}\in \SU,
  \end{split}
\end{align}
whereby $\{\w_\beta\}_{\beta\in I}\subseteq \Con$ is a net with $\{\iota_\Con(\w_\beta)\}_{\beta\in I}\rightarrow \ovl{\w}$. For the last line, observe that there are only finitely components $[h_\gamma^\nu]_{ij}$, each of them contained in $\aA_\alpha$.

Now, in the analytic case, let $\Delta\colon \qR \rightarrow \Homm_\Ge(\Paths_{\w},\SU)$
denote the  
injective composition $\Omega\cp \kappa\cp \varsigma$, for 
$\varsigma$  the map from {\bf Diagram III)}. Then,  
we have the following corollary to Lemma \ref{lemma:nullBohr}.
\begin{corollary}
  \label{cor:nullbohr}
  If $\Delta(\psi)(\gamma)=e$ holds for each $\gamma\in \Paths_\lin\subseteq \Paths_\w$,  
 we have $\psi\in\{0_{\mathrm{Bohr}},0_\RR\}$. 
\end{corollary} 
  \begin{proof} 
    If $\gamma\colon [0,l]\rightarrow \RR^3$ is a curve with $\dot\gamma(t)=\vec{e}_1$ for all $t\in [0,l]$, then 
    \begin{align}
      \label{eq:DeltavonRquer}
      \begin{split}
        \Delta(\psi)(\gamma)&\!\!\stackrel{\eqref{eq:hilfseq}}{=}\left(\varsigma(\psi)\left([h_\gamma^\nu]_{ij}\right)\right)_{ij}=\left(\big(\ovl{i_\w^*}\cp\ovl{\ishomc^*}\cp\xi\big)(\psi)\left([h_\gamma^\nu]_{ij}\right)\right)_{ij}
        =\left(\xi(\psi)\left([h_\gamma^\nu\cp \ishomc]_{ij}\right)\right)_{ij}\\
        &=\hspace{3pt}\left(\xi(\psi)\left(\Big[\pr_2\cp \parall{\gamma}{\ishomc(\cdot)}\Big]_{ij}\right)\right)_{ij}\stackrel{\eqref{eq:paralleukl}}{=}\begin{pmatrix} \xi(\psi)(c\mapsto \cos(l c)) &  \xi(\psi)(c\mapsto \I\sin(l c))  \\  \xi(\psi)(c\mapsto \I\sin(l c)) & \xi(\psi)(c\mapsto \cos(l c)) \end{pmatrix},
      \end{split}
    \end{align}
    hence $\xi(\psi)(\chi_l)=1$ for all $l\in \RR$. Thus, the claim is clear from  
    Lemma \ref{lemma:nullBohr}.
\end{proof}
Now, before we come to the final statement, let us first recall the equivalence relations 
$\psim$ and $\isim$ on $\Paths_\w$, defined in Appendix \ref{sec:HomPaths}. 
This is that for $\gamma, \gamma'\in \Paths_\w$ with $\dom[\gamma]=[a,b]$ and $\dom[\gamma']=[a',b']$, we have
\begingroup
\setlength{\leftmargini}{12pt}
\begin{itemize}
\item
$\gamma \psim \gamma'$ iff $\gamma=\gamma'\cp \rho$ holds for some analytic diffeomorphism $\rho\colon I\rightarrow I'\subseteq\RR$ with $\rho([a,b])=[a',b']$ and
$\dot\rho>0$.
\item
$\gamma \isim \gamma'$\hspace{3pt} iff $\im[\gamma]=\im[\gamma']$, $\gamma(a)=\gamma'(a')$, and $\gamma(b)=\gamma'(b')$ holds.
\end{itemize}
\endgroup
\noindent
Then, in the proof of the next theorem, we will make free use of the fact that
the equivalence relations $\csim, \psim$, and $\isim$ mutually coincide, cf.\ Lemma \ref{lemma:analytCurvesIndepetc}.\ref{lemma:analytCurvesIndepetc3}.
\begin{theorem}
  \label{th:propersubset}
  We have\quad$\qR\cong \wt{\Con_{\Ge,\w}}\subset \ovl{\Con}_{\Ge,\w}$\quad as well as\qquad $\RB\cong \wt{\Con_{\Ge,\lin}}=\ovl{\Con}_{E,\lin}$.
\end{theorem} 
  \begin{proof}  
  {\bf Statement 1:} 
  \vspace{4pt}

\noindent
    For the first statement, let $\tilde{\hommm} \colon \RR_{>0}\times (0,1)\rightarrow \RR$ be a function with
    \begin{align*}
      \tilde{\hommm}(r,x+y)=\tilde{\hommm}(r,x)+\tilde{\hommm}(r,y)\quad\mathrm{mod}\quad 2\pi,
    \end{align*}
    whenever $r\in \RR_{>0}$ and $x,y,x+y\in (0,1)$ holds. Then, we obtain an element $\hommm \in \Homm_\Ge(\Paths_\w,\SU)$ 
    if we define
    \[ 
    \hommm(\gamma):= 
    \begin{cases} 
      \exp(\tilde{\hommm}(\|\vec{r}\|,\tau)\cdot \murs(\vec{n})) &\quad\mbox{if } \gamma\csim \gc{n}{r}{x}{\tau}\in \Paths_{\mc} \\
      e & \quad \mbox{else}. 
    \end{cases}
    \] 
    In fact, since $\csim$ and $\isim$ coincide, we have
    \begin{align*}
    	\gc{n}{r}{x}{\tau}\csim \gamma_{\vec{n}',\vec{r}'}^{x',\tau'}\qquad\quad \Longleftrightarrow\qquad\quad \vec{r}=\vec{r}',\quad \vec{n}=\vec{n}',\quad x=x',\quad \tau=\tau',
    \end{align*}
    so that $\hommm$ is well defined as a map. Then, the homomorphism properties are easily verified for such $\gamma$ that are equivalent to some element in $\Paths_c$. For the other curves, it suffices to show that
	\begin{align*}
		\gamma\not\csim \gamma_c\qquad\forall\: \gamma_c\in \Paths_{\mc}\qquad\qquad\Longrightarrow\qquad\qquad \gamma|_{K}\not\csim \gamma_c\qquad\forall\: \gamma_c\in \Paths_{\mc},
	\end{align*}    
	for each compact interval $K\subseteq \dom[\gamma]$. For this, let us assume that $\gamma|_{K}\csim \delta|_{K'}$ holds for some  $\delta\colon \RR\rightarrow \RR^3$, defined as the elements of $\Paths_\mc$ by  (cf.\ \eqref{fggffdg})
\begin{align*}
      \delta\colon t\mapsto x + \cos(t)\cdot\vec{r} + \sin(t) \cdot\vec{n}\times \vec{r}.  
\end{align*}
Then, since $\sim_\Con$ and $\psim$ coincide, we have $\gamma|_K=\delta|_{K'}\cp \rho|_K$ for some analytic diffeomorphism $\rho\colon I\rightarrow I'\subseteq \RR$ with $\rho(K)=K'$ and $\dot\rho>0$. Let $\wt{\rho}\colon \wt{I}\rightarrow \wt{I}'\subseteq  \RR$ denote its maximal analytic immersive extension, and define $D:=\wt{I}\cap \dom[\gamma]$. 
\begingroup
\setlength{\leftmargini}{12pt}
\begin{itemize}
\item
If $D=\dom[\gamma]$ holds, we have done, because then analyticity implies $\gamma=\delta\cp\wt{\rho}|_{\dom[\gamma]}$; hence $\gamma\isim \delta|_L$ for $L:=\wt{\rho}(\dom[\gamma])$ compact. For this observe that $\delta|_L\in \Paths_\mc$ holds, because we must have $\sup(L)-\inf(L)<2\pi$, since otherwise $\gamma$ cannot be injective.
\item
In the other case, we find a boundary point $t$ of $D$, such that each open interval containing $t$ intersects $D$ and $\dom[\gamma]\backslash D$ non-trivially.

Let $\{t_n\}_{n\in \NN}\subseteq D\backslash\{t\}$ be a monotonous sequence with $\lim_n t_n=t$. 
 Then, $\lim_n\wt{\rho}(t_n)=t'\in \RR$ must exist, because $\sup[\wt{\rho}(D)]-\inf[\wt{\rho}(D)]\leq 2\pi$ must hold by injectivity of $\gamma$. Then, Lemma \ref{lemma:analytCurvesIndepetc}.\ref{lemma:analytCurvesIndepetc1} applied to a suitable extension $\wt{\gamma}$ of $\gamma$ and a suitable restriction of $\delta$, provides us with open intervals $J,J'$ with $t\in J$ and $t'\in J'$, such that $\wt{\gamma}|_J=\delta\cp\tau$ holds for some analytic diffeomorphism $\tau\colon J\rightarrow J'$ that necessarily coincides on $J\cap D\neq \emptyset$ with $\wt{\rho}$. Thus, $\wt{\rho}$ can be extended to $D\cup J\supset D$, which contradicts the maximality of $D$ as $J\cap [D\backslash \dom[\gamma]]$ is non-empty.
\end{itemize}
\endgroup
\noindent
Thus, to finish the proof of the first part, it remains to show the invariance \eqref{eq:InvGenConnRel} of $\hommm$, and that we can choose $\wt{\hommm}$ in such a way that $\hommm\notin \im[\Delta]$ holds. The first point is clear for such $\gamma$ that are not equivalent to any curve in $\Paths_\mc$, just because $\Co{\sigma}(e)=e$ holds for all $\sigma\in \SU$. Anyhow, if $\gamma\csim \gc{n}{r}{x}{\tau}$ holds, we obviously have
\begin{align*} 
   v+\sigma(\gamma)\csim \gcc{\sigma(\vec{n})}{\sigma(\vec{r})}{v+\sigma(x)}{\tau}\qquad\Longrightarrow\qquad \hommm(v+\sigma(\gamma))&=\exp(\tilde{\hommm}(\|\sigma(\vec{r})\|,\tau)\cdot \murs(\sigma(\vec{n})))\\
   &= \exp(\tilde{\hommm}(\|\vec{r}\|,\tau)\cdot \Ad(\sigma)(\murs(\vec{n})))\\
   &=\Co{\sigma}(\exp(\tilde{\hommm}(\|\vec{r}\|,\tau)\cdot\murs(\vec{n})))\\
   &= \Co{\sigma}\cp\hommm(\gamma).
\end{align*}
  Then, for $\tilde{\hommm}_0(r,\tau):=0$ and $\tilde{\hommm}_\pm(r,\tau):=\pm 2\pi \tau$, we have
  \begin{align*}
  \hommm_0\big(\gc{n}{r}{x}{\tau}\big)=e\qquad\quad\text{and}\qquad\quad \hommm_\pm\big(\gc{n}{r}{x}{\tau}\big)\stackrel{\eqref{eq:expSU2}}{=}\pm \tau_i\quad\text{for}\quad \textstyle \tau=\frac{1}{4}\quad\text{and} \quad\vec{n}=\vec{e}_i, 
  \end{align*}
  as well as $\hommm_0(\gamma)=\hommm_\pm(\gamma)=e$ for all $\gamma\in \Paths_\lin$. Thus, if $\wt{\Con_{\Ge,\w}}= \ovl{\Con}_{\Ge,\w}$ would hold, Corollary \ref{cor:nullbohr} would imply $\hommm_0,\hommm_\pm\in \Delta( \{ 0_{\mathrm{Bohr}},0_{\RR}\})$, which is impossible because $\hommm_0\neq \hommm_+\neq \hommm_{-}\neq \hommm_0$ holds. 
\vspace{8pt}

\noindent
{\bf Statement 2:}
\vspace{4pt}

\noindent
We need to construct a preimage of each $\hommm\in \Homm_\Ge(\Paths_\lin,\SU)$ under the map
\begin{align*}
	\Omega\cp\kappa\cp \varsigma_\lin\colon \RB \rightarrow \Hom_\Ge(\Paths_\lin,\SU)
\end{align*}   
for $\varsigma_\lin\colon \RB\rightarrow \ovl{\Con}_\lin$ as defined in {\bf Diagram II)} in Subsection \ref{subsec:LQC}. For this, let us first observe that each $\gamma\in \Paths_\lin$ is equivalent to a linear curve of the form $x+\gamma_{\vv,l}$ for some $x,\vv\in\RR^3$ and $l\in \RR_{>0}$ with $\|\vv\|=1$, cf.\ Definition \ref{def:lincirccurves}.\ref{def:lincirccurves1}. 

Then, for each $\hommm\in \Homm_\Ge(\Paths_\lin,\SU)$, we have $\hommm(x+\gamma_{\vv,l} )=\hommm(\gamma_{\vv,l})$, so that each such $\hommm$ is completely determined by its values on the curves $\gamma_{\vv,l}$. Moreover, since for each $\vv'\in \RR^3$ with $\|\vv'\|=1$, we find some $\sigma\in \SU$ with $\sigma(\vv)=\vv'$, \eqref{eq:InvGenConnRel} shows that $\hommm$ is even completely determined by its values on the curves $\gamma_{\vv,l}$ for some fixed $\vv$, and all $l\in \RR_{>0}$. Now, since $\hommm(\gamma_{\vv,l})\in H_{\vv}$ holds by \eqref{eq:linearinToruss}, we have
	\begin{align*}
    	\hommm(\gamma_{\vv,l})=\exp(\wt{\epsilon}(l)\cdot \murs(\vv))\qquad\forall\: l\in \RR_{>0},
    \end{align*}
    whereby $\wt{\epsilon}(l)\in [0,2\pi)$ is uniquely determined by the properties of the exponential map of $\SU$, cf.\ \eqref{eq:expSU2}. Then, $\wt{\epsilon}\colon \RR_{>0}\rightarrow [0,2\pi)$ defined in this way, necessarily fulfills
    \begin{align} 
    \label{eq:modEpsilon}
      \tilde{\hommm}(l_1+l_2)=\tilde{\hommm}(l_1)+\tilde{\hommm}(l_2)  \quad\mathrm{mod} \quad 2\pi\qquad\quad\forall\: l_1,l_2\in \RR_{>0}
    \end{align}
    again by the properties of the exponential map of $\SU$, because we have
    \begin{align*}
    	\exp(\wt{\epsilon}(l_l+l_2)\cdot \murs(\vv))&=\hommm(\gamma_{\vv,l_1+l_2})=\hommm(l_1\cdot \vv + \gamma_{\vv,l_1})\cdot \hommm(\gamma_{\vv,l_2})=\hommm(\gamma_{\vv,l_2})\cdot \hommm(\gamma_{\vv,l_1})\\
    	&=\exp(\wt{\epsilon}(l_1)\cdot \murs(\vv))\cdot \exp(\wt{\epsilon}(l_2)\cdot \murs(\vv))=\exp([\hspace{1pt}\wt{\epsilon}(l_1)+\wt{\epsilon}(l_2)]\cdot \murs(\vv)).
    \end{align*}
     Now, let us define $\psi_\hommm\in \RB$ by $\psi_\hommm(\chi_0):=1$ and $\psi_\hommm(\chi_\tau):=\e^{\I \sign(\tau)\:\tilde{\hommm}(|\tau|)}$ for $\tau\neq 0$. 
    Then, $\psi_\hommm$ is well defined, because
    \begin{align*}
      \psi_\hommm(\chi_{\tau}*\chi_{\tau'})
      =\e^{\I \sign(\tau+\tau')\: \tilde{\hommm}(|\tau+\tau'|)}
      =\e^{\I \sign(\tau)\:\tilde{\hommm}(|\tau|)} \e^{\I \sign(\tau')\:\tilde{\hommm}(|\tau'|)}
      =\psi_\hommm(\chi_{\tau})\:\psi_\hommm(\chi_{\tau'}).
    \end{align*}
    holds. Here, the second equality is clear for $\sign(\tau)=\sign(\tau')$, and 
    follows from 
    \begin{align*}
      \tilde{\hommm}(\tau)=\tilde{\hommm}(\tau-\tau'+\tau')\stackrel{\eqref{eq:modEpsilon}}{=}\tilde{\hommm}(\tau-\tau')+\tilde{\hommm}(\tau')\quad\mathrm{mod}\quad 2\pi  \qquad\quad\text{for }\tau>\tau'>0
    \end{align*}
    in the other cases. Then, for $\hommm':=(\Omega\cp \kappa\cp \varsigma_\lin)(\psi_\hommm)$ and $\gamma=\gamma_{\vec{e}_1,l}$, we have
    \begin{align*}
      \hommm'(\gamma)& \stackrel{\eqref{eq:hilfseq}}{=}\Big(
        \varsigma_\lin(\psi_\hommm)\big(\big[h^\nu_{\gamma}\big]_{ij}\big)\Big)_{ij}
      =\Big(
        \psi_\hommm\big(\big[h^\nu_{\gamma}\cp \ishomc\big]_{ij}\big)\Big)_{ij}
      \\
      &\stackrel{\eqref{eq:paralleukl}}{=}\begin{pmatrix} \psi_\hommm(c\mapsto \cos(l c)) &  \psi_\hommm(c\mapsto \I\sin(l c))  \\   \psi_\hommm(c\mapsto \I\sin(l c))& \psi_\hommm(c\mapsto \cos(l c)) \end{pmatrix}
      =\begin{pmatrix} \cos(\tilde{\hommm}(l)) & \I \sin(\tilde{\hommm}(l))  \\ \I  \sin(\tilde{\hommm}(l)) & \cos(\tilde{\hommm}(l)) \end{pmatrix}\\
      &\:\:= \exp(\tilde{\hommm}(l)\cdot \murs(\vec{e}_1))=\hommm(\gamma_{\vec{e}_1,l}),
    \end{align*}
    so that for $\gamma_{\vv,l}\in \Paths_\lin$ and $\sigma\in \SU$ with $\uberll{\sigma}{\vec{e}_1}=\vv$, we have
    \begin{align*}
      \hommm'(\gamma_{\vv,l})=\hommm'(\uberll{\sigma}{\gamma_{\vec{e}_1,l}})&=(\Co{\sigma}\cp \hommm')(\gamma_{\vec{e}_1,l})
      =(\Co{\sigma}\cp \hommm)(\gamma_{\vec{e}_1,l})=\hommm(\uberll{\sigma}{\gamma_{\vec{e}_1,l})}=\hommm(\gamma_{\vv,l}).
    \end{align*}
    Since this holds for all $l\in \RR_{>0}$, we must have $\hommm'=\hommm$, as both homomorphisms are completely determined by their images on the curves $\{\gamma_{\vv,l}\}_{l\in \RR_{>0}}$.
  \end{proof} 
Let us close this section with the following
\begin{Remark} 
  \label{rem:bohrmeasurezeugs}
  It is immediate from the proof of the second statement in Theorem \ref{th:propersubset} that the continuous group structure on $\RB$ corresponds to the group structure on $\Homm_\Ge(\Paths_\lin,\SU)$ defined by
  \begin{align*}
    (\hommm_1*\hommm_2)(\gamma):=\hommm_1(\gamma)\cdot \hommm_2(\gamma)\qquad\quad \hommm^{-1}(\gamma):=\hommm(\gamma)^{-1}\qquad\quad  e(\gamma):=e\qquad\qquad\forall\: \gamma\in \Paths_\lin.
  \end{align*}  
  These operations are well defined, because $[\hommm_1(\gamma),\hommm_2(\gamma)]=0$ holds for all $\hommm_1,\hommm_2 \in \Homm_\Ge(\Paths_\lin,\SU)$ and each $\gamma \in \Paths_\lin$, just because     
       $\hommm(\gamma_{\vv,l})\in H_{\vv}$ holds for all 
       $\hommm \in \Homm_\Ge(\Paths_\lin,\SU)$ by  \eqref{eq:linearinToruss}.
  Now, we are interested in measures on the spaces $\ovl{\Con}_{\Ge,\w}$ and $\ovl{\Con_{\Ge,\w}}$; and since
  \begin{equation*}
    \begin{split}
      \ovl{\Con_{\Ge,\w}} &\cong \wt{\Con_{\Ge,\w}}\subseteq \ovl{\Con}_{\Ge,\w} \cong \Homm_\Ge(\Paths_{\w},\SU)\\
      \ovl{\Con_{\Ge,\w}}\cong \ovl{\Con_{\Ge,\lin\mc}}&\cong \wt{\Con_{\Ge,\lin\mc}}\subseteq \ovl{\Con}_{\Ge,\lin\mc} \cong \Homm_\Ge(\Paths_{\lin\mc},\SU)
    \end{split}
  \end{equation*}
  holds,  one might ask whether the same group structure can also be defined on one of the spaces $\Homm_\Ge(\Paths_{\w},\SU)$ or $\Homm_\Ge(\Paths_{\lin\mc},\SU)$. 
  Here, the crucial question is, whether $[\hommm_1(\gamma),\hommm_2(\gamma)]=0$
  holds for all $\gamma \in \Paths_{\mc}$, and all $\hommm_1,\hommm_2 \in \Homm_\Ge(\Paths_{\lin\mathrm{c}},\SU)$, which is unfortunately not the case. In fact, let 
\begin{align*}
\gamma_\mm\colon [0,2\pi \mm]\rightarrow \RR,\quad t\mapsto r \cdot(\cos(t)\cdot \vec{e_1}+\sin(t)\cdot\vec{e}_2)
\end{align*}  
be as in \eqref{oppopasdsa}, as well as $\Homm_\Ge(\Paths_{\lin\mathrm{c}},\SU)\ni\hommm_c\colon \gamma\mapsto \pr_2\cp \parall{\gamma}{\w^c}(\gamma(0),e)$ be as in \eqref{eq:paralllc}. Then, for 
\begin{align*}
  	\hspace{-24.6pt}\textstyle x:=\frac{1}{r}\sqrt{\frac{\pi^2}{\tau^2}-\frac{1}{4}}\qquad\qquad\text{and}\qquad\qquad y:=\frac{1}{2r}\sqrt{\frac{\pi^2}{\tau^2}-1},
\end{align*}
  we have $\tau \cdot \beta_x=\tau\cdot \sqrt{x^2r^2+\frac{1}{4}}=\pi$\quad and\quad $\tau\cdot \beta_y=\tau\cdot \sqrt{y^2r^2+\frac{1}{4}}=\frac{\pi}{2}$,
  and obtain from \eqref{eq:paralllc} that
  \begin{align*}
    \hommm_{x}(\gamma_\mm)=\begin{pmatrix} -\e^{-\I\frac{\tau}{2}} & 0 \\[3pt] 
     0 &  -\e^{\I\frac{\tau}{2}} \end{pmatrix}\qquad\qquad\text{and}\qquad\qquad\hommm_{y}(\gamma_\mm)=
    \begin{pmatrix} \frac{\I \tau}{\pi}\e^{-\I\frac{\tau}{2}} & \frac{2yr\tau}{\pi}\e^{-\I\frac{\tau}{2}} \\[3pt] 
    - \frac{2yr\tau}{\pi}\e^{\I\frac{\tau}{2}} & - \frac{\I \tau}{\pi}\e^{\I\frac{\tau}{2}}  \end{pmatrix}
  \end{align*} 
  holds, hence $[\hommm_{x}(\gamma_\mm),\hommm_{y}(\gamma_\mm)]\neq 0$. \hspace*{\fill} {$\ddagger$} 
\end{Remark}

\section{Conclusions}
As we have seen in Lemma \ref{lemma:CylSpecActionvorb}, quantum reduction is always possible if the set of curves $\Paths$ used to define the quantum configuration space of the full theory is invariant under the action induced on the base manifold. This is the case in the standard situation where $\Paths=\Paths_\w$ is the set of embedded analytic curves, provided that the action under consideration is analytic. For instance, this is the case for homogeneous isotropic, spherically symmetric, and homogeneous LQC; and in each of these situations, invariance also holds  for the set $\Paths_\lin\subseteq \Paths_\w$ of linear curves. For homogeneous isotropic LQC, this  made it possible to compare quantum-reduction with the traditional approach for both $\Paths_\lin$ and $\Paths_\w$. More precisely,  
in Theorem \ref{th:propersubset}, we have shown that quantization and reduction commute in the linear, but not in the analytic case. 

Moreover, quantum-reduced configuration spaces are always embedded into the quantum configuration space of the full theory, as well as determined by the algebraic relation \eqref{osdpsdfpsd} if the set of curves under consideration is additionally independent and closed under decompositions and inversions.\footnote{By Lemma \ref{lemma:analytCurvesIndepetc}.\ref{lemma:analytCurvesIndepetc5}, each subset $\Paths\subseteq \Paths_\w$ is automatically independent if it is  closed under decompositions and inversions.} In fact, then the quantum configuration space of the full theory can be identified with a certain space of homomorphisms of paths, whereby the quantum-reduced configuration space then consists of exactly such homomorphism, which fulfil condition \eqref{osdpsdfpsd}. 

In Theorem \ref{theorem:noGroupStruc}, we have shown that no Haar measure exists on the cosmological configuration space $\RR\sqcup \RB$; and in Lemma \ref{lemma:Radon}, we have characterized the finite Radon measures on this space. 
\vspace{6pt}

\noindent
{\it Comment: During the evaluation of this article, in \cite{MAXTH}, it was shown that quantization and reduction do not commute in (semi)-homogeneous LQC as well. Moreover, using Proposition \ref{prop:autspec}, in \cite{MEAS}, the Bohr measure has been singled out on $\RB$ and $\RR\sqcup\RB$ by means of the same invariance condition; and in \cite{MeasuresOnRquer} further Radon measures have been constructed on $\RR\sqcup\RB$ by means of projective structure techniques. 
}

\section*{Acknowledgements}
The author thanks Christian Fleischhack for numerous discussions
and many helpful comments on a draft of the present article. 
He is grateful for general support by Benjamin Schwarz and discussions with various members of the math faculty of the University of Paderborn. Moreover, he thanks Alexander Stottmeister and Benjamin Bahr for providing him with physical background.
The author has been supported by the Emmy-Noether-Programm of
the Deutsche Forschungsgemeinschaft under Grant FL~622/1-1.

\appendix
\section*{Appendix}

\section{Technical Proofs}
This section contains the proof of Lemma \ref{lemma:CylSpecActionvorb}. Moreover, we determine the set of homogeneous isotropic connections \eqref{eq:euklconns} introduced in Example \ref{ex:LQC}, and prove a claim made in the introduction.
\label{sec:homisoconns}
\subsection{The proof of a lemma in Subsection \ref{sec:InvGenConnes}} 
\label{subsec:InvCyl}
\noindent\textbf{Proof of Lemma \ref{lemma:CylSpecActionvorb}}: 
Let $\Paths\ni \gamma\colon [a,b]\rightarrow M$, $\w\in \Con$, and $g\in G$. Moreover, denote by $\gamma_p^\w$ the horizontal lift of $\gamma$ w.r.t.\ $\w$ in $p\in \pi^{-1}(\gamma(a))$, and define
\begin{align*}
  \vg:=\IndA(g,\gamma),\qquad\quad\vg_p^\w:= \theta(g,\gamma_p^\w),\qquad\quad \vw:=\phi(g,\w), \qquad\quad \vp:=\theta(g,p).
\end{align*}
Then, $\vg_p^\w(a)=\vp$ holds, as well as $\pi\cp \vg_p^\w(t)=\pi \cp \theta(g,\gamma_p^\w(t))=\IndA(g,\gamma(t))=\vg(t)$ for each $t\in[a,b]$, and
\begin{align*}
  \vw_{\vg_p^\w(t)}\big(\vpdotw\big)=(\theta_{g^{-1}}^*\w)_{\vg_p^\w(t)}\big(\vpdotw\big)=\w_{\gamma_p^\w(t)}(\dot{\gamma}_p^\w(t))=0.
\end{align*}
This shows that $\vg_p^\w(t)$ is the horizontal lift of $\vg$ w.r.t.\ $\w$ in $\vp$, hence
\begin{align}
\label{gdfgfgdgf}
  \parall{g\cdot \gamma}{\phi(g,\w)}(\theta(g,p))=\parall{\vg}{\vw}(\vp)=\vg_p^\w(b)=\theta\big(g,\parall{\gamma}{\w}(p)\big).
\end{align}
Then, if we substitute $p$ by $p':=\theta(g^{-1},p)$, as well as $\gamma$ by $\gamma':=\IndA(g^{-1},\gamma)$, we obtain 
$\parall{\gamma}{\vw}(p)=\theta(g,\parall{\gamma'}{\w}(p'))$.

Now, let $\nu=\{\nu_x\}_{x\in M}\subseteq P$ and $\psi_x$, be as in Definition \ref{def:Connectionss}. Then, for $p=\nu_{\gamma(a)}$ and $\gamma'$, $p'$ as above, we obtain
from the morphism properties of the maps $\theta(g,\cdot)$, $\parall{\gamma'}{\w}$, and $\psi_{\gamma(b)}$ that
\begin{align}
  \label{eq:trafogenerators}
  \big(\phi_g^*\hspace{1pt}h_\gamma^\nu\big)(\w)&=h_\gamma^\nu(\phi(g,\w))=\psi_{\gamma(b)}\cp \parall{\gamma}{\vw}\big(\nu_{\gamma(a)}\big)\nonumber\\
  &=\psi_{\gamma(b)}\cp \theta\left(g,\parall{\gamma'}{\w}\left(\theta\big(g^{-1},\nu_{\gamma(a)}\big)\right)\right)\nonumber\\
  &=\psi_{\gamma(b)}\cp \theta\left(g,\parall{\gamma'}{\w}\left(\nu_{\gamma'(a)}\right)\cdot\psi_{\gamma'(a)}\left(\theta\big(g^{-1},\nu_{\gamma(a)}\big)\right)\right)\\
  &=\psi_{\gamma(b)}\left[\theta\left(g,\parall{\gamma'}{\w}\left(\nu_{\gamma'(a)}\right)\right)\cdot\psi_{\gamma'(a)}\left(\theta\big(g^{-1},\nu_{\gamma(a)}\big)\right)\right]\nonumber\\
  &=\psi_{\gamma(b)}\left[\theta(g,\nu_{\gamma'(b)})\cdot\psi_{\gamma'(b)}\left(\parall{\gamma'}{\w}\left(\nu_{\gamma'(a)}\right)\right)\right]\cdot\psi_{\gamma'(a)}\left(\theta\big(g^{-1},\nu_{\gamma(a)}\big)\right)\nonumber\\
  &=\psi_{\gamma(b)}\left[\theta(g,\nu_{\gamma'(b)})\right]\cdot\psi_{\gamma'(b)}\left(\parall{\gamma'}{\w}\left(\nu_{\gamma'(a)}\right)\right)\cdot\psi_{\gamma'(a)}\left(\theta\big(g^{-1},\nu_{\gamma(a)}\big)\right)\nonumber\\
  &=\underbrace{\psi_{\gamma(b)}\left[\theta(g,\nu_{\gamma'(b)})\right]}_{\delta_1}\cdot\: h^\nu_{\gamma'}(\w)\cdot\underbrace{\psi_{\gamma'(a)}\left(\theta\big(g^{-1},\nu_{\gamma(a)}\big)\right)}_{\delta_2}.\nonumber
\end{align}
Thus, for the generators $\rho_{ij}\cp h_\gamma^\nu$ of $\Cylk$, we have
\begin{align*}
 \textstyle \phi^*_g\left(\rho_{ij}\cp h_\gamma^\nu\right)=\rho_{ij}\left(\delta_1\cdot h_{\gamma'}^\nu \cdot\delta_2\right)=\sum_{p,q} \rho_{ip}(\delta_1)\cdot \rho_{qj}(\delta_2)\cdot \rho_{pq}\cp h_{\gamma'}^\nu \in \Cylsk,
\end{align*}
so that $\phi_g^*\left(\Cylsk\right)\subseteq \Cylk$ holds. Then, for $f\in \Cylk$, we let $\Cylsk\supseteq \{f_n\}_{n\in \mathbb{N}}\rightarrow f$ be a converging sequence, and obtain $\phi_g^*f_n\rightarrow \phi_g^*f\in \Cylk$, because $\phi_g^*$ is an isometry. \hspace*{\fill} {\scriptsize $\blacksquare$}

\subsection{The proof of a claim made in Section \ref{Intro}}
\begin{lemma}
  \label{lemma:Circular}
  If $\gamma\in \Paths_c$ is fixed, the parallel transport functions $c\mapsto \left(\pr_2\cp \parall{\delta}{\w^c}\right)_{ij}$ for $1\leq i,j\leq 2$ along the curves $\delta \in \{\gamma,\Paths_\lin\}$, already generate the $\Cstar$-algebra $C_0(\RR)\oplus \CAP(\RR)$.
  \end{lemma}
  \begin{proof}
  We have $\gamma=x+ \sigma(\gamma_\tau)$ for some $x\in \RR^3$, some $\sigma\in \SU$, and (cf.\ Remark \ref{rem:paralltransp}.\ref{rem:paralltransp2})
\begin{align*}
	\Paths_\mc\ni \gamma_\mm\colon [0,\mm]\rightarrow \RR,\quad t\mapsto r\cdot (\cos(t)\cdot\vec{e}_1+\sin(t)\cdot\vec{e}_2)
\end{align*}    
     for some $0<\mm<2\pi$. Thus, by invariance, the parallel transports along $\gamma$, give rise to the function
     \begin{align*}
    f\colon c\mapsto \big(\pr_2\cp \parall{\gamma_\tau}{\w^c}(\gamma_\tau(0),e)\big)_{11},
    \end{align*} 
    explicitly given by
    \begin{align*}
      \textstyle f(c)\stackrel{\eqref{eq:paralllc}}{=} \e^{-\frac{\I}{2}\mm}\left[\cos(\beta_c \mm)+\frac{i}{2\beta_c}\sin(\beta_c \mm)\right]\qquad \text{for}\qquad \beta_c=\sqrt{c^2r^2+\frac{1}{4}}.
    \end{align*}
    The unique decomposition of $f$ into the vector space direct sum $C_0(\RR) \oplus \CAP(\RR)$, is given by
    \begin{align*}
      \textstyle f(c)= \e^{-\frac{\I}{2}\mm}\left[\cos(\beta_c \mm)+\frac{i}{2\beta_c}\sin(\beta_c \mm)- \cos(r\tau c)\right] \oplus \e^{-\frac{\I}{2}\mm}\cos(r\tau c).
    \end{align*} 
    Now, since the first summand $f_0\in C_0(\RR)$ vanishes nowhere, the functions $\{f_0\cdot \chi_l\}_{l\in \RR}$ separate the points in $\RR$; hence, generate $C_0(\RR)$ by  the complex version of the Stone-Weierstrass theorem for locally compact Hausdorff spaces. Thus, since the functions $\chi_l$ arise from parallel transports along linear curves, the claim follows, cf. Remark \ref{rem:paralltransp}.\ref{rem:paralltransp1}.
  \end{proof}

\subsection{Homogeneous isotropic connections}
\label{subsec:InvIsoHomWang}
Let us recall the following result from Wang. \cite{Wang}
\vspace{6pt}

\noindent
{\bf Theorem (Hsien-Chung Wang)}
\vspace{2pt}

\noindent
Let $(\theta,G)$ be a Lie group of automorphisms of the principal fibre bundle $(P,\pi,M,S)$, such that the induced action $\IndA$ is transitive. Then, for each $p\in P$, the map $\w\mapsto \theta_p^*\w$ is a bijection between the $G$-invariant connections on $P$, and the linear maps $L\colon \mathfrak{g}\rightarrow \mathfrak{s}$ that fulfil
\begingroup
\setlength{\leftmargini}{16pt}
\begin{enumerate}
\item
\label{opdfopdfodpfdf1}
  $L(\vec{j}\:)=\dd_e\psi_p(\vec{j}\:)$ \hspace{55pt}for all $\vec{j}\in \mathfrak{g}_{\pi(p)}$,
\item
\label{opdfopdfodpfdf2}
  $L\cp\Ad(j)=\Ad(\psi_p(j))\cp L$\hspace{10pt} for all $j\in G_{\pi(p)}$.
\end{enumerate}
\endgroup
\noindent
Here, the homomorphism $\psi_p\colon  G_{\pi(p)}\rightarrow S$ is determined by $\theta(j,p)=p\cdot \psi_p(j)$, and $\mathfrak{g}_{\pi(p)}$ denotes the Lie algebra of the stabilizer $G_{\pi(p)}$ of $\pi(p)$ w.r.t.\ the induced action $\IndA$ on $M$.
\hspace*{\fill} {\scriptsize $\blacksquare$}\newline
\vspace{-2ex}
\newline
Now, assume that we are in the situation of Example \ref{ex:LQC}, i.e., that we have $P=\RR^3\times \SU$ and $G=\Ge=\Gee$. We now show that the (obviously smooth) connections 
\begin{align*}
  \w^c_{(x,s)}(\vv_x,\vec{\sigma}_s)= c\: \Ad(s^+)[\hspace{1pt}\murs(\vv_x)]+s^+\vec{\sigma}_s \qquad (\vv_x,\vec{\sigma}_s)\in T_{(x,s)}P 
\end{align*}
for $c\in \RR$, are exactly the $E$-invariant ones.

First, to verify their $E$-invariance, let $(x,s)\in P$, $(v,\sigma)\in E$, and $(\vv_x,\vec{\sigma}_s)\in T_{(x,s)}P$. Then, $\dd_{(x,s)}\theta_{(v,\sigma)}(\vv_x,\vec{\sigma}_s)= (\uberll{\sigma}{\vv_x},\sigma\cdot \vec{\sigma}_s)$ holds, so that we have
\begin{align}
\label{pioisdfiopfdsoifpd}
\begin{split}
  (\theta_{(v,\sigma)}^*\w^{c})_{(x,s)}(\vv_x,\vec{\sigma}_s)&=\w^{c}_{(v+\uberll{\sigma}{x},\sigma\cdot s)}(\dd_{(x,s)}\theta_{(v,\sigma)}(\vv_x,\vec{\sigma}_s))\\
  &=c \Ad(s^+\sigma^+)\left[\hspace{1pt}\murs\cp \uberll{\sigma}{\vv_x}\right]+ s^+\sigma^+\sigma\cdot\vec{\sigma}_s\\
  &=c\hspace{2pt} (\Ad(s^+)\cp \Ad(\sigma^+)\cp \Ad(\sigma)\cp \murs)(\vv_x) + s^+ \vec{\sigma}_s\\
  &=c \Ad(s^+)[\hspace{1pt}\murs(\vv_x)]+s^+ \vec{\sigma}_s\\
  &=\hspace{0.5pt}\w^{c}_{(x,s)}(\vv_x,\vec{\sigma}_s).
  \end{split}
\end{align}
It remains to show that each $E$-invariant connection equals $\w^c$ for some $c\in\RR$. For this observe that 
\begin{align*}
	\theta_{(0,e)}^*\w^c(\vv,\vec{s})=c\:\murs(\vv)+\vec{s}
\end{align*}
holds by \eqref{pioisdfiopfdsoifpd}, so that we only have to show that there are no other linear maps $L\colon \RR^3\times \mathfrak{su}(2)\rightarrow \mathfrak{su}(2)$ that fulfil \ref{opdfopdfodpfdf1})  and \ref{opdfopdfodpfdf2}) in Wang's theorem. 

For this, observe that $\pi(p)=0$ and $\Ge_{0}=\{0\} \times \SU$ holds, so that for $j=(0,\sigma)\in \Ge_{0}$, we have 
\begin{align*}
j\cdot p=(0,\sigma)\cdot(0,e)=(0, \sigma)=p\cdot \sigma,
\end{align*}
hence $\psi_p(j)=\sigma$. 
Thus, if $L\colon \RR^3\times \mathfrak{su}(2)\rightarrow \mathfrak{su}(2)$ is a linear map as in Wang's theorem, then,
together with Condition \ref{opdfopdfodpfdf1}), this gives $L(\vec{s})=\dd_e\psi_p(\vec{s})=\vec{s}$. Thus, it remains to show that $L(\vv)=c\:\murs(\vv)$ holds for some $c\in \RR$, and each $\vv\in\RR^3$. Due to Condition \ref{opdfopdfodpfdf2}), we have
\begin{align*}
	L(\vv)=[\Ad(\sigma^+)\cp L\cp\Ad(\sigma)](\vv)\qquad\forall\: \sigma\in \SU,
\end{align*}
whereby 
\begin{align*}
  \Ad(\sigma)(\vv)&=\dt{t}{0}(0,\sigma)\cdot_\uberl(t \cdot\vv,e)\cdot_\uberl(0,\sigma)^{-1} =\dt{t}{0}(\uberll{\sigma}{t\cdot \vv},\sigma)\cdot_\uberl(0,\sigma^+)\\ &=\dt{t}{0}(\uberll{\sigma}{t\cdot \vv},\sigma\cdot\sigma^+)=(\uberll{\sigma}{\vv},0),
\end{align*}
hence $L(\vv)=(\Ad(\sigma^+)\cp L)(\uberll{\sigma}{\vv})$ for each $\vv\in \RR^3$, and each $\sigma\in \SU$. 
Then, for $\sigma_t:=\exp(t\cdot \vec{s})$ with $\vec{s}\in \mathfrak{su}(2)$, it follows from linearity of $L$ that
\begin{align*}
  0&=\dt{t}{0}L(\vv)
  =\dt{t}{0}(\Ad(\sigma^+_t)\cp L)(\uberll{\sigma_t}{\vv})
  =\dt{t}{0}\sigma^+_t\cdot(L\cp \murs^{-1})(\sigma_t\cdot\murs(\vv)\cdot\sigma^+_t)\cdot\sigma_t\\[3pt]
  &\hspace{-1.8pt}\stackrel{\text{lin.}}{=}-\vec{s}\cdot L(\vv)+ (L\cp \murs^{-1})\left(\vec{s}\cdot\murs(\vv)-\murs(\vv)\cdot\vec{s}\right)  +L(\vv)\cdot\vec{s}.
\end{align*}
Consequently, $[\vec{s},L(\vv)]=(L\cp \murs^{-1})\left([\vec{s},\murs(\vv)]\right)$ holds for all $\vv\in \RR^3$, and all $\vec{s}\in \mathfrak{su}(2)$, so that for $1\leq i,j,k\leq 3$, we have
\begin{align*}
  [\tau_i,L(\vec{e}_j)]=(L\cp \murs^{-1})([\tau_i,\tau_j])=2\epsilon_{ijk}(L\cp \murs^{-1})(\tau_k)=2\epsilon_{ijk}L(\vec{e}_k).
\end{align*} 
This enforces $L(\vv)=c\:\murs(\vv)$ for some $c\in\RR$, and all $\vv\in \RR^3$,
hence
\begin{align*}
  L(\vv,\vec{s})=L(\vv)+L(\vec{s})=c\:\murs(\vv)+\vec{s}. 
\end{align*}

\section{Homomorphisms of Paths}
\label{sec:HomPaths}
In this section, we will prove slight variations of statements from  \cite{Ashtekar2008} that we need in the main text. 
So, for the rest of this section, let $(P,\pi,M,S)$ be a principal fibre bundle with compact structure group $S$, and $\nu=\{\nu_x\}_{x\in M}$ a section as in Definition \ref{def:Connectionss}. 
\begin{definition}
  Let $\gamma\colon [a,b]\rightarrow M$ be a curve.
  \begingroup
\setlength{\leftmargini}{12pt}
  \begin{itemize}
  \item
    The inverse curve of $\gamma$ is defined by $\gamma^{-1}\colon [a,b]\ni t\mapsto \gamma(b+a-t)$.  
  \item
    A decomposition of $\gamma$ is a family of curves $\{\gamma_i\}_{0\leq i\leq k-1}$, 
    such that $\gamma|_{[t_i,t_{i+1}]}=\gamma_i$ holds for $0\leq i\leq k-1$, for real numbers $a=t_0<{\dots}<t_k=b$. 
  \end{itemize}
  \endgroup
\end{definition}
In the following, let $\Paths$ be a set of $\CC{k}$-paths, such that $\gamma\in \Paths$ implies $\gamma^{-1}\in \Paths$, and that for each decomposition $\{\gamma_i\}_{0\leq i\leq k-1}$ of $\gamma$, we have $\gamma_i\in \Paths$ for all $0\leq i\leq k-1$. Then,
\begin{definition}
  \label{def:hompaths}
    \begingroup
\setlength{\leftmargini}{12pt}
  \begin{itemize}
  \item
    For $x,x'\in M$, let $\Iso(x,x')$ denote the set of morphisms $\varphi\colon F_x\rightarrow F_{x'}$, and $\AF:=\bigsqcup_{x,x'\in M}\Iso(x,x')$.
  \item
    Define the equivalence relation $\csim$ on $\Paths$ by $\gamma\csim \gamma'$ 
    iff $\parall{\gamma}{\w}=\parall{\gamma'}{\w}$ holds for all $\w\in \Con$; and observe that $\gamma\not\csim \gamma'$ holds if the start and end points of $\gamma$ and $\gamma'$ do not coincide. 
  \item
    Let $\Homm(\Paths,\AF)$ denote the set of all maps $\homm\colon \Paths \rightarrow \AF$, such that for all $\gamma\in \Paths$, we have:
        \begingroup
\setlength{\leftmarginii}{18pt}
    \begin{enumerate}
    \item[\textrm{(a)}]
      $\homm(\gamma)\in \Iso(\gamma(a),\gamma(b))$, 
      and $\homm(\gamma)=\id_{F_{\gamma(a)}}$ holds if $\gamma$ is constant,
    \item[\textrm{(b)}]
      $\homm(\gamma)= \homm(\gamma_{k-1})\cp \dots\cp \homm(\gamma_{0})$ if $\{\gamma_i\}_{0\leq i\leq k-1}$ is a decomposition of $\gamma$,
    \item[\textrm{(c)}]
      $\homm(\gamma^{-1})=\homm(\gamma)^{-1}$,
    \item[\textrm{(d)}]
      $\homm(\gamma')=\homm(\gamma)$ holds if 
      $\gamma'\csim \gamma$.
    \end{enumerate}
    \endgroup
  \end{itemize}
  \endgroup
  \end{definition}
  Moreover,
  \begin{definition}
  \label{dfiodsfiuosdf}
      \begingroup
\setlength{\leftmargini}{12pt}
    \begin{itemize}
    \item
      A refinement of a finite subset $\{\gamma_1,\dots,\gamma_l\}\subseteq \Paths$ is a finite collection $\{\delta_1,\dots,\delta_n\}\subseteq \Paths$, such that each $\gamma_j$ admits a decomposition $\{(\gamma_j)_i\}_{1\leq i\leq k_j}$, such that each $(\gamma_j)_i$ is equivalent to one of the paths $\delta_r,\delta_r^{-1}$ for some $1\leq r\leq n$.
    \item
      A family $\{\delta_1,\dots,\delta_n\}\subseteq \Paths$ is said to be independent iff for each collection $\{s_1,\dots,s_n\}\subseteq S$, there exists some $\w\in \Con$, such that $h^\nu_{\w}(\delta_i)= s_i$ holds for all $1\leq i\leq n$. Due to \eqref{eq:indep}, this definition does not depend on the explicit choice of $\nu$.
    \item
      $\Paths$ is said to be independent iff each finite collection $\{\gamma_1,\dots,\gamma_l\}\subseteq \Paths$ 
      admits an independent refinement.
    \end{itemize}
    \endgroup
  \end{definition}
Let $\cC_\Paths$ denote the $\Cstar$-algebra of cylindrical functions that corresponds to the set of curves $\Paths$, and denote it spectrum by $\A$. Then, we have
\begin{lemma}
  \label{lemma:SpeczuHomm}
  Let $\ovl{\w}\in \ovl{\Con}$, and $\{\w_\alpha\}\subseteq \Con$ be a net with $\{\iota_\Con(\w_\alpha)\}_{\alpha\in I}\rightarrow \ovl{\w}$. Then,\footnote{Here, we have $\lim_\alpha \parall{\gamma}{\w_\alpha}(\cdot)\colon p\mapsto \lim_\alpha \parall{\gamma}{\w_\alpha}(p)$ for $p\in F_{\gamma(a)}$ and $\dom[\gamma]=[a,b]$.}
  \begin{align*}
  	\kappa\colon \ovl{\Con} &\rightarrow \Homm(\Paths,\AF)\\
  	\ovl{\w}&\mapsto \textstyle\left[\gamma \mapsto \lim_\alpha \parall{\gamma}{\w_\alpha}(\cdot)\right] 
\end{align*}
is well defined and injective, as well as surjective if $\Paths$ is independent.
\end{lemma}
  \begin{proof}
    Let $\rho$ be a faithful\footnote{By compactness of $S$, such a representation exists. \cite{BroeckerDiek1985}} matrix representation of $S$, and $B:=\im[\rho]\subseteq M_n(\CCC)$.  
    Then, $\rho$ is a homeomorphism to $B$ equipped with the relative topology, because $S$ is compact and $B$ is Hausdorff. 
    Moreover, for $[h_\gamma]:=\rho\cp h_\gamma^\nu$, we have $\lim_\alpha(\iota_\Con(\w_\alpha)([h_\gamma]_{ij}))_{ij}=(\ovl{\w}([h_\gamma]_{ij}))_{ij}\in B$ by compactness (closedness) of $B$. Then, using continuity of $\rho^{-1}$ and that of the right multiplication in the bundle $P$, we obtain
    \begin{align}      
      \label{eq:algpro}
      \begin{split}
        \textstyle\lim_\alpha \parall{\gamma}{\w_\alpha}(p)&
        =\textstyle\lim_\alpha \nu_{\gamma(b)}\cdot (\psi_{\gamma(b)}\cp\parall{\gamma}{\w_\alpha})(\nu_{\gamma(a)})\cdot \psi_{\gamma(a)}(p)\\  
        &=\textstyle\lim_\alpha \nu_{\gamma(b)}\cdot h_\gamma^\nu(\w_\alpha)\cdot \psi_{\gamma(a)}(p)\\
        &=\textstyle\lim_\alpha\nu_{\gamma(b)}\cdot \rho^{-1}\big((\iota_\Con(\w_\alpha)([h_\gamma]_{ij}))_{ij}\big)\cdot \psi_{\gamma(a)}(p)\\
        &=\textstyle\nu_{\gamma(b)}\cdot \rho^{-1}\big((\ovl{\w}([h_\gamma]_{ij}))_{ij}\big)\cdot \psi_{\gamma(a)}(p).
      \end{split}
    \end{align}
    This shows that the limes exists, and that it is independent on the choice of the net $\{\w_\alpha\}_{\alpha\in I}$. 
    By the same arguments, we conclude from $\kappa(\im[\iota_\Con])\subseteq \Homm(\Paths,\AF)$ that $\kappa(\ovl{\w})\in \Homm(\Paths,\AF)$ holds.
    
    Let $\dD$ denote the unital $^*$-subalgebra of $\cC_\Paths$ that is generated by the functions $[h_\gamma]_{ij}$. By injectivity of $\rho$ and the Stone-Weierstrass theorem, the functions $\rho_{ij}$ are dense in $C(S)$, and it is then straightforward to see that this implies denseness of $\dD$ in $\cC_\Paths$. Consequently, if $\kappa(\ovl{\w}_1)=\kappa(\ovl{\w}_2)$ holds for $\ovl{\w}_1,\ovl{\w}_2\in \Spec(\cC)$, we have $\ovl{\w}_1|_\dD=\ovl{\w}_2|_\dD$ by \eqref{eq:algpro}, hence $\ovl{\w}_1=\ovl{\w}_2$ by continuity of these maps. 
     
    Now, if $\Paths$ is independent, and $\homm\in \Homm(\Paths,\AF)$, we define its  preimage $\ovl{\w}_\homm$ under $\kappa$, by
\begin{align*}   
       \ovl{\w}_\homm(1):=1\qquad\quad \ovl{\w}_\homm([h_\gamma]_{ij}):=(\rho_{ij}\cp \psi_{\nu_{\gamma(b)}}\cp\homm(\gamma))(\nu_{\gamma(a)})\qquad\quad \ovl{\w}_\homm([h_\gamma]^*_{ij}):=\ovl{\ovl{\w}_\homm([h_{\gamma}]_{ij})}. 
 \end{align*}
  Moreover, for arbitrary $f\in \dD$, we choose a representation as a finite sum of products of the form 
  \begin{align*}
    \textstyle F=\sum_i \lambda_i\cdot h_{j_1}^i\cdot {\dots}\cdot h_{j_{n_i}}^i
    \end{align*} 
    with $\lambda_i \in \CCC$, and where each
    $h^i_{j_l}$ equals $1$, a generator $[h_\gamma]_{ij}$, or the complex conjugate $[h_\gamma]^*_{ij}$ of a generator. Then, we assign to $f$ the value
    \begin{align*}
    \textstyle\ovl{\w}_\homm(F):=\sum_i\lambda_i\cdot \ovl{\w}_\homm(h^i_{j_1})\cdot {\dots}\cdot \ovl{\w}_\homm(h^i_{j_{n_i}}).
    \end{align*}
    Obviously, $\ovl{\w}_\homm$ is a $^*$-homomorphism iff it is well defined, and extends by linearity to an element in $\Spec(\cC)$ iff it is continuous. Now, for well-definedness, let $F_1$,$F_2$ be two representations of $f$,
    and denote by $[\gamma_1],\dots, [\gamma_m]$ the equivalence classes of the paths occurring in both expressions. 
    Let $\{\delta_1,\dots,\delta_n\}\subseteq \Paths$ be a refinement
    of $\{\gamma_1,\dots,\gamma_m\}$, as well as $\w\in\Con$ with\footnote{For simplicity, assume that $\dom[\delta_r]=[0,1]$ holds for $1\leq r\leq n$.} $h^\nu_{\delta_r}(\w)=(\psi_{\nu_{\delta_r(1)}}\cp\homm(\delta_r))(\nu_{\delta_r(0)})$ for $1\leq r\leq n$. Then, it follows from the algebraic properties of parallel transports, and that of $\homm$ that 
    \begin{align*}    
      \ovl{\w}_\homm(F_1)=F_1(\w)=f(\w)=F_2(\w)=\ovl{\w}_\homm(F_2)
    \end{align*}    	
    holds. 
    This shows that $\ovl{\w}_\homm$ is well defined, and its continuity is now clear from $|\ovl{\w}_\homm(f)|=|f(\w)|\leq \|f\|_\infty$.
   Since, by construction we have $\kappa(\ovl{\w}_\homm)=\homm$,  surjectivity of $\kappa$ follows.
\end{proof}
For the rest of this section, let $M$ be analytic, and $\Paths_\w$ the set of embedded analytic curves in $M$ defined on some compact interval. Then, for $\gamma\in \Paths_\w$, its inverse in the sense of mappings will be denoted by $[\gamma]^{-1}$. Moreover, for $\gamma,\gamma'\in \Paths_\w$, we define
  \begingroup
\setlength{\leftmargini}{12pt}
  \begin{itemize}
  \item
	  $\gamma\psim \gamma'$\quad iff\quad$\gamma=\gamma'\cp\rho|_{[\dom[\gamma]}$ holds for some analytic diffeomorphism $\rho\colon I\rightarrow I'\subseteq \RR$ with $\dot\rho>0$, and 
	  
	  \hspace{50pt}\qquad$\rho(\dom[\gamma])=\dom[\gamma']$. 
  \item
$\gamma\isim\gamma'$\quad iff\quad $\im[\gamma]=\im[\gamma']$, as well as $\gamma(a)=\gamma(a')$ and $\gamma(b)=\gamma(b')$ holds for $\dom[\gamma]=[a,b]$, and 

	  \hspace{56pt}\quad$\dom[\gamma']=[a',b']$. 
  \end{itemize}
  \endgroup
  \noindent
  Then,
\begin{lemma}
  \label{lemma:analytCurvesIndepetc}
  \begingroup
\setlength{\leftmargini}{16pt}
  \begin{enumerate}
  \item
    \label{lemma:analytCurvesIndepetc1}
Let $\gamma\colon I\rightarrow M$, $\gamma'\colon I'\rightarrow M$ be analytic embeddings, and $I\backslash\{t\}\supseteq \{t_n\}_{n\in \NN}\rightarrow t\in I$ as well as $I'\backslash\{t'\}\supseteq \{t'_n\}_{n\in \NN}\rightarrow t'\in I'$ sequences with $\gamma(t_n)=\gamma'(t_n')$ for each $n\in \NN$. Then, $\gamma|_{J}=\gamma'\cp \rho$ holds for some analytic diffeomorphism $\rho\colon J\rightarrow J'\subseteq I'$ with $\rho(t)=t'$.    
  \item
    \label{lemma:analytCurvesIndepetc2}
    For $\gamma,\gamma' \in \Paths_\w$, let $x$ be an accumulation point of $\im[\gamma]\cap \im[\gamma']$ with\footnote{Since $\im[\gamma]\cap \im[\gamma']$ is compact, $x$ is contained therein.} $\gamma(t)=x=\gamma'(t')$ for $t\in \dom[\gamma]$  and  $t'\in \dom[\gamma']$. Then, $\gamma=\gamma'\cp\rho|_{\dom[\gamma]\cap J}$ holds for some analytic diffeomorphism $\rho\colon J\rightarrow J'$ with $\rho(t)=t'$.  
  \item
    \label{lemma:analytCurvesIndepetc3}
    If $\dim[S]\geq 1$ holds, we have\qquad $\gamma\csim \gamma'$\qquad $\Longleftrightarrow$\qquad  $\gamma\isim \gamma'$\qquad $\Longleftrightarrow$\qquad  $\gamma\psim \gamma'$.
  \item
    \label{lemma:analytCurvesIndepetc4}
    For $\gamma,\gamma'\in \Paths_\w$, the preimage $[\gamma]^{-1}(\im[\gamma]\cap \im[\gamma'])$ is the disjoint union of finitely many isolated points, and $0\leq m\leq 2$ compact intervals $\{L_k\}_{1\leq k\leq m}$.
  \item
    \label{lemma:analytCurvesIndepetc5}
   If $S$ is connected with $\dim[S]\geq 1$, and  $\Paths\subseteq \Paths_\w$ is closed under decomposition and inversion of its elements, then $\Paths$ is independent.
  \end{enumerate}
  \endgroup
\end{lemma}  
  \begin{proof} 
  \begingroup
\setlength{\leftmargini}{16pt}
    \begin{enumerate}
    \item
	Let $(U,\psi)$ be an analytic submanifold chart of $\im[\gamma']$, which is centered at $x:=\gamma'(t')=\gamma(t)$, and maps $\im[\gamma']\cap U$ into the $x_1$-axis. We choose an open interval $J\subseteq I$ with $t\in J$ and $\gamma(J)\subseteq U$, and consider the analytic functions $f_{\zd}:=\psi^{\zd}\cp\gamma|_{J}$ for $k=2,\dots,\dim[M]$. Then, $t$ is an accumulation point of zeroes of each $f_{\zd}$, so that $f_{\zd}=0$ holds by analyticity. Thus, we have $\psi(\gamma(J))\subseteq \psi(\im[\gamma']\cap U)$; and since $[\gamma']^{-1}\cp \psi^{-1}|_{\psi(U\cap\im[\gamma'])}$ and $\psi\cp \gamma|_{J}$ are analytic immersive, 
we can just define $\rho:=[\gamma']^{-1}\cp \gamma|_{J}$.    
    \item
    This is clear from the first part, when applied to embedded analytic extensions of $\gamma$ and $\gamma'$.
    \item
    Write $\dom[\gamma]=[a,b]$ and $\dom[\gamma']=[a',b']$. Then, the second of the implications
    \begin{align*}
    	 \gamma\isim \gamma'\qquad\Longrightarrow\qquad \gamma\psim \gamma' \qquad\qquad\text{and}\qquad\qquad
    	\gamma\psim \gamma'\qquad\Longrightarrow\qquad \gamma\csim \gamma'
    \end{align*}
    is clear, because parallel transports are invariant under $C^1$-reparametrizations of curves. Moreover, if $\gamma\isim \gamma'$ holds, we can apply Part \ref{lemma:analytCurvesIndepetc1}) to embedded analytic extensions $\wt{\gamma}$ and $\wt{\gamma}'$ of $\gamma$ and $\gamma'$, respectively, in order to find open intervals $J,J'$ with $\dom[\gamma]\subseteq J$ and $\dom[\gamma']\subseteq J'$, such that $\wt{\gamma}(J)=\wt{\gamma}'(J')$ holds. Then, $\rho:=[\wt{\gamma}']^{-1}\cp \wt{\gamma}|_J$ is an analytic diffeomorphism  with $\gamma=\gamma'\cp\rho|_{\dom[\gamma]}$.
    
    Finally, if $\gamma\csim \gamma'$ holds, we have $\gamma(a)=\gamma'(a')$ and $\gamma(b)=\gamma'(b')$ by definition. Then, 
       $\im[\gamma]\neq \im[\gamma']$ implies $\gamma(t)\notin \im[\gamma']$ or $\gamma'(t')\notin \im[\gamma]$ for some $t\in [a,b]$ or some $t'\in [a',b']$,  respectively. In the first case, (the second one follows analogously) by compactness of $\im[\gamma']$, we find $\epsilon >0$, such that $\gamma(B_\epsilon(t))\cap \im[\gamma']=\emptyset$ holds. We fix $\w\in \Con$, define $s:=h_{\gamma'}^\nu(\w)$, and  choose $s'\neq s$ (we have $\dim[S]\geq 1$ by assumption).     
        Then, by Proposition A.1 in \cite{ParallTranspInWebs}, we find $\w'\in\Con$ with $h_\gamma^{\nu}(\w')=s'$, such that $\w'$ equals $\w$ outside $B_\epsilon(\gamma(t))$. Thus, we have 
\begin{align*}
	  h_{\gamma}^\nu(\w')=s'\neq s=h_{\gamma'}^\nu(\w)=h_{\gamma'}^\nu(\w')\qquad \Longrightarrow\qquad \gamma \nsim_\Con \gamma',
\end{align*}        
      which contradicts the assumption.
    \item
      The statement is clear if $T:=\im[\gamma]\cap \im[\gamma']$ is finite. In the other case, there are accumulation points $\{x_\alpha\}_{\alpha \in I}\subseteq T$ by compactness of $T$. By the second part, for each $\alpha\in I$, we find an interval $L_\alpha\subseteq \dom[\gamma]$ with $x_\alpha\in \gamma(L_\alpha)\subseteq T$. Then, if we let $L_\alpha$ be maximal w.r.t.\ this property, it must be closed in $\dom[\gamma]$ by compactness of $T$ and continuity of $\gamma$. Obviously, then the claim is clear if $L_\alpha=\dom[\gamma]$ holds. 
      
      In the other case, 
   the compact interval $[\gamma']^{-1}(\gamma(L_\alpha))$ must contain at least one boundary point of $\dom[\gamma']$, which is immediate from  Part \ref{lemma:analytCurvesIndepetc2}) as well as maximality of $L_\alpha$. Then, since $\dom[\gamma']$ only admits two boundary points, and since 
   \begin{align*}
    L_{\alpha'}\cap L_\alpha\neq \emptyset\qquad\Longrightarrow\qquad L_{\alpha'}=L_\alpha\qquad \forall\:\alpha'\in I
   \end{align*}
   holds by maximality of $L_\alpha$, the claim is clear.
    \item
      It follows from Proposition A.1 in \cite{ParallTranspInWebs} that a set $\{\delta_1,\dots,\delta_n\}\subseteq \Paths_\w$ is independent if $\im[\delta_i]\cap \im[\delta_j]$ is finite for all $1\leq i\neq j\leq n$. Thus, it suffices to show that each collection of the form $\{\gamma , \sigma_1,\dots,\sigma_l\} \subseteq \Paths$ with $\im[\sigma_i]\cap \im[\sigma_j]$ finite for all $1\leq i\neq j\leq l$, admits a refinement $\{\delta_1,\dots,\delta_n\}$, such that $\im[\delta_i]\cap \im[\delta_j]$ is finite for all $1\leq i\neq j\leq n$ as well. 

In fact, if  $\{\gamma_1,\dots,\gamma_d\}$ is given, then we can apply this to $\gamma:=\gamma_1$ and $\sigma_1:=\gamma_2$, in order to obtain a refinement of $\{\gamma_1,\gamma_2\}$ with the desired intersection property. Thus, we can assume that  for $1<c< d$, we are given a refinement 
      $\{\delta_1,\dots,\delta_l\}$ of $\{\gamma_1,\dots, \gamma_c\}$, such that $\im[\delta_i]\cap\im[\delta_j]$ is finite for all $1\leq i\neq j\leq l$. Now, by assumption, there exists a refinement $\{\delta'_1,\dots ,\delta'_{n}\}$ of $\{\gamma_{c+1},\delta_1,\dots \delta_l\}$, such that $\im[\delta'_i]\cap\im[\delta'_j]$ is finite for all $1\leq i \neq j\leq n$. Then, using the fact that the equivalence relations $\csim$ and $\psim$ coincide, it is easy to see that $\{\delta'_1,\dots \delta'_{n}\}$ is a refinement of $\{\gamma_{c+1},\gamma_1,\dots,\gamma_{c}\}$ as well. 
      
      Now, if $\{\gamma , \sigma_1,\dots,\sigma_l\} \subseteq \Paths$ is as above,  for each pair $(\gamma,\sigma_p)$, we let $\{L^p_k\}_{1\leq k\leq m_p}$ denote the intervals in $\dom[\gamma]=[a,b]$, provided by Part \ref{lemma:analytCurvesIndepetc4}). Moreover, we let $K_p^k\subseteq \dom[\sigma_p]$ denote the respective intervals, for which $\gamma(L^p_k)=\sigma_p(K^p_k)$ holds. 
      
      We split each $\sigma_p$ at the boundary points of the $K_p^k$ for $1\leq k\leq m_p$, and denote by $\Delta$ the, necessarily finite, collection of subcurves of the $\sigma_p$, obtained in this way. Moreover, we choose $a=t_0<{\dots}<t_s=b$, such that each $L^p_k$ occurs exactly once as an interval $A_i=[t_i,t_{i+1}]$ for some $0\leq i\leq s-1$. This is possible, because $\sigma_p(K_k^p)\cap \sigma_q(K_{k'}^{q})$ can only be infinite if $p=q$ and $k=k'$ holds. We let $\Gamma$ denote the set of such restrictions $\gamma|_{A_i}$, for which $A_i$ is non-equal to each $L_p^k$, and observe that $\Gamma\sqcup \Delta$ is the refinement we were searching for.  
    \end{enumerate}
    \endgroup
  \end{proof}

\end{document}